\pdfoutput=1
\PassOptionsToPackage{table,dvipsnames}{xcolor}

\documentclass[12pt]{article}

\usepackage{amsmath,amssymb}
\usepackage{mathtools}

\usepackage[utf8]{inputenc}
\usepackage[T1]{fontenc}
\usepackage{newpxtext,newpxmath}

\usepackage{geometry}
\usepackage{authblk}

\usepackage{setspace}
\usepackage{parskip}
\usepackage{titlesec}

\usepackage{graphicx}
\usepackage{array}
\usepackage{tabularx}
\usepackage{threeparttable}
\usepackage{longtable}
\usepackage{booktabs}
\usepackage{multirow}
\usepackage{xcolor}          
\usepackage[most]{tcolorbox}
\usepackage{listings}
\usepackage{pifont}
\usepackage{enumitem}
\usepackage{wrapfig}
\usepackage{caption}
\usepackage{subcaption}
\usepackage{etoolbox}
\usepackage{silence}
\usepackage[super,sort&compress,numbers]{natbib}

\usepackage{url}
\usepackage{hyperref}
\hypersetup{
  colorlinks = true,
  linkcolor  = blue,
  citecolor  = blue,
  urlcolor   = blue
}
\usepackage[capitalise]{cleveref}

\newcommand{\hl}{\textcolor{black}}
\newcommand{\revise}[1]{{\leavevmode\color{black}\relax#1}}
\newenvironment{revblock}{\par\color{black}}{\par}
\AtBeginDocument{\pdfstringdefDisableCommands{\def\revise#1{#1}}}
\AtBeginDocument{\colorlet{red}{black}}

\newcommand{\radio}[1]{\text{\Large$\circ$}~#1}
\newcommand{\checkbx}[1]{$\square$~#1}

\newcommand{\ratingscalenine}[2]{%
\begin{center}
\begin{tabularx}{\textwidth}{*{9}{>{\centering\arraybackslash}X}}
\footnotesize #1 & \footnotesize 2 & \footnotesize 3 & \footnotesize 4 &
\footnotesize 5 & \footnotesize 6 & \footnotesize 7 & \footnotesize 8 &
\footnotesize #2 \\
$\circ$ & $\circ$ & $\circ$ & $\circ$ & $\circ$ & $\circ$ & $\circ$ & $\circ$ & $\circ$ \\
\end{tabularx}
\end{center}
}

\newcommand{\ratingscaleseven}[2]{%
\begin{center}
\begin{tabularx}{\textwidth}{*{7}{>{\centering\arraybackslash}X}}
\footnotesize #1 & \footnotesize 2 & \footnotesize 3 & \footnotesize 4 &
\footnotesize 5 & \footnotesize 6 & \footnotesize #2 \\
$\circ$ & $\circ$ & $\circ$ & $\circ$ & $\circ$ & $\circ$ & $\circ$ \\
\end{tabularx}
\end{center}
}

\newcommand{\MSE}{\mathrm{MSE}}
\newcommand{\Bsel}{B_{\mathrm{sel}}}
\newcommand{\neff}{n_{\mathrm{eff}}}

\newif\ifshowdisplayitems
\showdisplayitemstrue
\newcommand{\displayitem}[1]{\ifshowdisplayitems#1\fi}

\makeatletter
\newcount\refseg@count
\newcount\refseg@k
\newcount\refseg@lo
\newcount\refseg@hi
\newbox\refseg@box
\newif\ifrefseg@sink
\def\citation#1{\@for\refseg@tmp:=#1\do{%
  \@ifundefined{refseg@seen@\refseg@tmp}{%
    \global\@namedef{refseg@seen@\refseg@tmp}{}%
    \global\advance\refseg@count\@ne}{}}}
\def\refsegmark#1{\expandafter\xdef\csname refseg@N@#1\endcsname{\the\refseg@count}}
\def\refseg@prev{0}
\def\refseg@bibitem#1{%
  \ifrefseg@sink\egroup\global\refseg@sinkfalse\fi
  \global\advance\refseg@k\@ne
  \ifnum\refseg@k<\refseg@lo
    \global\advance\c@NAT@ctr\@ne\setbox\refseg@box\vbox\bgroup\global\refseg@sinktrue\@newlistfalse\everypar{}
  \else\ifnum\refseg@k>\refseg@hi
    \global\advance\c@NAT@ctr\@ne\setbox\refseg@box\vbox\bgroup\global\refseg@sinktrue\@newlistfalse\everypar{}
  \else
    \refseg@origbibitem{#1}%
  \fi\fi}
\AtBeginDocument{\pretocmd{\endthebibliography}{\ifrefseg@sink\egroup\global\refseg@sinkfalse\fi}{}{}}
\newcommand{\printrefsegment}[2]{%
  \immediate\write\@auxout{\string\refsegmark{#1}}%
  \@ifundefined{refseg@N@#1}{}{%
    \refseg@lo=\refseg@prev\relax\advance\refseg@lo\@ne
    \refseg@hi=\csname refseg@N@#1\endcsname\relax
    \xdef\refseg@prev{\the\refseg@hi}%
    \ifnum\refseg@lo>\refseg@hi\else
      \IfFileExists{\jobname.bbl}{%
        \begingroup
          \renewcommand{\refname}{#2}%
          \global\refseg@k\z@
          \let\refseg@origbibitem\bibitem
          \let\bibitem\refseg@bibitem
          \@input{\jobname.bbl}%
        \endgroup}{}%
    \fi}}
\AtBeginDocument{%
  \immediate\write\@auxout{\string\bibstyle{naturemag}}%
  \immediate\write\@auxout{\string\bibdata{main_references}}}
\AtEndDocument{\IfFileExists{\jobname.bbl}{}{\typeout{No file \jobname.bbl.}}}
\makeatother

\title{\Large\bfseries \revise{Thousands of scientists find AI rarely says no in science}}

\author[1,2]{Honglin Bao\textsuperscript{\ding{41}}}
\author[1,2]{Siyang Wu}
\author[1,2]{Xiao Liu}
\author[1]{Sida Li}
\author[3]{\\Shiyun Cao}
\author[1,2]{James A.~Evans\textsuperscript{\ding{41}}}
\affil[1]{Data Science Institute, University of Chicago}
\affil[2]{Knowledge Lab, University of Chicago}
\affil[3]{Department of Psychology, Harvard University}

\begin{document}
\date{\vspace{-1em}}
\maketitle

\begin{quote}
\ding{41} Correspondence to: Honglin Bao
(\href{mailto:honglinbao@uchicago.edu}{honglinbao@uchicago.edu});
James A.~Evans (\href{mailto:jevans@uchicago.edu}{jevans@uchicago.edu}).
\end{quote}

\vspace{1em}

\noindent
\textbf{\revise{Claims that artificial intelligence (AI) will accelerate discovery outpace evidence from working scientists \citep{wang2023scientific,evans2025after,wang2026ai}. Here we invited authors of 121,640 preprints in biology, medicine, chemistry and social science to evaluate hypotheses that large language models (LLMs) generated from puzzles in their own papers. From 6,749 participants and 25,139 rating sets on novelty, feasibility, probability of being true and willingness to pursue, three broad patterns emerge. First, every model proposed substantive null hypotheses—negations of expected effects—far less often than humans did. Non-reasoning LLMs converged on similar hypotheses \citep{jiang2025artificial}, while reasoning models broadened them, indistinguishable from human authors in diversity. The wider literature mirrors this: across 39 million papers from 2010 to 2025, the share lacking null statements rose after ChatGPT's release, and the field's center drifted further while its spread contracted. Second, scientists favored probable, familiar hypotheses over novel ones, and senior respondents were less willing to pursue any LLM hypothesis regardless of its rated quality \citep{cui2022aging}. LLM hypotheses scored lowest in the social sciences \citep{ruan2026evaluating}, likely reflecting weaker performance in pluralistic fields \citep{lamers2021disagreement}. Third, widely used novelty measures agreed only weakly with experts, and LLM judges' ratings clustered narrowly in the upper-moderate range, strongly favoring LLM-generated hypotheses for novelty. A Qwen3-14B reward model trained on expert preferences outperformed state-of-the-art judge models by up to 27\%, exceeded inter-reviewer agreement, and revealed field-specific nuances in scientific taste. Today's AI rarely says no in science: it seldom proposes absence or rejects weak ideas, so its ideation and judgment still benefit from human grounding.}}
\newpage
\bigskip
\noindent
The scientific community is being asked, with unusual urgency, to reorganise
discovery around artificial intelligence. Frontier laboratories now use LLMs
to mine the literature, generate hypotheses, design experiments, and
prioritise research directions \citep{wang2023scientific}. The promise is
that machines can navigate vast structured and unstructured knowledge to propose novel, feasible, and probable ideas at scales no human can match. The risk is that fluent text
substitutes for genuine insight: an idea may sound original because the
phrasing is unfamiliar, while still failing to extend, contradict or
clarify what is known \citep{zhang2025noveltybench,kusumegi2025scientific}. Whether AI is
accelerating discovery, or merely its appearance, cannot
be settled by automated benchmarks alone, because those benchmarks
themselves remain unvalidated against \revise{expert judgment, which is itself one of several forms of evidence, together with actual adoption, empirical validation and longer-term scholarly impact, by which a contribution should be assessed}. Existing algorithmic studies have been evaluated on small groups of scientists\revise{ \citep{si2025ideas}}, and a large-scale diagnostic study is still lacking.

\begin{figure}[htbp]
    \centering
    \displayitem{\includegraphics[width=1\textwidth]{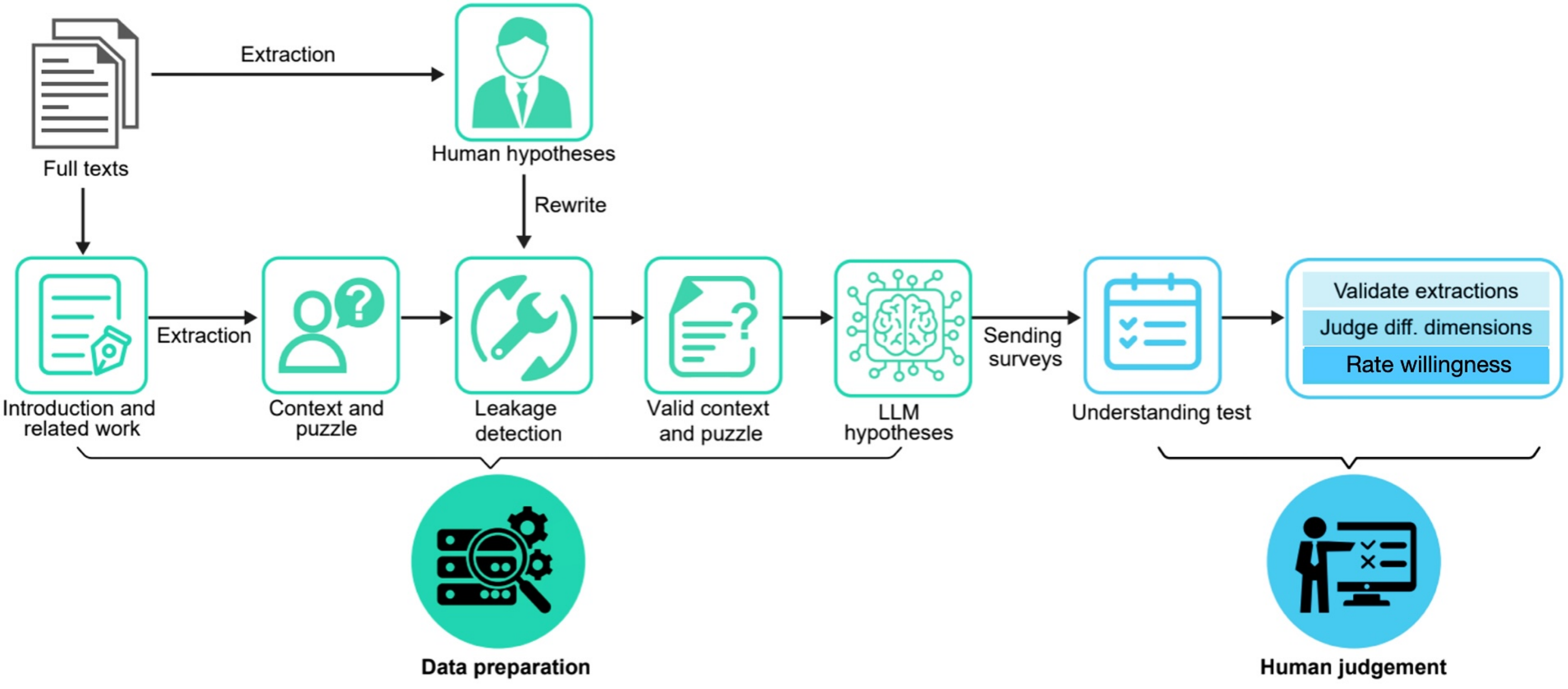}}
    \caption{\textbf{An expert-audit pipeline for AI-generated research
    ideas.} Full-text preprints ($n$ = 121{,}640) from six non-arXiv platforms
    feed an extraction stage that recovers (i) the author's hypotheses,
    (ii) the surrounding factual context, and (iii) the core scientific
    puzzle, with paraphrase-based leakage detection between (i), (ii), and (iii). LLMs propose hypotheses from the context and puzzle alone\revise{, each proposing as many hypotheses as the authors stated}; a custom set of hypotheses is sent to the authors by email, who pass
    an understanding test before rating each on the quality of extraction, novelty, empirical
    feasibility, probability of being true, and \revise{willingness to pursue it. Of 6,749 participants, 5,259 contributed 25,139 usable evaluation sets}.}
    \label{fig:pipeline}
\end{figure}

To resolve this empirically, we returned to the source. We asked scientific authors
to evaluate AI-generated hypotheses derived from their own recent papers, on the
premise that authors are the most informed---and most motivated---judges of
ideas that immediately extend their own work. We assembled 121,640 full-text preprints posted after 2023 across BioRxiv (biology, 68\%), MedRxiv (medicine, 20\%), SocArXiv,
PsyArXiv, EdArXiv (social science, jointly 9\%) and ChemRxiv (chemistry, 3\%)\revise{; preprints are the only practical way to reach researchers about their own recent work at scale, because publisher copyright restricts the use of published versions}. \revise{We} deliberately
excluded arXiv\revise{ for two reasons. Its full text is a standard component of LLM pretraining corpora}
\citep{touvron2023llama,soldaini2024dolma,gao2020pile,weber2024redpajama,kandpal2025common,wolfram2025layers,wu2025mapping}\revise{, so models could reproduce authors' hypotheses from memory rather than generate them independently, a risk shared by much prior work on hypothesis generation \citep{lu2026rethinking,sprueill2024chemreasoner}; and most arXiv papers are in computing and mathematics, where work is rarely framed as hypothesis testing \citep{cockburn2020threats,denning2013science}} (\hyperref[sec:matmethods]{Methods} and Supplementary Information \ref{subsec:data_collection} for pretraining corpus inspection and contamination analysis). For each paper, we used the reasoning model o3-mini to extract the core
scientific puzzle, the surrounding factual context, and the author's
own hypotheses. To prevent leakage of human hypotheses into LLMs, we built a paraphrase-based detector and discarded any context and puzzle containing sentences whose embedding similarity \citep{song2020mpnet} to the corresponding human hypotheses exceeded a threshold calibrated on 20{,}000 controlled rephrasings; 99.7\% of authors confirmed the
extracted context and puzzle, and 98.6\% confirmed the extracted hypotheses in our
survey. Full details can be found in \hyperref[sec:matmethods]{Methods} and Supplementary Information \ref{2.2}. We then prompted 26 LLMs to propose hypotheses based solely on the context and puzzle\revise{, each proposing as many hypotheses as the authors had stated}. These models included nineteen non-reasoning chat
models and seven reasoning models across eight providers (Supplementary Information \ref{full_list}). Each author received a custom set of the five most semantically distinguishable hypotheses for their paper, balancing responses from reasoning and chat models, and then rated each on conceptual novelty, empirical feasibility, probability of being true, and \revise{their willingness to pursue it in future work}. \hl{To ensure data reliability, we administered a comprehension test confirming that all participants understood and correctly applied our rubrics, and we excluded potentially careless responses (e.g., submissions completed in under 2 minutes with identical ratings across all questions, against an average completion time of 15 minutes). These processes excluded 22.1\% of low-quality responses.} \revise{Of the 6,749 scientists who participated, 5,259 contributed 25,139 usable evaluation sets.} Figure \ref{fig:pipeline} summarizes the pipeline (\hyperref[sec:matmethods]{Methods} and Supplementary Information \ref{subsec:survey_design}). 

\revise{Our central finding is that today's AI rarely says no in science, in two senses. As a generator, it rarely proposes that an expected effect is absent: every model we tested formulated substantive null hypotheses far less often than the original authors did. As a judge, it rarely rejects an idea: LLM evaluators placed nearly every hypothesis in the upper-middle of the scale, whereas experts rated weak ideas harshly, and the LLM evaluators always scored LLM-generated hypotheses as more novel than the authors' own. Both kinds of "no" do essential scientific work. The first marks the limits of a theory; the second separates strong ideas from weak ones. In both roles, today's AI is defaulting to yes where science needs a no. At scale, this kind of sycophancy \citep{sharma2024towards} threatens science more broadly: nulls disappear from the literature, fields risk contraction, and when AI judges AI, we are left overly optimistic about what AI can do.}

\revise{We organize the study around one question: how AI-generated hypotheses differ from human ones, and how these differences could enter science. We first characterize what AI proposes, namely how often models propose substantive nulls and how diverse their hypotheses are (Figure~\ref{fig:ideation} and Extended Data Figure~\ref{fig:diverse}). We then examine how experts rate these hypotheses and how willing they are to pursue them (Figure~\ref{fig:judgment} and Extended Data Figure~\ref{fig:percep}). We next test how closely state-of-the-art (SOTA) automated assessments reproduce researchers' ratings (Figure~\ref{fig:autoeval} and Extended Data Table~\ref{tab:llm_judge_accuracy}), and train a reward model to predict these ratings and assess the learnability of distinctive scientific tastes. Finally, we show how the survey patterns appear in macro-level trends across the scientific literature, and propose a stylized model to illustrate possible consequences of uncritical AI use in science (Figure~\ref{macro}).}

\section*{\revise{Negation and diversity across AI hypotheses}}
\label{sec:ideation2}

We trained a
high-precision ensemble classifier\revise{, which we call the null-statement classifier} (99.5\% cross-validated accuracy; \hyperref[sec:matmethods]{Methods} and Supplementary Information \ref{classify}), to detect null hypotheses, which contain explicit claims of no relationship,
no effect, or no difference. \revise{The extraction prompts focus on substantive null hypotheses, in which the predicted outcome is itself the absence of an effect (for example, ``A has no effect on B’’), in contrast to a possible expectation that the effect exists, so the researcher's theoretical bet is on the null (more discussion in Supplementary Information \ref{sec:nulldef}).} Humans formulate such hypotheses sparingly\revise{ (0.8\% of papers)}, but every LLM in our panel formulates them much less (Figure~\ref{fig:ideation}a). Even agentic deep research systems with live web search rarely articulate one\revise{, although retrieval could surface the prior literature that establishes the expectation of an effect, against which a null hypothesis becomes informative. The gap persists with richer inputs (the Introduction, with or without Related Work) as well as with contexts and puzzles from journal articles published in 2026 without preprints (Supplementary Information \ref{sec:nullrobust}). When the authors proposed a null, models were more likely than usual to do the same; even so, in 90.6\% of these cases, no model proposed a null. Moreover, humans and models show little agreement on nulls (Supplementary Information \ref{sec:nulldef}).} \revise{One possible explanation of null suppression lies in the training data. Scientists are much more likely to publish, and to theorize post hoc about, empirical associations and positive findings than non-associations and null findings. A well-documented selection bias is often referred to as the ``file drawer'' problem, in which scientists file away results from failed investigations and remain unmotivated to publish them \citep{rosenthal1979file, chen2025geographical}. This bias enters LLMs' training corpus. Our results suggest that the bias also operates earlier, at the stage of hypothesis generation with uncritical use of AI, where null-oriented research is more rarely initiated}. While the logical move from ``A is associated with B'' to ``A is \emph{not} associated with B'' is obviously a cognitive primitive of scientific thought \citep{adams2021people,briggs2026must}, LLMs compress their training data, and rare forms of reasoning such as negation \revise{may} become even more attenuated \citep{sun2024head,jaiswal2023compressing,wu2025missing, shumailov2024ai}. \revise{Another source is post-training alignment with human preferences, which can make LLMs prone to sycophancy \citep{li2026narrative, sharma2024towards}, but base models did not consistently propose more nulls than instruction-tuned ones (Supplementary Information \ref{sec:nullrobust})}.

\revise{Beyond null hypotheses, we examined total diversity in AI hypotheses. Hypotheses from non-reasoning LLMs cluster in a narrow region of the idea space, a homogeneity that has been termed an ``artificial hivemind'' \citep{jiang2025hivemind}.}
Pairwise cosine similarity \citep{song2020mpnet} between hypotheses
generated for the same paper is highest within the non-reasoning group,
and higher than that of any other groups (Figure~\ref{fig:ideation}b). To make this
geometric, we \revise{projected texts to two and three dimensions and treat each generated hypothesis as a displacement vector
from a shared origin defined by its context-and-puzzle embedding}. This enables us to define and visualize an “idea space,” as shown in Figure~\ref{fig:ideation}c, where we use Google models and human scientists as an example. Reasoning models occupy a more diverse region than non-reasoning models, measured by average standard deviations across dimensions (Figure~\ref{fig:ideation}d)\revise{, whereas we find no significant difference between reasoning models and human authors. We confirmed these conclusions with a variety of different representations, such as different scientific-domain embedding models (SPECTER2 \citep{Singh2022SciRepEvalAM}) and dimension reduction methods (Supplementary Information \ref{sec:si_space} and Extended Data Figure \ref{fig:diverse})}. The gain is not driven solely by parameter counts; it tracks whether the
model deliberates internally, consistent with the view that reasoning
elicits an implicit ensemble of perspectives \citep{kim2026reasoning}.


\section*{\revise{Willingness to pursue AI hypotheses}}
\label{sec:judgment}

We modeled \revise{respondents' reported willingness to pursue each hypothesis in future work (hereafter, ``willingness to pursue'')} as a function of \revise{its rated characteristics of novelty, feasibility and probability, along with four
candidate properties}---human--AI idea similarity, field, seniority, and prior
AI \revise{engagement}---using a Mundlak (correlated random-effects) specification that separates \revise{within- from between-scientist variation; what makes a particular author favour one of \textit{their} ideas over another from persistent differences among author favourites \citep{schunck2013within}.} Author-fixed-effect
and ordinal-logit specifications give the same answers (\hyperref[sec:matmethods]{Methods}, Supplementary Information \ref{stats1}, and SI Table \ref{tab:adoption_combined}). \hl{The distribution of scientist characteristics and ratings can be found in Extended Data Figure \ref{fig:distribution}. A non-response analysis indicates that respondents are representative of the full sample: we find no evidence that respondents and non-respondents differ systematically on any observed characteristic. Across all four fields, the distribution of respondents over 117 subtopics closely matches the distribution of research across those subtopics in the full corpus (Pearson $r$ = 0.98). See Extended Data Figure \ref{fig:nonresponse} and Supplementary Information \ref{app:label_nonres} for more details. Re-estimating the specifications with weights given by the inverse of the estimated probability of response yields coefficients that are similar in sign, magnitude, and statistical significance (Supplementary Information \ref{app:selection}).} All \textit{p}-values are two-sided. Across the paper, several claims rest on the \emph{absence} of an effect rather than its presence. Because a non-significant coefficient is on its own uninformative about whether an effect is small or merely imprecisely estimated, we also supplement these claims with two one-sided tests (TOST) of equivalence. All results are robust (Supplementary Information \ref{sec:equivalence}). 

From these analyses, four regularities emerge. (i)~\textbf{\revise{Respondents were more willing to pursue hypotheses resembling their own work}} (within-author coefficient $=1.28$, $p<10^{-3}$;
Figure~\ref{fig:judgment}a, Extended Data Table \ref{tab:ols_combined}). Higher resemblance (captured by the cosine embedding similarity \citep{song2020mpnet} between human and AI ideas) is associated with higher perceived feasibility
(coefficient $=2.29$, $p<10^{-3}$) and probability (coefficient $=3.42$, $p<10^{-3}$) but \emph{lower} perceived novelty
(coefficient $=-3.15$, $p<10^{-3}$). As a result, \revise{willingness to pursue} rises anyway, a pattern consistent with prior work on
self-similar citation \citep{bao2024simulation, katz1999self} (Extended Data Figure \ref{fig:percep}). \revise{The association holds when similarity is measured with SPECTER2 embeddings or with displacement vectors that remove the shared context and puzzle (SI Table \ref{tab:ols_pursue}).}

(ii) \textbf{\revise{Senior scientists reported less willingness to pursue the presented AI-generated hypotheses regardless of rated characteristics}}. Using within-field yearly citation percentile (coefficient $=-0.16$, $p<0.001$), within-field yearly publication percentile (coefficient $=-0.10$, $p=0.047$), and log-transformed academic age (base $e$, coefficient $=-0.04$, $p=0.031$) to represent seniority/status yields consistent conclusions (Figure~\ref{fig:judgment}b,c, Extended Data Table \ref{tab:ols_combined}). All three measures and prior AI engagement are right-skewed (Extended Data Figure \ref{fig:distribution} and Supplementary Information \ref{stats1}). \revise{We use citations to represent seniority, which is associated only with willingness to pursue, not rated novelty, feasibility or probability (all $p>0.05$; Extended Data Figure \ref{fig:percep}b), so the lower willingness of senior respondents reflects a residual reluctance to pursue rather than a judgment that the hypotheses are weaker.}

Medicine is the most receptive field
(8.05\% above social science, $p<10^{-3}$, Figure~\ref{fig:judgment}e), consistent with our observation that medical scientists have the highest \revise{prior AI engagement} (Figure~\ref{fig:judgment}d), measured by the proportion of prior publications \revise{classified as AI-engaged research, with AI as either a method or topic} \citep{hao2026artificial} (\hyperref[sec:matmethods]{Methods} and Supplementary Information \ref{aiuse}), although prior \revise{engagement} does not itself
predict greater \revise{willingness} after controlling for \revise{rated characteristics}
(Extended Data Table \ref{tab:ols_combined}). \revise{The lower willingness in social science is concentrated among senior respondents.} Once a
field-by-seniority interaction enters the model, field effects
vanish (Figure~\ref{fig:judgment}f, Extended Data Table \ref{tab:ols_results34}). 

(iii)~\textbf{Probability dominates among the three \revise{rated characteristics}} (coefficient $=0.31$, $p<10^{-3}$). Feasibility (coefficient $=0.13$, $p<10^{-3}$) and novelty (coefficient $=0.12$, $p<10^{-3}$) matter less and roughly equally, suggesting that scientists are risk-averse, preferring to undertake likely-to-succeed
ideas over surprising ones (Figure~\ref{fig:judgment}g, Extended Data Table \ref{tab:ols_combined}). The aversion is
strongest in biology (coefficient $+0.06$, $p<10^{-3}$) and medicine (coefficient $+0.06$, $p=0.003$), weakest in social science, captured by a
\revise{field-by-characteristic} interaction in the model (Figure~\ref{fig:judgment}h, Extended Data Table \ref{tab:ols_results}, and SI Table \ref{tab:ols_results_m2}). This pattern likely reflects the funding-driven model of the life sciences, where the high cost of failed experiments makes risky ideas particularly expensive to pursue.

(iv)~\textbf{\revise{LLM-generated hypotheses received the lowest ratings in the social sciences across dimensions, a pattern that may reflect the field's theoretical pluralism}}. With each field receiving a balanced
sample from all LLMs, AI ideas in social science were rated as less novel (Figure~\ref{fig:judgment}i),
less feasible (Figure~\ref{fig:judgment}j) and less probable (Figure~\ref{fig:judgment}k) than in any other domain\revise{, although the absolute differences on the 1--9 scales are small}. \revise{These ratings do not simply reflect subjective characteristics: bias such as evaluator seniority is associated with willingness to pursue, but not with the ratings themselves} (Extended Data Figure \ref{fig:percep}). \revise{Two rival explanations are not supported by our data. First, social-science inputs could be underrepresented in pretraining data: their lexical coverage in a 10-billion-token sample of the Dolma corpus actually exceeds that of biology and chemistry inputs (Supplementary Information \ref{sec:si_dolma}). Second, a folk-knowledge account, in which models produce obvious hypotheses \citep{mohseni2026hard}, predicts high probability ratings, while LLM hypotheses are actually rated as having low probability (Figure~\ref{fig:judgment}k). We therefore point to theoretical pluralism as an interpretation, but not a precise finding.} \revise{In} pluralistic fields like the social sciences, where targets of research are themselves contested — competing interpretations, context-dependent findings, \revise{theories that change partly because people respond to the categories used to describe them \citep{hacking1995looping}, and the highest rates of expressed disagreement \citep{lamers2021disagreement}} — scientists prize the surprising (Figure~\ref{fig:judgment}h), and an AI system optimized to reproduce the modal pattern of its training data has little to converge on. \revise{An independent benchmark likewise finds the lowest idea-generation scores in the social sciences \citep{ruan2026evaluating}.} 

\section*{\revise{Automated judges track experts weakly}}
\label{sec:autoeval}

A vast methodological literature treats LLM-as-a-judge\revise{ (a model prompted to act as an evaluator)}, $n$-gram novelty,
semantic distance, conditional perplexity, cross-entropy, natural-language inference, and
literature-grounded novelty checkers as proxies for expert assessment of research
ideas \citep{shahid2025literature,moussa2025scholareval}. \revise{These proxies operate at different levels: $n$-gram measures capture lexical divergence; embedding, perplexity, and entropy capture semantic or distributional surprise; literature-grounded novelty checkers capture LLM-assessed novelty relative to prior work. Weak agreement can therefore reflect a mismatch of constructs as well as limits of the metrics.} With 25,139 expert ratings as ground truth, we can ask whether any of these proxies
recover expert judgment.

None performs well. We first prompted Gemini~2.5~Flash, DeepSeek~R1 and OpenAI's
o4-mini Deep Research model with the same rubric and rating scale shown to
human authors, with and without an injected scientist persona, operationalized by scientists' own ideas\revise{; such personas are widely used to build scientist agents in multi-agent auto-research systems \citep{ueda2025exploring}}. The
\revise{Deep Research model, which uses retrieval augmentation (supplying externally retrieved information, such as web search results, to the assessment),} achieved the highest correlation
with human judgment, but no model exceeded $r=0.35$ on any dimension
(Figure~\ref{fig:autoeval}a). Persona prompting helped in most settings, but not significantly for the retrieval-based Deep Research model. It even reduced novelty accuracy. \revise{The main failure is a strong central tendency: LLM ratings cluster in the upper-middle of the scale and rarely reach the extremes that experts use, so that experts rate weak ideas harshly whereas LLMs rate almost everything as acceptable.} Adding explicit encouragement to assign extreme scores only increases the standard deviation of ratings by 0.33 on a 1–9 scale (Figure~\ref{fig:autoeval}b and Supplementary Information \ref{prompts}). Across all settings, matching human experts' probability of assigning truth to hypotheses was the easiest dimension on which LLMs performed. \revise{AI also rates AI-generated ideas as more novel than human-generated ones. Rating the authors' own hypotheses and AI-generated ones from the context and puzzle alone, the LLM judges scored LLM hypotheses as markedly more novel (7.30 versus 6.20 under DeepSeek-R1 and 6.67 versus 4.36 under Gemini 2.5 Flash; $p<0.001$; Supplementary Information \ref{sec:si_judgehuman}), a conclusion unsupported by any expert-anchored evidence from our study.}

To probe a broader range of evaluators, including SOTA reward models\revise{ (models trained to score outputs according to human preferences)}, we follow standard practice and convert human ratings into within-scientist pairwise preferences. We construct a held-out test set of 5,000 pairs with transitive relations removed (i.e., we keep hypothesis pairs h1–h2 and h2–h3, but exclude h1–h3). On this set, we evaluate the judgment of novelty \revise{(ties removed)} across a wide range of established novelty assessment models and metrics, including LLM-as-a-judge (as discussed above), 2-gram and 3-gram novelty, length-normalized cross-entropy, conditional perplexity, semantic distance, the natural language inference-based entailment score, the SOTA score-based idea judge GraphEval-GNN \citep{feng2025grapheval}, and the top three reward models that learn human preferences available on RewardBench\revise{, a benchmark for evaluating reward models} \citep{lambert2025rewardbench,malik2025rewardbench}. Each method follows the same protocol: score both items in a pair, determine the predicted winner, and count the prediction as correct if its direction matches human judgment. All methods perform near chance (Figure \ref{fig:autoeval}c). LLM-as-a-judge fares especially poorly, as its central-tendency bias \revise{often} produces near-identical ratings of mediocrity. \revise{Asked instead to compare a pair directly, LLM judges score higher (0.53--0.61; Extended Data Table~\ref{tab:llm_judge_accuracy}), but their preferences shift under logically equivalent rephrasings of the same question (Supplementary Information~\ref{reward}).}

\section*{\revise{Reward models exceed reviewer agreement}}
\label{sec:reward}

On the within-scientist pairwise preference dataset, we
post-trained a Qwen3-14B model under a Bradley--Terry objective\revise{, which learns a score for each hypothesis from pairwise preferences,}
augmented with a margin term proportional to the rating gap (Supplementary Information \ref{reward}). \revise{A single model trained jointly on novelty, feasibility and probability performed as well as dimension-specific models, so one model suffices for all three dimensions (Extended Data Table \ref{tab:weight_config_all}). Models trained on a single dimension transfer poorly to the others (for example, novelty-only models score 0.41--0.49 on feasibility), indicating that the dimensions are non-redundant, although not necessarily independent.}

Scientific domain-specific models slightly outperformed a single general model when in-domain data were abundant (Extended Data Table \ref{tab:accuracy_all_dims}), suggesting that fields
share a common evaluative core and carry nuances of taste that are not ineffable, but can be statistically captured\revise{ \citep{tong2026ai}}. In the held-out test set, the biology
model reached 69\% pairwise accuracy on novelty, 62\% on feasibility and
67\% on probability, and the social-science model reached 64\%, 62\% and 67\%.
For comparison, the top three models available on RewardBench \citep{malik2025rewardbench} hover at chance on this task across dimensions: the best,
\texttt{Skywork-Reward-V2-Qwen3-8B}, scores 53\%, 55\% and 49\%
across feasibility, probability, and novelty, and the o4-mini Deep
Research judge scores 41\%, 55\% and 43\% (Extended Data Table \ref{tab:llm_judge_accuracy}). Our models improve on the
strongest baselines by up to 27\% (+14 percentage points on average across three dimensions).

We benchmark these numbers against the realistic reference of human
agreement. From 26,731 OpenReview submissions to 46 conferences (2017--2025;
spanning computer science, physics, medicine and the social sciences; \hyperref[sec:matmethods]{Methods} and Supplementary Information \ref{humanalign}), we constructed within-conference pairwise comparisons\revise{ and measured how often the reviewers listed in one slot (for example, Reviewer 2) agreed with the preference of the reviewers in another slot (for example, Reviewer 1). Because the reviewers in a slot differ across papers, this measures agreement between independent reviewers.} Averaged over $1{,}000$ randomised \revise{choices of slot pairs}, agreement was $61.0 \pm 0.1\%$. Our reward models exceed it. Existing automated evaluators do not.

\section*{\revise{Field-level consequences with intensive AI use}}
\label{sec:human_knowledge}

\revise{We next ask whether our survey findings could generalize to the field as a whole, using a descriptive corpus analysis and a stylized simulation (Figure \ref{macro}).} Focusing on medicine and social science, fields our survey shows differ strongly in \revise{respondents' willingness to pursue AI-generated hypotheses}, we assembled all 38,863,847 papers published between 2010 and 2025 from OpenAlex. Applying the \revise{AI-engaged research classifier} \citep{hao2026artificial} to titles and abstracts (Figure \ref{macro}a), we find AI is indeed more popular in medicine, either as a method or topic, and this predates LLMs. \revise{Earlier AI methods had already reshaped parts of biomedicine, for example protein structure prediction \citep{jumper2021alphafold}.} A similar pattern appears in the social sciences. AI-focused research began expanding in 2017, coinciding with the emergence of the Transformer architecture. This is a conservative estimate, insofar as researchers may use AI for research quietly (e.g., a writing tool) without reporting it explicitly as a method or topic.

We then track papers reporting zero null \revise{statements} by applying our \revise{null-statement} classifier to every sentence in an abstract (Figure \ref{macro}b). We define all-significant papers as those containing zero detected null \revise{statements}. Social science initially had more all-significant papers than medicine, but \revise{this ordering reversed after ChatGPT's release in November 2022: after November 2022, the all-significant share rose sharply in both fields, and more so in medicine, the field with more AI-engaged research.} Our results align well with prior work using LLM/human annotation methods: where earlier estimates\citep{briggs2026must} put significant-only papers at 77\% in political science (2010–2024), we find on average 74\% in medicine and 77\% in social science (2010–2025)\revise{; rates of negative statements estimated with other methods from independent work are of similar magnitude \citep{fanelli2010positive,aczel2018quantifying} (more comparisons see Supplementary Information \ref{sec:nulldef})}. Notably, this share was \emph{declining} before LLMs, plausibly due to reforms involving pre-registration, the replication-crisis reckoning, and the open-science movement. \revise{After 2022 the share rose again, reversing this decline. A detected null statement may express a prior expectation of no effect, a non-significant finding, or a reporting convention. Because abstracts are written after the results are known \citep{rubin2017does}, human annotators cannot reliably recover which of these a null statement reflects from its surrounding wording (Supplementary Information \ref{sec:nulldef}). The corpus trend, which is consistent with the survey finding, may originate at the ideation stage, reflecting a lack of negative thinking in AI-assisted research, or may arise because LLMs steer researchers toward particular kinds of questions (e.g., data-intensive ones \citep{hao2026impact}) for which non-significant findings are naturally rarer owing to greater statistical power \citep{lucas2013too}. It may also reflect a combination of mechanisms; we find no evidence, however, that LLM-assisted polishing \textit{alone} drives the change: when we asked the LLM to ``polish without changing the content'' based on a random sample of 100,000 arXiv abstracts from 2019, the proportion of null statements remained unchanged (Supplementary Information \ref{sec:nullrobust}).} %

Embedding all papers with SPECTER2 \citep{Singh2022SciRepEvalAM}, the SOTA embedding model specifically designed for scientific knowledge, we operationalize a field as a circle with radius given by the average Euclidean distance from all papers to the center in the 768 model dimensions. Then we compute two quantities \citep{hao2026artificial}: center drift, the distance of the center in each year from its 2010 position (Figure \ref{macro}c), and knowledge extension, the year-over-year growth ratio of the radius (Figure \ref{macro}d). \revise{Field centers drifted more rapidly after 2022, more so in medicine, plausibly toward data-rich topics \citep{hao2026artificial}; in medicine, the field radius briefly expanded and then contracted after 2022. These measures are coarse proxies (\hyperref[sec:matmethods]{Methods}), but the pattern agrees with independent evidence that AI-associated research concentrates on data-rich topics and established targets \citep{varoquaux2022machine,jayatunga2022ai,lu2026stronger, doshi2024generative, tang2026ai}. It also mirrors the survey: medicine, the field with the highest willingness to pursue AI-generated hypotheses and the strongest weight on probability, shows a strong rise in all-significant abstracts, field radius contraction, and center drift.}

\revise{To illustrate what these patterns could mean for a field, we built a stylized simulation, intended as a conceptual illustration rather than as evidence (\hyperref[sec:matmethods]{Methods} and Supplementary Information \ref{simlation}). In the model, AI acts as a fast but error-correlated instrument: it produces more studies, which averages away random error, but makes their errors more alike and suppresses null statements, which biases visible (e.g., published) estimates of small effects upwards. Under these assumptions, net knowledge peaks at intermediate AI use, and mature fields should use less AI (Figure~\ref{macro}e--h).}

\section*{Discussion}
\label{sec:discussion}

\revise{Our expert audit of AI-generated hypotheses returns a sober verdict. No model proposes substantive null hypotheses as often as humans do, and non-reasoning LLMs converge on similar hypotheses.} Expert evaluators
bring their own systematic biases. \revise{LLM-generated hypotheses received the lowest ratings in the social sciences, likely because they struggle in pluralistic fields.} The automated infrastructure built
to substitute for expert judgment cannot yet do so. \revise{Field-level trends after 2022 are consistent with these survey findings.} \revise{Four} implications
follow.

\begin{figure}[htbp]
    \centering
    \displayitem{\includegraphics[width=1\textwidth]{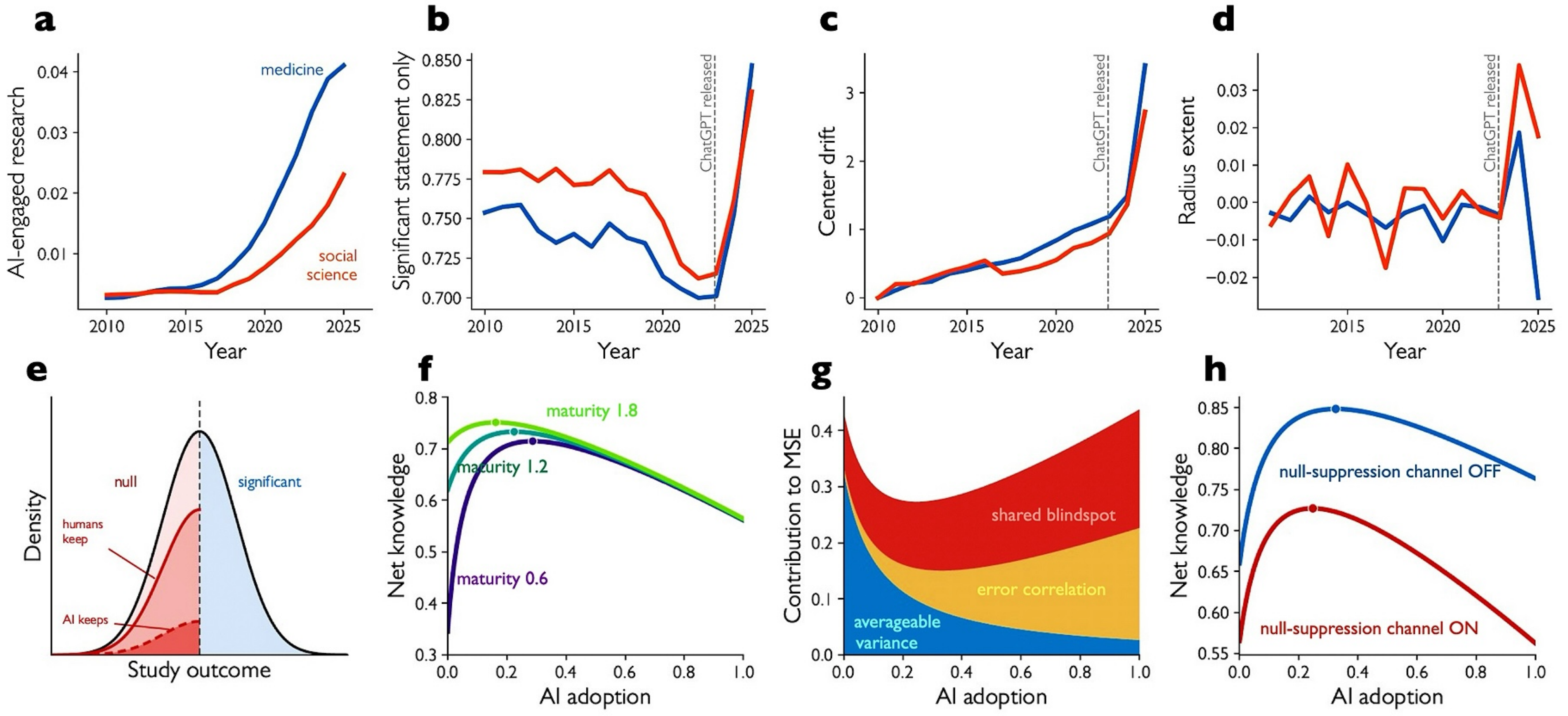}}
    \caption{\hl{\textbf{\revise{Field-level trends after the release of ChatGPT, and a stylized simulation.}} Based on 39 million scientific papers in medicine and social science from 2010 to 2025: \textbf{a}, medicine \revise{has more AI-engaged research}. \textbf{b}, after the release of ChatGPT, the share of papers without any null \revise{statements} rises sharply. \textbf{c}, after the release of ChatGPT, the field's center moves rapidly away from its 2010 position. \textbf{d}, \revise{in medicine, the field radius briefly expands and then contracts after 2022}. Based on simulation studies\revise{ (Supplementary Information \ref{simlation})}: \textbf{e}, an illustration of model design showing how \revise{suppressing null statements shifts visible (e.g., published) estimates}. \textbf{f}, the more mature a field, the smaller the optimal level of AI use. \textbf{g}, in AI-assisted research, errors become correlated and blind spots shared, while higher productivity offsets random noise, though at diminishing marginal returns as AI use increases. \textbf{h}, toggling \revise{null suppression} on and off shows that the loss of null findings substantially \revise{lowers} net knowledge.}}
    \label{macro}
\end{figure}

\revise{First, AI-generated ideas enter science through researchers’ willingness to pursue them, and this channel is itself biased. Respondents favoured familiar, probable hypotheses over novel ones, whereas all models already avoid null hypotheses; two biases likely to reinforce one another. This may resolve an apparent tension: reasoning models prompted independently generate hypotheses whose diversity is indistinguishable from that of human authors, yet AI-associated science contracts across the total field (Figure~\ref{macro}) \citep{kusumegi2025scientific,hao2026artificial, tang2026ai}. If scientists filter a pool of AI suggestions toward directions that are probable, familiar, or where AI most lowers the cost of research, such as data-rich topics \citep{hao2026artificial}, and AI literature search tools further surface highly cited work \citep{algaba2026deep}, contraction can arise from how AI is adopted rather than from AI alone. Nor do these averages imply that models never propose exceptional ideas: although AI cannot yet reliably produce top ideas, its occasional strong proposals may still contribute value.}

\revise{Second, claims of AI-driven acceleration should be assessed against expert judgments of novelty, feasibility, and probability, not benchmark performance. Expert judgment is necessary but fallible, as our findings on seniority show, so the impact of AI on discovery should be evaluated through expert assessment, actual adoption, subsequent empirical validation and longer-term scholarly impact; many existing evaluations relying on a few non-author raters warrant caution.}

\begin{figure}[htbp]
    \centering
    \displayitem{\includegraphics[width=0.9\textwidth]{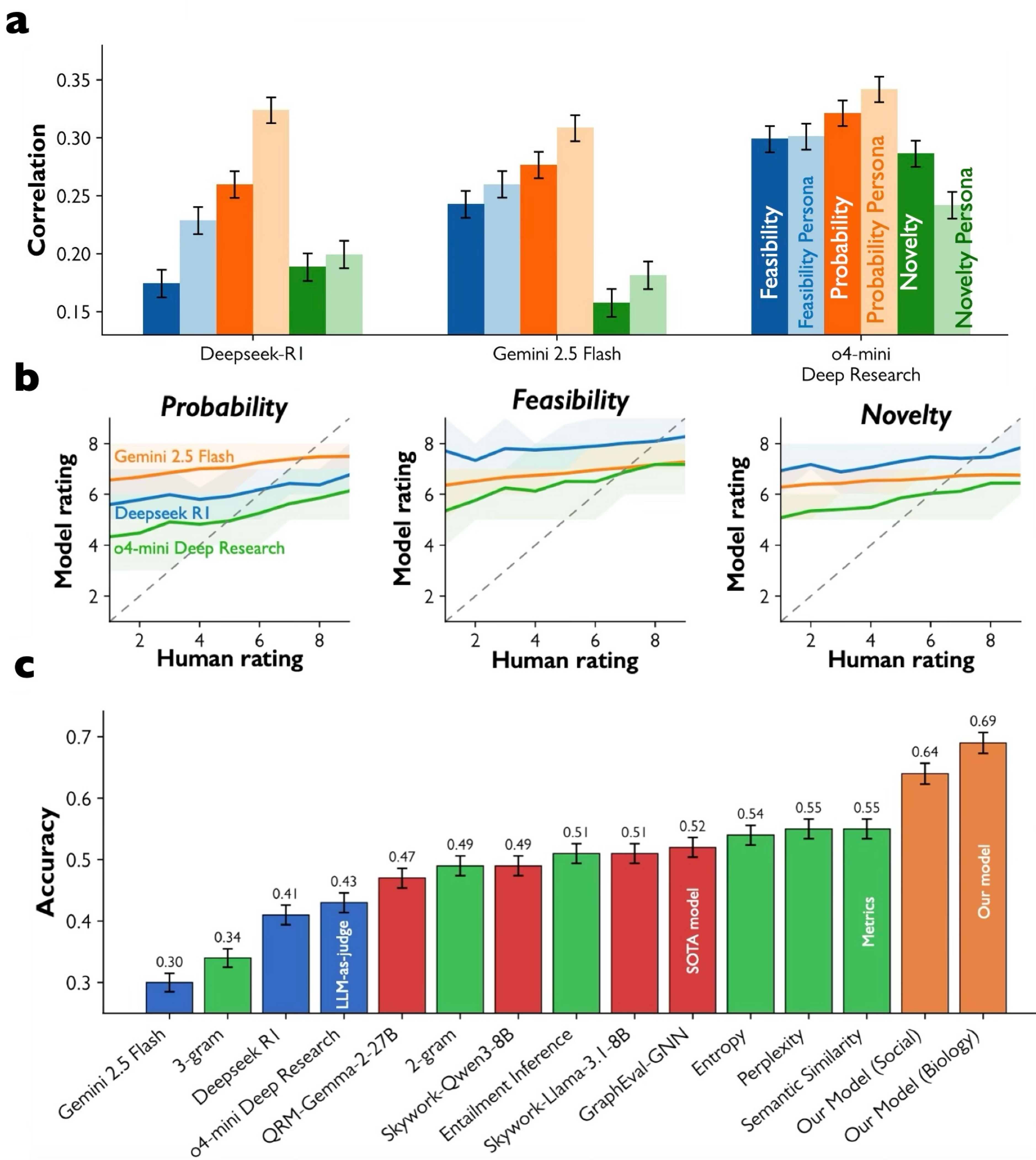}}
    \caption{\textbf{\revise{Automated evaluators agree weakly with expert ratings.}} \textbf{a},~Pearson correlation between LLM-judge ratings
    and human ratings on the full dataset, by dimension and judge, with and without injected
    scientist persona. The retrieval-augmented Deep Research judge is the best. Persona injection slightly helps. Yet no setting exceeds $r=0.35$. \textbf{b},~Calibration of LLM
    judges against human ratings on the full dataset. If LLM ratings track humans, points
    would lie on the gray diagonal; instead, all three judges exhibit a strong central tendency with narrow inter-quartile (25\% and 75\%) ranges, and rarely issue the extreme scores that human raters routinely assign. \textbf{c}, Accuracy of novelty judgment on the held-out pairwise test set\revise{ ($n$ = 5,000 pairs)}: LLM-as-a-judge, SOTA models, and popular metrics perform poorly. Our models perform better ($p<0.001$). Error bars = 95\% CI.}
    \label{fig:autoeval}
\end{figure}

\revise{Third, if automated judges replace expert evaluation, assessments of AI-generated science may become systematically optimistic. When the evaluator shares the generator's training distribution, agreement between them may reflect shared priors instead of shared standards. LLM judges rated LLM hypotheses as more novel than the authors' own and clustered in the upper-middle of the scale (Figure~\ref{fig:autoeval}b). Reward models trained on expert preferences, such as ours, are a more expert-aligned alternative, but they align only moderately with experts and encode current taste that will need recalibration as fields evolve.}

\revise{Fourth}, the shortfall in \revise{substantive} null-hypothesis formulation \revise{matters even though humans rarely propose them. A theory's empirical content lies in what it forbids \citep{popper1959logic}; progress comes from stating rival explanations, ``no effect'' among them, and designing tests to exclude them \citep{chamberlin1890multiple,platt1964strong}; and some of science's deepest claims are claims of absence \citep{michelson1887motion,gallistel2009null}. Humans already lean away from absence \citep{klayman1987confirmation,wason1960failure,greenwald1975prejudice}; AI does not introduce this asymmetry, but in our data it substantially amplifies it.} The human asymmetry that Adams and colleagues call our blindness to subtractive change \citep{adams2021people} is sharper in machines: ours is a bias of attention, theirs is a bias of evidence. Pretraining corpora are overwhelmingly populated by claims of what \emph{is} the case, amplified by a publication system that suppresses null results~\cite{briggs2026must}. A claim that ``A is \textit{not} associated with B'' is informationally richer than its positive counterpart: it presupposes a backdrop of prior expectation that makes the absence surprising~\cite{shi2023surprising}. Generating it, therefore, requires not merely plausible continuation but a theory of the field's expected contrast, a piece of meta-knowledge underdetermined by flat text. The deeper portion of negative knowledge, including what has been ruled out, what fails to replicate, and what does not work, accumulates in the unwritten residues of embodied practice: failed experiments, rejected grants, abandoned notebooks, and \revise{informal knowledge, shared within laboratories, of approaches known not to work. Little of this enters training data.} Reasoning \revise{may help} because deliberation simulates the contrast structure that flat text omits~\cite{kim2026reasoning}, but reasoning over a corpus already pruned of negation \revise{may not} recover what was never recorded. \revise{These forces may form a self-reinforcing loop: because researchers rarely publish null statements, LLMs seldom encounter them and may learn to avoid them further in the compression process \citep{sun2024head}; researchers then use these models and favour probable and familiar over novel ideas, and each turn could narrow the space of inquiry.} Given AI-assisted publishing, the human–AI differences we report in diversity and null suppression are likely conservative lower bounds. Until pretraining ingests these absences through deliberate elicitation of expert null intuitions, registered reports, replication archives, and laboratory notebooks of dead ends, the asymmetry \revise{is likely to} persist.

A deeper limit, however, is not negation but appetite. Today's models are
\revise{largely} optimised to predict, not to wonder. Their generative behaviour is
anchored to the distribution of what has been said, not to the marginal
value of what could be learned next. As we have argued
elsewhere~\cite{evans2025after}, the next phase of scientific
automation will require encoding \emph{computational curiosity}\revise{, long studied in machine learning as intrinsic motivation and autotelic learning \citep{schmidhuber1991curious,schmidhuber2010formal,barto2013intrinsic,oudeyer2007what,pathak2017curiosity,colas2022autotelic}}---an
intrinsic drive that prizes anomaly, contradiction, and absence, and
that pulls a system toward data it has not yet seen rather than toward
the centre of data it has~\cite{loewenstein1994}. \revise{A curious system \citep{schmidhuber2010formal} would also propose null hypotheses, which test whether an expected relationship exists at all.} It would probe the edges of consensus rather than
its centre. And it would treat the construction of new measurements,
instruments, interventions and natural experiments as part of the
hypothesis space, not as exogenous to it. Until AI is routinely directed toward violations of its expectations and to strategically accumulate data to facilitate abductive, surprise-driven discovery, it \revise{is unlikely to generate sustained} advances~\cite{shi2023surprising}. Hypothesis generators that
draw only from the published literature are \revise{largely limited to}
recombination~\cite{farrell2025large}. \revise{Systems that can gather new evidence, rather than only recombine published findings, are better placed to extend the literature. Our survey offers an expert-anchored diagnostic for methods that aim to increase the diversity and creativity of LLM outputs \citep{wang2025optimizing,ismayilzada2025creative}.}

\revise{Our reward models exceed inter-reviewer agreement, but closing the remaining gap is not the goal. Expert communities revise the criteria by which ideas are judged as science evolves.} What experts
contribute is not only a verdict on individual ideas but the 
collective construction of criteria by which ideas come to be
judged. Novelty, feasibility, and probability are not fixed features
of a hypothesis; they are functions of a shifting disciplinary
horizon that communities continuously discover, contest,
and revise as their fields progress. Distilling \emph{current}
taste into a model is feasible, as our results show. Distilling the
practice that lets taste \emph{evolve} is harder, and is the work for
which expert communities exist. 

\revise{Our study has limitations. First, it evaluates general-purpose, instruction-following LLMs rather than bespoke scientific systems, such as graph-reasoning multi-agent frameworks \citep{ghafarollahi2025sciagents}, and uses a single round of generation rather than the interactive, iterative collaboration of systems such as Co-Scientist \citep{gottweis2026coscientist}; our findings characterize the models and settings tested, not every form of AI-assisted research. Second, respondents knew the hypotheses were AI-generated. All received the same disclosure, so comparisons within the survey are informative, but the design cannot separate a hypothesis's characteristics ratings from reactions to its disclosed origin; blinded comparisons with independent experts are a direction for future work. Third, cross-field contrasts may partly reflect selection into preprinting, although a robustness analysis restricted to journal papers without preprints yields consistent results (Supplementary Information \ref{sec:si_preprint} and \ref{sec:nullrobust}). A further journal-level assessment would require partnerships with publishers. Fourth, the human--AI comparison of null hypotheses is observational: authors drew on knowledge, experience and preliminary results that no input can give a model, although the gap persists with richer inputs and web-enabled systems (Supplementary Information \ref{sec:nullrobust}). Fifth, the corpus analysis is descriptive. It reads abstracts only, sentence by sentence, and may miss null statements reported deep within the full text; causal claims would require designs that identify actual LLM use, such as (credible) generative-AI disclosure statements. Sixth, our sample is weighted toward biology and medicine, and we have not established that ratings are measured equivalently across fields, so cross-field differences in raw ratings may reflect evaluative standards as well as the hypotheses, although statistical design features such as field fixed effects mitigate much of this concern (Supplementary Information \ref{sec:si_comparability}). Seventh, embedding similarity does not separate research objects, relations and mechanisms, and our safeguards reduce but cannot rule out pretraining contamination, particularly for closed models. Finally, our dependent variable is respondents’ stated willingness to pursue a hypothesis. This measure may reflect personal resources and constraints, and responses might differ under a reviewer-style framing.} 

\revise{Science advances by saying no. A hypothesis draws its content from what it forbids, nulls close a path that others would otherwise walk, and an evaluation informs only when the judge is willing to reject. On present evidence, AI does this poorly. Its sycophancy defaults to yes where science demands a no. It fails to participate in the `creative destruction' of science by seldom proposing that an expected effect is absent or calling out a weak idea. Prompting models to say no more often is unlikely to suffice (explicit encouragement to assign extreme scores barely widened judges' ratings; Figure~\ref{fig:autoeval}b), because saying no demands higher quality than saying yes. A hypothesis or claim that confirms a field's priors, like those that confirm a person's, pass with little scrutiny, whereas a substantive null that contradicts them invites far more and must withstand it. Likewise, a judge that separates strong hypotheses from weak ones needs greater discriminating power, and scientists will judge its errors more harshly. Human refusal is fallible, as the blanket reluctance of senior respondents shows, but it is exercised. AI for discovery therefore remains a collaborator: it expands the search, surfaces candidate directions, and assists ideation. Framing the question, recognising what kind of result would matter, contesting the categories themselves, and rejecting what does not hold still remain the province of human scientists. Until models learn to intelligently say no, both to expected effects and their own ideas, scientists must say it for them.}


\printrefsegment{main}{References}


\clearpage

\begin{figure}[htbp]
    \centering
    \displayitem{\includegraphics[width=1\textwidth]{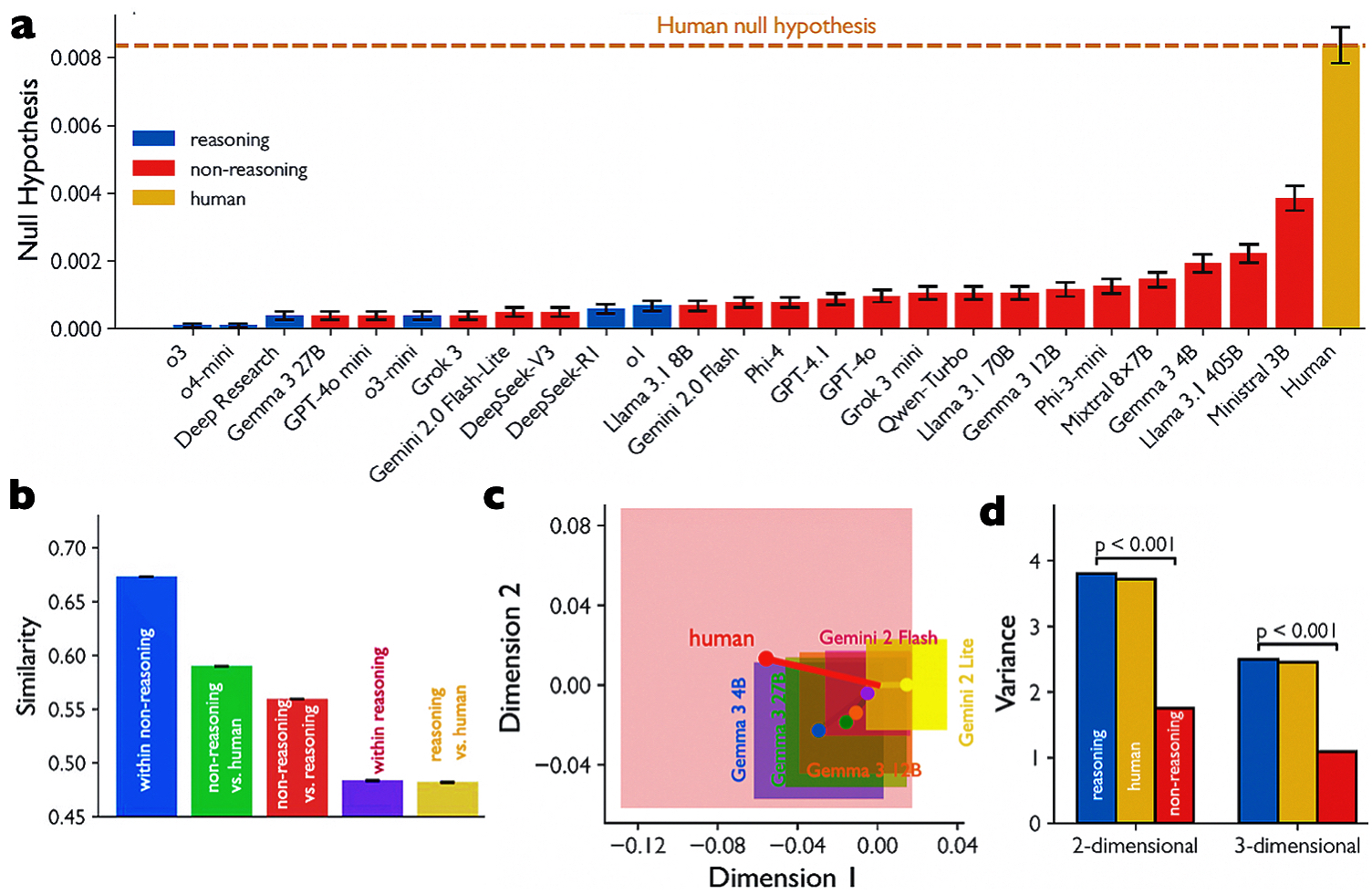}}
    \caption{\textbf{\revise{Humans negate more than any model, and non-reasoning models converge.}} \textbf{a},~Rate of explicit \revise{substantive} null-hypothesis formulation, by source\revise{; bars are coloured by reasoning model, non-reasoning model and human. A response counts as null if at least one of its hypotheses is classified as null. Each model is compared with the human rate on the same papers by a one-sided exact McNemar test with Holm--Bonferroni adjustment}. Humans are imperfect at negative reasoning, but every LLM is markedly worse (p<0.001), including the agentic deep-research system with live web access. Error bars = 95\% CI\revise{ (normal approximation)}. \textbf{b},\revise{~Pairwise cosine similarity
    of hypotheses generated for the same paper, by group. Non-reasoning
    LLMs are more similar to each other (the ``artificial
    hivemind''\citep{jiang2025hivemind}, p<0.001), robust to displacement-vector-based similarities and other embedding models (Extended Data
Figure \ref{fig:diverse}).} \textbf{c}, geometrically, we treat each hypothesis as a displacement from a common context-and-puzzle origin in the embedding space (We use Google models based on 10000 random papers as an example, and the length and width of a model are represented by the corresponding confidence interval of the mean for each dimension; UMAP for dimension reduction). \textbf{d}, reasoning models cover a substantially more diverse region than non-reasoning models in both 2D and 3D spaces. UMAP for dimension reduction, robust to other dimension reduction methods and embedding models (Extended Data Figure \ref{fig:diverse}). \revise{Reasoning models and humans do not differ significantly.}}
    \label{fig:ideation}
\end{figure}

\begin{figure}[htbp]
    \centering
    \displayitem{\includegraphics[width=1\textwidth]{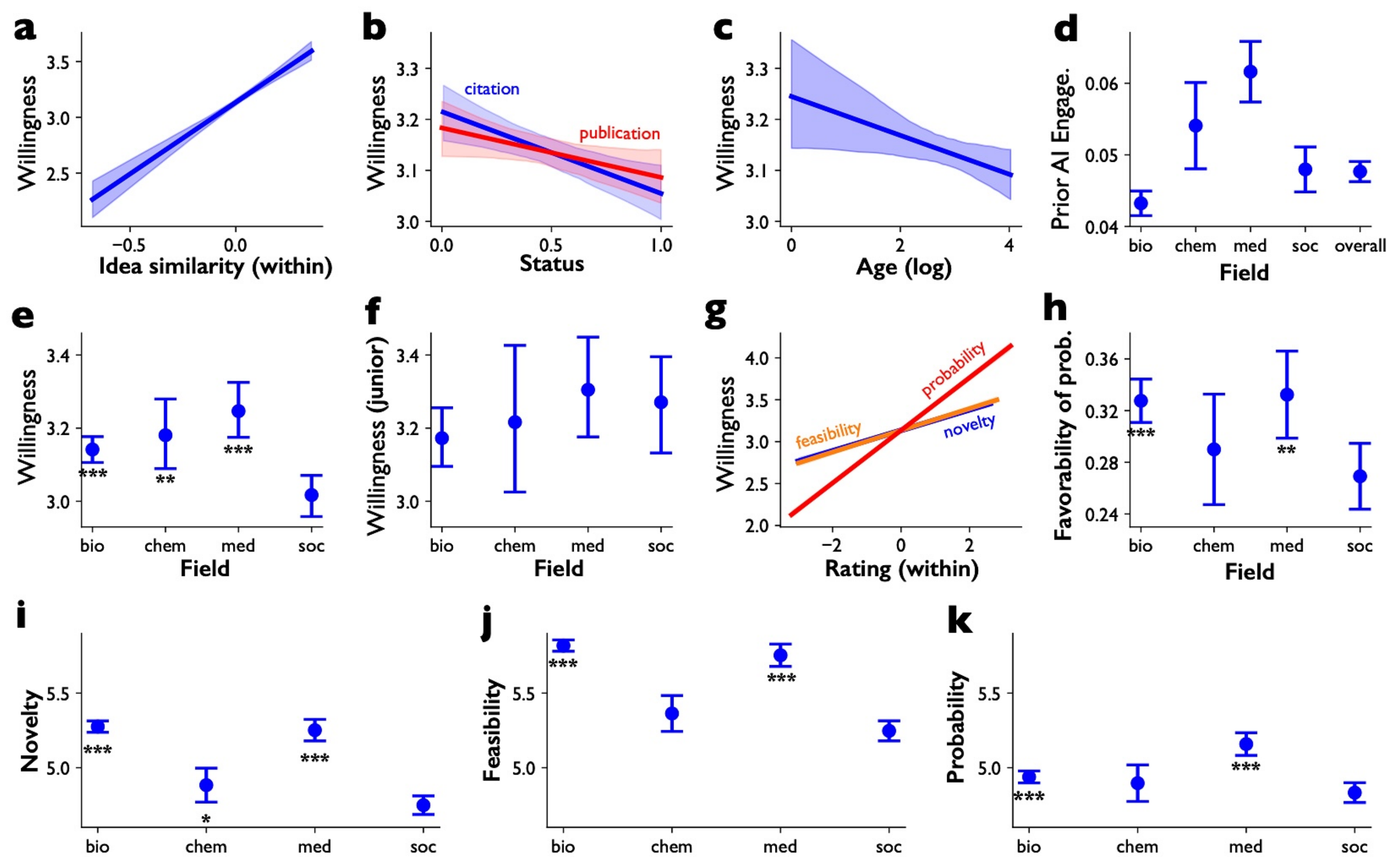}}
    \caption{\textbf{\revise{Factors associated with the willingness to pursue AI-generated hypotheses.}} Marginal predictions from the Mundlak \revise{willingness-to-pursue} model are shown, controlling for rated \revise{characteristics}. \textbf{a}, within-scientist similarity to the author's own ideas is the strongest single \revise{correlate of willingness}. \textbf{b}, status, represented by within-field citation/publication percentile, lowers \revise{willingness}. \textbf{c}, seniority, represented by log-transformed academic age, lowers \revise{willingness}. \textbf{d}, prior-AI \revise{engagement} across fields. It is not a significant predictor of \revise{willingness}. \textbf{e}, social scientists \revise{are the least willing}. \textbf{f}, once a field-by-seniority interaction is included in the model, the field main effects vanish. Among junior scientists at the baseline citation percentile of 0 (i.e., without seniority-induced decay), no statistically significant differences \revise{in willingness} are observed across fields. Senior scientists in every field are \revise{less willing}, with the decay being most pronounced in the social sciences. \textbf{g},~Effect of within-scientist deviations in rated \revise{characteristics on willingness}: probability $\gg$ feasibility $\approx$ novelty. \textbf{h}, biology and medicine show larger probability slopes than social science. \textbf{i} to \textbf{k},~\revise{ratings} of LLM ideas\revise{, collected on 1--9 scales}. Social science is the weakest domain on novelty (\textbf{i}), feasibility (\textbf{j}) and probability (\textbf{k}). All fields' \revise{ratings} and \revise{willingness} remain mid-scale. Baseline = social science. Uncertainty intervals: 2.5/97.5 percentiles of 300 draws from the asymptotic distribution of the estimated coefficients $\mathcal{N}(\hat\beta,\hat V)$ with cluster-robust $\hat V$. $^{*}p<0.05$; $^{**}p<0.01$; $^{***}p<0.001$. Error bars/bands = 95\% CI.}
    \label{fig:judgment}
\end{figure}

\clearpage

\section*{\revise{Methods}}
\label{sec:matmethods}

Full details of \revise{the} Methods are reported in Supplementary Information~\ref{sec:methods}.

\textbf{Corpus and human-in-the-loop survey.} From six non-arXiv preprint platforms (BioRxiv 68\%, MedRxiv 20\%, ChemRxiv 3\%, plus a 9\% social-science cluster of PsyArXiv/EdArXiv/SocArXiv), we assembled 121{,}640 post-2023 empirical papers; arXiv was excluded because 73\% of its papers fall outside the classical hypothesis-testing tradition~\citep{cockburn2020threats,denning2013science} and arXiv full text dominates standard LLM pretraining corpora (LLaMA, Dolma, The Pile, RedPajama, Common Pile) while other platforms are largely excluded -- 10-gram overlap against a 10B-token Dolma~v1.6 pretraining sample matched only 4 of 8{,}000 random paragraphs, all false positives on inspection. For each paper, o3-mini extracted the author's explicit hypotheses and, separately, the scientific puzzle and surrounding context rewritten from the introduction and related-work sections (parsed via GROBID); leakage was suppressed by dropping any extracted context/puzzle as long as one sentence exceeded an MPNet~\citep{song2020mpnet} cosine similarity threshold of 0.82 (calibrated on 20{,}000 GPT-4.1 paraphrases of 1{,}000 human hypotheses, 95\% recall) against the paper's human hypotheses, with post-hoc author satisfaction of 98.62\% on hypotheses and 99.70\% on context/puzzle. \revise{Two annotators judged that 92\% and 82\%, respectively, of 300 random context--puzzle pairs revealed nothing about the direction of the authors' hypotheses; their flagged leakage concerned goal-oriented puzzles (``how can we enhance\ldots?''), which signal a direction but not the hypothesis's variables (Supplementary Information~\ref{sec:nulldef}).} We then prompted 26 representative LLMs from eight providers, spanning OpenAI, LLaMA, Gemma, Phi, Mistral, DeepSeek, Qwen, Grok, and Gemini model families, to generate ideas\revise{; human hypotheses were extracted as a numbered list and each model was asked to generate the same number of hypotheses, so that both sides had the same ceiling}. For each paper we sent the authors five distinguishable AI-generated hypotheses and the authors rated each on novelty, feasibility, probability, and \revise{willingness to pursue} after passing a comprehension check of the rubrics. In total 6{,}749 scientists participated, \revise{and 5,259 of them contributed} 25{,}139 valid four-dimensional evaluations. Full procedures in Supplementary Information~\ref{subsec:data_collection}, \ref{subsec:name_disambiguation},~\ref{full_list},~\ref{2.2}, and~\ref{subsec:survey_design}.

\textbf{\revise{Null-statement and AI-engaged classifiers}.} Two auxiliary classifiers support the analysis. The \revise{null-statement} classifier \revise{(used both for hypotheses and for abstract sentences)} feeds TF-IDF tokens (NLTK Treebank tokenization; stopwords retained to preserve negation cues such as ``no effect'') into an ensemble of two linear (support vector machine, logistic regression) and two nonlinear (random forest, gradient boosting) base learners, trained on 2{,}000 o3-mini-generated labels (null vs. not null) validated at 100\% agreement against two human annotators on a 200-instance manual check; 5-fold cross-validation accuracy is 99.5\%. \revise{Two annotators also agreed in all cases on whether 100 null statements from abstracts were indeed null statements, although they could not reliably tell from the surrounding wording whether the statement originated at ideation or at reporting (Supplementary Information~\ref{sec:nulldef}).} \revise{Prior AI engagement} of each participating scientist is computed by the proportion of prior papers \revise{classified as AI-engaged research (AI as a method or a topic)} from their 848{,}750 papers using the fine-tuned BERT title--abstract classifier\citep{hao2026artificial} (F1~$\approx$~0.875 against expert annotation with Fleiss' $\kappa = 0.96$); a Gemini 2.5 Flash re-annotation of AI use yields 98.80\% agreement. See Supplementary Information~\ref{classify} and~\ref{aiuse}.

\textbf{\revise{Willingness-to-pursue} regression.} We estimate \revise{willingness to pursue} as a function of scientist-rated novelty, feasibility, probability, and potential biases using a correlated random-effects (Mundlak) linear specification~\citep{schunck2013within} that decomposes each rating into a within-scientist deviation $(\text{rating}_{ij} - \overline{\text{rating}}_i)$ and a scientist-level mean $\overline{\text{rating}}_i$, isolating which drives a given scientist's own \revise{willingness} from persistent cross-scientist-level differences. Sources of bias include seniority (within-field yearly citation percentile; robust to within-field yearly publication-count percentile and log-academic-age proxies), prior AI \revise{engagement} (see above), human--AI idea cosine similarity (using the MPNet embedding model \citep{song2020mpnet}\revise{; robust to SPECTER2 embeddings and to displacement vectors}), and field. Average marginal predictions are reported with clustered robust standard errors and conclusions replicate under an author-fixed-effect specification, a Mundlak-decomposed ordered logistic model, and three additional specifications including interaction effects between field, seniority, and \revise{rated characteristics} (Supplementary Information~\ref{stats1}). \hl{Non-response does not appear to bias the sample. Respondents and non-respondents are statistically indistinguishable on every observed characteristic, and in all four fields the distribution of respondents across the 117 subtopics closely tracks the distribution of research in the full corpus (Pearson $r$=0.98; Extended Data Figure~\ref{fig:nonresponse} and Supplementary Information~\ref{app:label_nonres}). Re-estimating the main specification with weights equal to the inverse estimated probability of response yields coefficients similar in sign, magnitude, and statistical significance (Supplementary Information~\ref{app:selection}). Null results are supported by two one-sided tests of equivalence in Supplementary Information \ref{sec:equivalence}.}

\textbf{Novelty metrics.} We benchmark seven operationalizations of novelty against author novelty ratings\revise{; they capture lexical (the $n$-gram measures), semantic or distributional (embedding distance, cross-entropy, perplexity and NLI) and learned (GraphEval) notions of novelty}: semantic distance $1-\cos(\mathbf{e}_g, \mathbf{e}_c)$ where $\cos(\mathbf{e}_g, \mathbf{e}_c)$ is the cosine similarity between MPNet embeddings of the generated idea $g$ and the puzzle+context $c$; Jaccard-complement bi-gram and tri-gram divergence of $g$ and $c$; length-normalized cross-entropy of $g$'s tokens under the empirical token distribution of $c$; natural language inference-based derivation $p_e - p_c$ from DeBERTa~\citep{he2020deberta} (the probability of entailment minus contradiction)---lower, more novel; conditional model perplexity of $g$ given $c$ averaged across four well aligned 7B base models (Qwen-7B, Mistral-7B-v0.1, LLaMA-2-7B, and DeepSeek-LLM-7B-base); and GraphEval~\citep{feng2025grapheval}, a graph-based LLM idea evaluator that links $g$ via BERT-similarity edges into an idea graph, and predicts the decision probability distribution via aggregated neighbor node representations; we collapse its four-way output to an overall score $P(\text{Spotlight}) + P(\text{Oral}) + P(\text{Poster})$. Full formulations are shown in Supplementary Information~\ref{novelty}.

\textbf{\revise{Reward model and reviewer benchmark}.} We trained a multi-dimensional reward model $r(y) = (r_{\text{nov}}, r_{\text{feas}}, r_{\text{prob}})$ on the four consecutive pairs $(h_1,h_2),(h_2,h_3),(h_3,h_4),(h_4,h_5)$ from each scientist's five-hypothesis ranking; the non-consecutive pairs are deducible by transitivity and induce severe overfitting (training accuracy $\approx$~90\%, test $<$~60\%). The Bradley--Terry loss incorporates both preference direction and score-gap margin. An 80-trial grid search over learning rate and margin weight selects hyperparameters as well as the backbone (Qwen3-14B). As an interpretive baseline, we benchmark against the agreement between non-overlapping human reviewers across 26{,}731 OpenReview submissions to 46 venues 2017--2025~\citep{bao2025language}: a randomized pairwise direction check \revise{across reviewer slots} yields 61.0\%~$\pm$~0.1\% consistency. LLM-as-a-judge, SOTA models, and popular metrics fall below this human baseline on our task (Supplementary Information~\ref{humanalign} and~\ref{reward}).

\revise{\textbf{Corpus analysis.} We retrieved all 38,863,847 medicine and social-science papers published between 2010 and 2025 from OpenAlex. AI-engaged research was identified from titles and abstracts with the classifier above, and each abstract sentence was scored with the null-statement classifier; a paper is all-significant if no sentence is classified as null. For the knowledge-space measures, each paper's title and abstract were embedded with SPECTER2~\citep{Singh2022SciRepEvalAM} (768 dimensions). For each field and year, the center is the mean embedding and the radius is the mean Euclidean distance of papers to the center. Center drift is the Euclidean distance between a year's center and the 2010 center, and knowledge extension is the year-over-year ratio of radii, following ref.~\citep{hao2026artificial}.}

\hl{\textbf{Simulation model.} \revise{The simulation is a conceptual illustration whose parameters are motivated by, but not calibrated to, our empirical findings.} We model a field as pooling many noisy estimates of a whole \emph{space} of true effects $\{\theta_j\}$ that trace the many questions a field could ask, each with a non-negative true magnitude, most small and a few large. We then track the mean-squared error (MSE) of the field's pooled estimate, which for each question decomposes into an averageable variance term, a shared bias, and an error-correlation floor. AI adoption is a single dial $a\in[0,1]$ that moves four primitives: it raises throughput and productivity $n$ (beneficial) but also the cross-study error correlation $\rho$, the shared blind-spot bias $B_h$, and the rate $s$ at which null results get suppressed, all of which are harmful for scientific understanding. \revise{The last of these acts through selective visibility (e.g., publication). The modelled distribution is the sampling distribution of study estimates, $\hat\theta_{i,j}\sim\mathcal{N}(\theta_j,\sigma^2)$; a statement is null when it falls below a decision threshold, and nulls are visible to other scientists (e.g., published) with probability $1-s$ ($s=0.75$ for human research and $0.85$ under full automation). The threshold is set at zero without loss of generality, because a nonzero significance threshold is equivalent to shifting true effects relative to the boundary (Supplementary Information~\ref{simlation}). Selection inflates visible estimates (the exaggeration, or Type~M, error) most for small effects, by far the most common kind, and barely affects large ones.} The key asymmetry is that throughput is the only channel reaching the averageable term, and it does so with diminishing returns ($\sim\!1/n^2$), while the three harms sit in the irreducible part. Net knowledge $V=1-\MSE$ is therefore a trade-off between benefits and harms. This model \revise{illustrates} three \revise{patterns}: (1) the optimal AI share $a^\ast$ is interior, so a human--AI mix dominates full automation; (2) $a^\ast$ is lower in mature fields because AI's throughput gains matter most in young, study-poor fields and least where evidence is abundant; and (3) \revise{net knowledge} at high adoption \revise{falls further} as null suppression intensifies. Details can be found in Supplementary Information~\ref{simlation}.}

\revise{\textbf{Ethics.} The survey was exempted by the University of Chicago Institutional Review Board (protocol IRB25-1372, ``The Capabilities and Potential of AI for Automating Scientific Idealization: A Large-Scale Human-in-the-Loop Study''). All research was performed in accordance with relevant guidelines and regulations. Informed consent was obtained from all participants. Participants could restrict the use of their papers (for example, to open-source models only), and ratings outside the scope of their consent were removed.}

\revise{\textbf{Use of AI tools.} LLMs were the objects of study and were also used as research instruments, as described above: to extract hypotheses, contexts and puzzles, to generate hypotheses, as judges, to generate classifier training labels, and, in a control experiment, to polish abstracts. We used AI to polish the language of the manuscript and to assist in writing code. All AI-generated text and code were reviewed, edited and verified by the authors before use.}

\printrefsegment{methods}{\revise{Methods references}}

\section*{\revise{Data availability}}
Aggregated, de-identified ratings\revise{ and the derived corpus-level measures} have been released at the \href{https://github.com/listar2000/science-reward-model}{project repository}. Per IRB25-1372, individual identifiers will not be released. \revise{Preprints are available from the respective preprint servers and bibliographic records from OpenAlex.}

\section*{\revise{Code availability}}
\revise{The} trained reward-model checkpoints\revise{ and} pipeline code have been released at the \href{https://github.com/listar2000/science-reward-model}{project repository}\revise{ and will be archived in a DOI-minting repository.}

\section*{Acknowledgements}
We thank Misha Teplitskiy, Julia Koschinsky, Pengda Wang, Beichen Lu, Kim Weeden, Robert Ward, Zhen Zhang, Junsol Kim, Gio Choi, Austin Kozlowski, Philip N. Cohen, Laura K.~Nelson, Juan Pablo Pardo-Guerra, Alexander C.~Furnas, Yian Yin, Alex Yan, and participants at the Yale AI for Social Science Methods Workshop for discussion. We thank Knowledge Lab members and the many scientists who participated in, and wrote to us about, the survey, as well as the Bluesky community, whose debates sharpened the study's design. Computation used the University of Chicago Midway / Data Science Institute Research Computing Cluster.

\section*{\revise{Funding}}
\revise{This research received no specific grant from any funding agency in the public, commercial, or not-for-profit sectors.}

\section*{Author contributions}
H.B.\ and J.A.E.\ designed the study. H.B.\ drafted the survey. S.C.\ and
H.B.\ implemented the survey, with all authors contributing to its
distribution. S.W., S.L., X.L.\ and H.B.\ designed the reward-model
experiments. H.B. conducted data analysis, simulation models, and wrote the first draft. S.W., X.L., S.L., and J.A.E.\ contributed to revising the manuscript.
J.A.E.\ supervised the work.

\section*{Competing interests}
J.A.E.\ is affiliated with Google. \revise{The other authors declare no competing interests.}

\section*{\revise{Ethics statement}}
\revise{The study was exempted by the University of Chicago Institutional Review Board (IRB25-1372) and informed consent was obtained from all participants.}

\section*{\revise{Correspondence}}
\revise{Correspondence and requests for materials should be addressed to Honglin Bao (\href{mailto:honglinbao@uchicago.edu}{honglinbao@uchicago.edu}) or James A.~Evans (\href{mailto:jevans@uchicago.edu}{jevans@uchicago.edu}).}

\clearpage
\setcounter{figure}{0}
\setcounter{table}{0}
\renewcommand{\figurename}{Extended Data Figure}
\renewcommand{\tablename}{Extended Data Table}
\renewcommand{\theHfigure}{ED.\arabic{figure}}
\renewcommand{\theHtable}{ED.\arabic{table}}

\captionsetup{labelfont=bf,labelsep=period}
{\Huge \textbf{Extended Data \revise{Figures and Tables}}}

\ifshowdisplayitems\else
\revise{\noindent Extended Data legends are listed below; the Extended Data figures and tables themselves are supplied as separate image files.}
\fi

\vspace{5em}

\begin{figure}[htbp]
    \centering
    \displayitem{\includegraphics[width=0.75\textwidth]{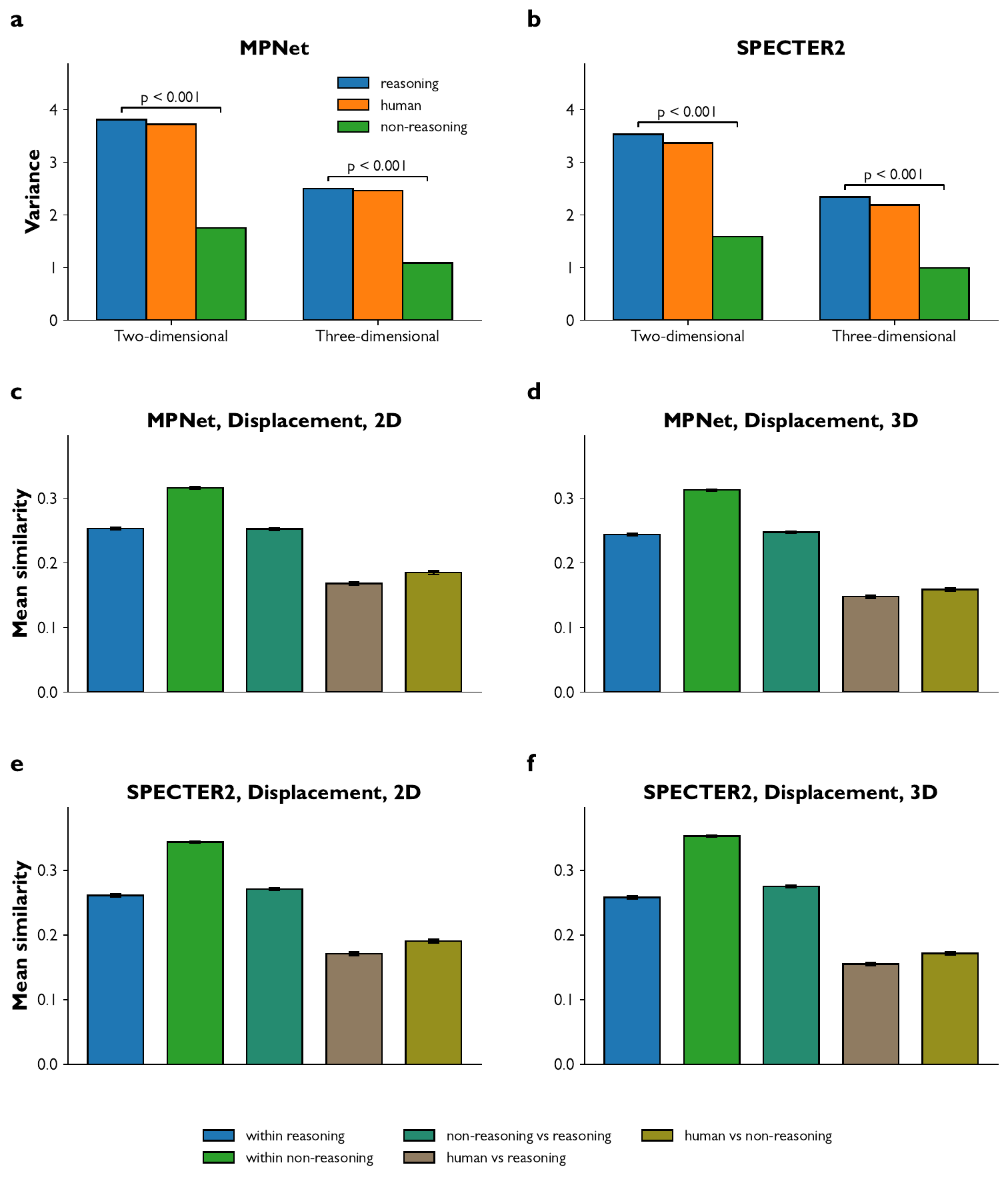}}
    \caption{\revise{\textbf{Robustness checks for hypothesis-space comparisons.} Embeddings are projected into two or three dimensions with UMAP, and each hypothesis is represented as a displacement vector relative to its context-and-puzzle origin. We then compare the diversity and pairwise similarity of these vectors across source groups. \textbf{a}, Diversity under MPNet embeddings. Reasoning models traverse more of the hypothesis space than non-reasoning models, and this conclusion is robust to the choice of dimensionality-reduction methods not only UMAP: PCA (2D: 0.15 vs.\ 0.09, $p=0.001$; 3D: 0.15 vs.\ 0.09, $p<0.001$) and t-SNE (2D: 38.1 vs.\ 17.8, $p=0.004$; 3D: 16.1 vs.\ 7.5, $p<0.001$). We find no substantial or significant difference between reasoning models and humans. \textbf{b}, The same conclusions hold under SPECTER2 embeddings. Each bar represents the mean across models and dimensions. \textbf{c}--\textbf{f}, Within-group similarity across 2D and 3D projections, using both SPECTER2 and MPNet embeddings. Non-reasoning models consistently show the strongest within-group homogeneity. When only displacement vectors are considered, within-group similarity is substantially lower than in Fig.~\ref{fig:ideation}b. More details can be found in Supplementary Information \ref{sec:si_space}.}}
    \label{fig:diverse}
\end{figure}

\newpage

\begin{figure}[htbp]
    \centering
    \displayitem{\includegraphics[width=1\textwidth]{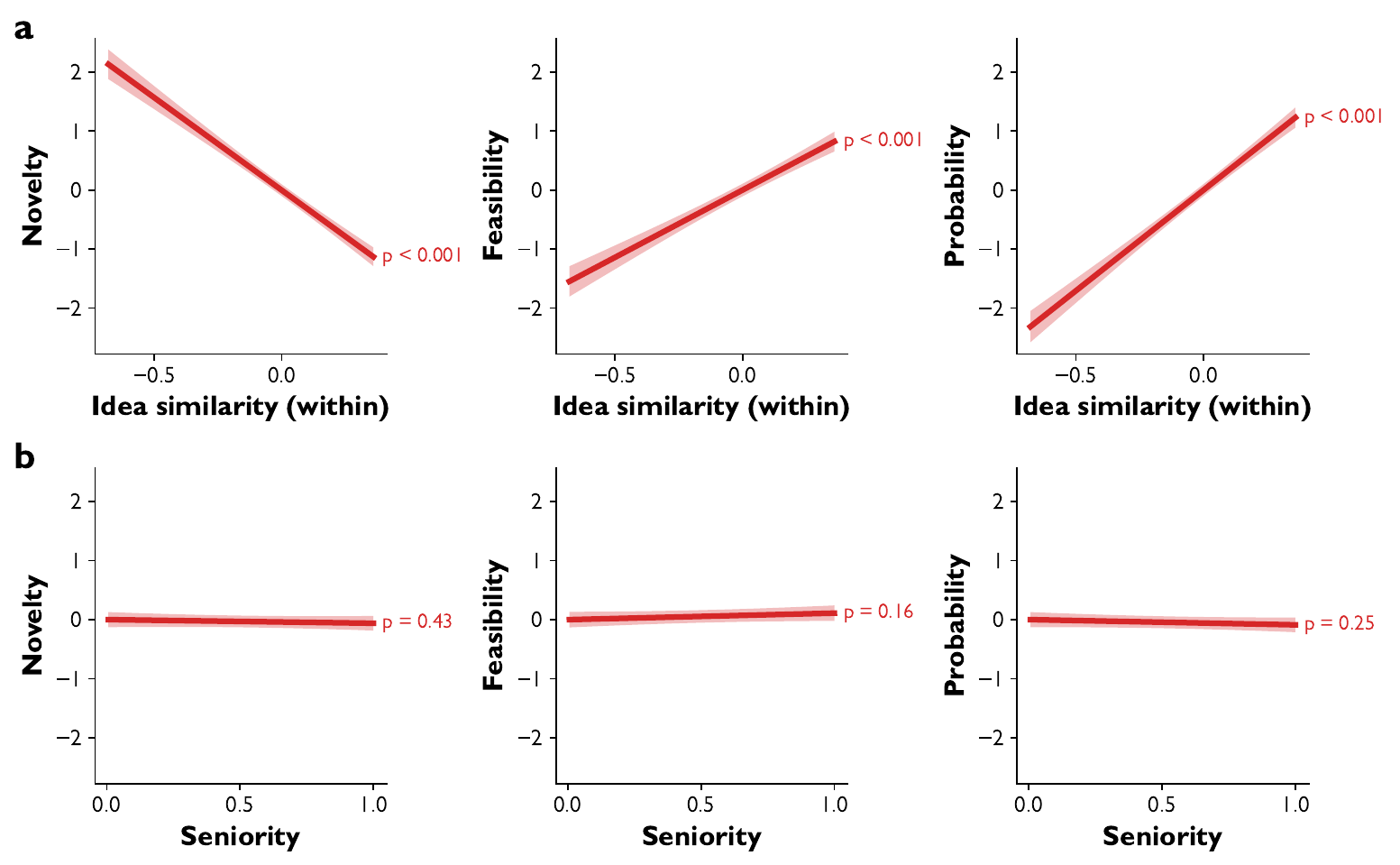}}
    \caption{\textbf{Human-AI idea alignment shapes \revise{rated characteristics}, but seniority \revise{shapes only willingness to pursue}}. Regressing the potential sources of AI-idea bias from humans against \revise{rated novelty, feasibility and probability} (marginal predictions are shown): \textbf{a}, AI ideas that resemble what evaluators have produced themselves are rated as more feasible and more likely to be true, yet less novel. Authors who have conducted similar work are more likely to view the idea as both implementable and valid. However, alignment with one's own ideas penalizes perceived novelty, since what overlaps with one's own thinking is, by construction, less surprising. \textbf{b}, seniority \revise{is not significantly associated with any rated characteristic (all $p>0.05$), although it is associated with lower willingness to pursue (main text Figure~\ref{fig:judgment}b,c)}. The effect of human bias toward AI ideas reported in the main paper is the additional impact after controlling for \revise{rated characteristics}. Top row: the $x$-axis is human--AI idea similarity, demeaned within scientists. Bottom row: the $x$-axis is evaluator seniority, the percentile rank (0--1) of the evaluator's yearly citation count within the field. In all panels, the $y$-axis is the rating demeaned within evaluator, which is equivalent to including evaluator fixed effects and removes systematically harsh or lenient rating tendencies. Error bands = 95\% CI.}
    \label{fig:percep}
\end{figure}
\clearpage

\begin{figure}[htbp]
    \centering
    \displayitem{\includegraphics[width=1\textwidth]{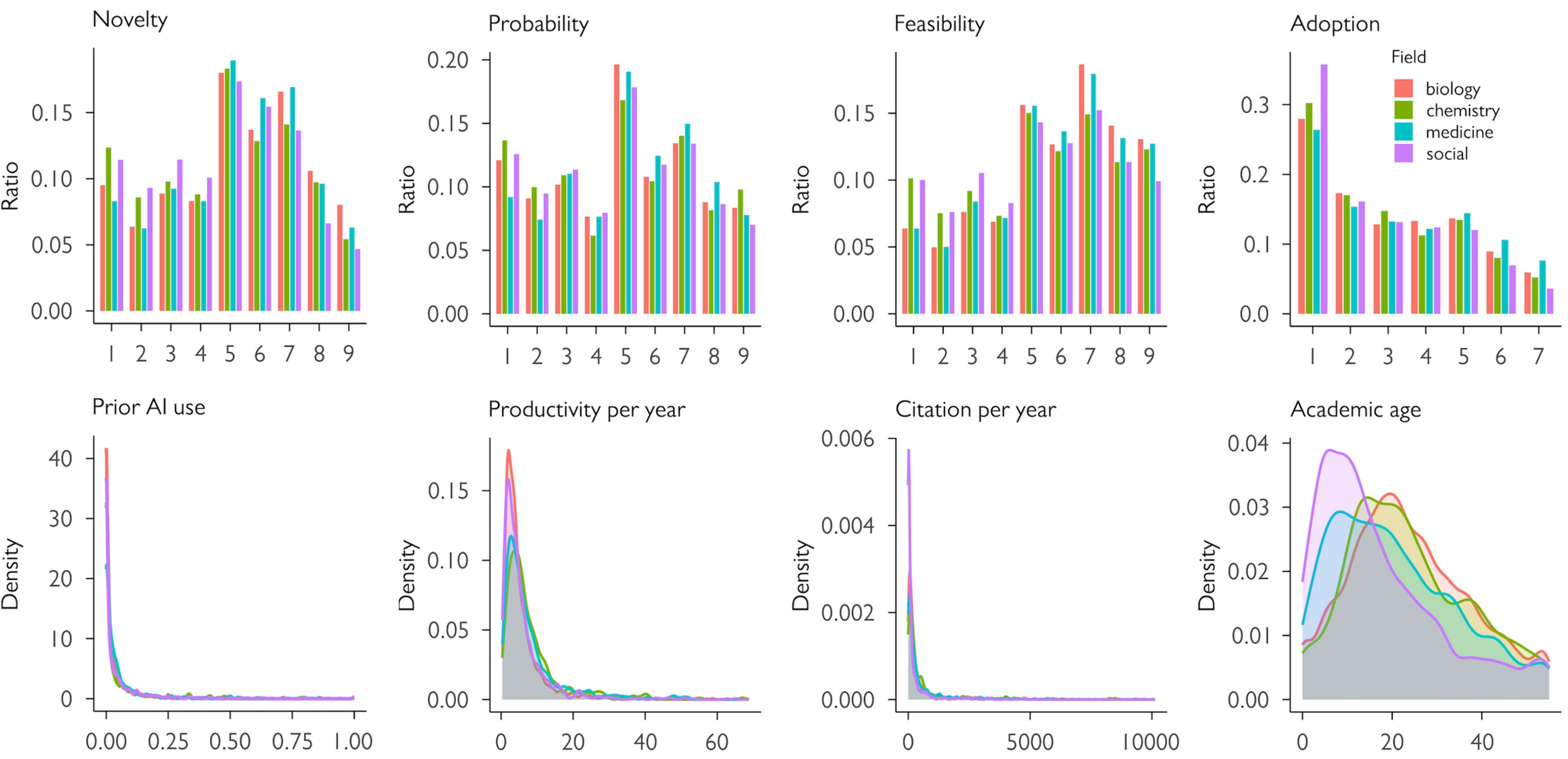}}
    \caption{\hl{\textbf{Distribution of variables}. Across fields, the \revise{rated characteristics} (novelty, feasibility, and probability) share a similar shape that is approximately normal, whereas \revise{willingness to pursue} departs from normality and is right-skewed. The seniority measures (yearly citations, yearly publications, and age) and prior AI \revise{engagement} likewise exhibit right-skewed distributions, visualized via Gaussian kernel density estimation. These patterns motivate the statistical transformations applied to these variables in the main text.}}
    \label{fig:distribution}
\end{figure}
\clearpage

\begin{figure}[htbp]
    \centering
    \displayitem{\includegraphics[width=1\textwidth]{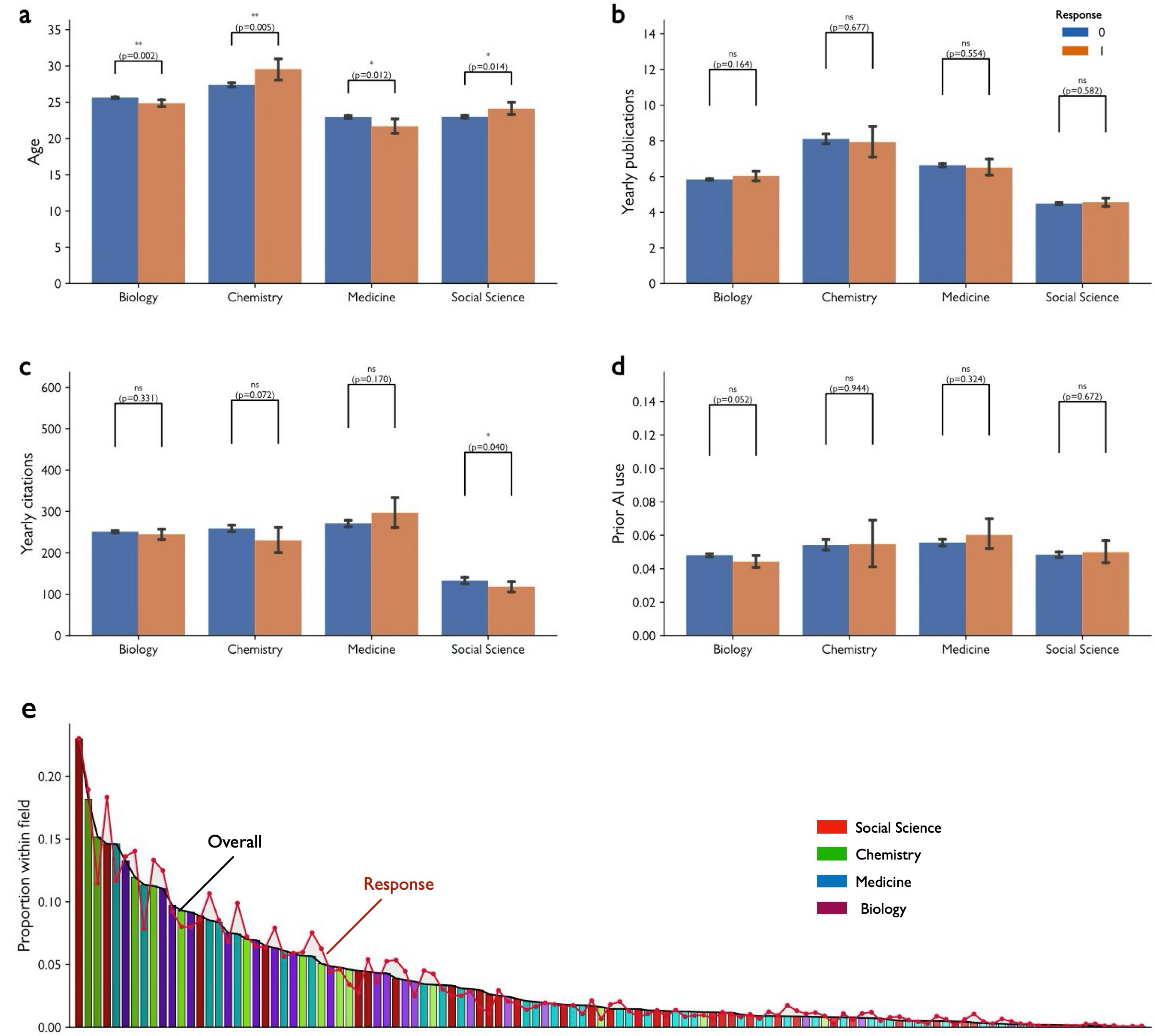}}
    \caption{\hl{\textbf{No significant differences between respondent and nonrespondent researchers}. Across biology, medicine, chemistry, and social science, we find no evidence that survey respondents differ systematically from nonrespondents in age (\textbf{a}), productivity (\textbf{b}), citations (\textbf{c}), or prior AI \revise{engagement} (\textbf{d}). Error bars = 95\% CI. Respondents are also representative at the subtopic level: in all four fields, the distribution of respondents across the 117 subtopics closely tracks the distribution of research in the overall corpus (Pearson $r$=0.98, panel e, and Supplementary Information~\ref{app:label_nonres}). The 117 subtopics are ordered left to right by their within-field percentile in the overall corpus (black line); the corresponding percentile among respondent scientists is shown in red.}}
    \label{fig:nonresponse}
\end{figure}
\clearpage

\begin{table}[htbp]
\centering
\caption{\textbf{Pairwise judgment accuracy across models and settings}. On a held-out set of 5,000 human preference pairs, we evaluate state-of-the-art reward models and two LLM-as-a-judge configurations: (i) direct comparison pairwise prompting, in which the judge is asked which idea is better, and (ii) rating-based pairwise judgment, in which the judge assigns individual scores that are subsequently converted into a pairwise preference. All reward models, including ours, follow the rating-based (score-then-convert) protocol (that is how they are trained). \revise{Our general model (Extended Data Table~\ref{tab:accuracy_all_dims}) outperforms every rating-based baseline in every field.} Notably, the direct-comparison LLM-as-a-judge exhibits strong framing bias: its predictions are unstable under logically equivalent rephrasings of the same query. \revise{Values are pairwise accuracies (share of pairs whose predicted direction matches the scientist's preference). ``Converted ratings'' indicates whether scores were converted into pairwise preferences (Y) or the judge compared the pair directly (N).}}
\label{tab:llm_judge_accuracy}
\ifshowdisplayitems
\renewcommand{\arraystretch}{1.2}
\newcolumntype{C}{>{\centering\arraybackslash}X}
\newcolumntype{L}{>{\raggedright\arraybackslash}X}
\begin{tabularx}{\linewidth}{L L c c c C}
\toprule
Model & Label & Feasibility & Probability & Novelty & Converted ratings\\
\midrule
nicolinho/QRM-Gemma-2-27B      & SOTA Reward model & 0.51 & 0.51 & 0.47 & Y \\
Skywork-Reward-V2-Qwen3-8B     & SOTA Reward model & 0.53 & 0.55 & 0.49 & Y \\
Skywork-Reward-V2-Llama-3.1-8B & SOTA Reward model & 0.50 & 0.53 & 0.51 & Y \\
OpenAI o4-mini Deep Research   & LLM-as-a-judge & 0.41 & 0.55 & 0.43 & Y \\
DeepSeek R1                    & LLM-as-a-judge & 0.40 & 0.48 & 0.41 & Y \\
Gemini 2.5 Flash               & LLM-as-a-judge & 0.36 & 0.43 & 0.30 & Y \\
OpenAI o4-mini Deep Research   & LLM-as-a-judge & 0.59 & 0.61 & 0.61 & N (direct comparison) \\
DeepSeek R1                    & LLM-as-a-judge & 0.56 & 0.60 & 0.55 & N (direct comparison) \\
Gemini 2.5 Flash               & LLM-as-a-judge & 0.58 & 0.60 & 0.53 & N (direct comparison)\\
\bottomrule
\end{tabularx}
\fi
\end{table}

\newpage

\begin{table}[htbp]\centering
\caption{\textbf{Factors \revise{associated with willingness to pursue}}. Major conclusions are robust across seniority measures (within-field yearly citation/publication percentile and academic age). We use the citation measure to represent seniority by default. Across the paper, prior AI \revise{engagement} is also log-transformed due to its long-tailed distribution. \revise{Each coefficient is followed by its cluster-robust standard error in parentheses. ``Within'' terms are deviations from the scientist's own mean rating and ``between'' terms are the scientist's mean rating. Social science is the reference field. *** $p<0.001$, ** $p<0.01$, * $p<0.05$.}}
\label{tab:ols_combined}
\ifshowdisplayitems
\begin{tabular}{lccc}
\toprule
 & (1) Citation pct. & (2) Productivity pct. & (3) log(Age) \\
\midrule
Intercept                          & -0.9717*** & -1.0100*** & -0.9505*** \\
                                   & (0.117)    & (0.116)    & (0.123)    \\
Biology                            & 0.1245***  & 0.1244***  & 0.1426***  \\
                                   & (0.037)    & (0.037)    & (0.038)    \\
Chemistry                          & 0.1635**   & 0.1635**   & 0.1813**   \\
                                   & (0.061)    & (0.061)    & (0.061)    \\
Medicine                           & 0.2295***  & 0.2287***  & 0.2397***  \\
                                   & (0.050)    & (0.050)    & (0.050)    \\
Novelty (within)                   & 0.1227***  & 0.1227***  & 0.1226***  \\
                                   & (0.006)    & (0.006)    & (0.006)    \\
Novelty (between)                  & 0.2191***  & 0.2198***  & 0.2199***  \\
                                   & (0.010)    & (0.010)    & (0.010)    \\
Feasibility (within)               & 0.1296***  & 0.1296***  & 0.1296***  \\
                                   & (0.006)    & (0.006)    & (0.006)    \\
Feasibility (between)              & 0.0730***  & 0.0722***  & 0.0719***  \\
                                   & (0.012)    & (0.012)    & (0.012)    \\
Probability (within)               & 0.3143***  & 0.3143***  & 0.3142***  \\
                                   & (0.006)    & (0.006)    & (0.006)    \\
Probability (between)              & 0.4468***  & 0.4481***  & 0.4474***  \\
                                   & (0.012)    & (0.012)    & (0.012)    \\
Human-AI idea similarity (within)  & 1.2835***  & 1.2835***  & 1.2825***  \\
                                   & (0.120)    & (0.120)    & (0.120)    \\
Human-AI idea similarity (between) & 0.4726***  & 0.4717***  & 0.4684***  \\
                                   & (0.148)    & (0.148)    & (0.147)    \\
Seniority                          & -0.1605*** & -0.0970*   & -0.0386*   \\
                                   & (0.049)    & (0.049)    & (0.018)    \\
\revise{Prior AI engagement}       & -0.1697    & -0.0793    & -0.2315    \\
                                   & (0.382)    & (0.382)    & (0.390)    \\
\midrule
Observations  & 23,615 & 23,615 & 23,635 \\
$R^2$         & 0.395  & 0.395  & 0.395  \\
Adj. $R^2$    & 0.395  & 0.395  & 0.395  \\
F-statistic   & 874.5  & 873.5  & 876.0  \\
\bottomrule
\end{tabular}
\fi
\end{table}

\clearpage

\begin{table}[htbp]\centering
\caption{\textbf{Regression results with interaction effects between seniority and field}. Once a field-by-seniority interaction is included in the model, the field main effects vanish. Senior scientists in every field are \revise{less willing to pursue AI-generated hypotheses}, with the decay being most pronounced in the social sciences. \revise{The outcome is willingness to pursue. Cluster-robust standard errors in parentheses. Field-level comparisons use social science as the baseline.} *** $p<0.001$, ** $p<0.01$, * $p<0.05$.}
\label{tab:ols_results34}
\ifshowdisplayitems
\footnotesize
\begin{tabular}{lc}
\toprule
 & \revise{Willingness to pursue} \\
\midrule
Intercept & -0.8070*** \\
 & (0.126) \\
\midrule
Biology & -0.0983 \\
 & (0.076) \\
Chemistry & -0.0545 \\
 & (0.125) \\
Medicine & 0.0340 \\
 & (0.101) \\
\midrule
Novelty (within) & 0.1227*** \\
 & (0.006) \\
Novelty (between) & 0.2178*** \\
 & (0.010) \\
\midrule
Feasibility (within) & 0.1296*** \\
 & (0.006) \\
Feasibility (between) & 0.0742*** \\
 & (0.012) \\
\midrule
Probability (within) & 0.3143*** \\
 & (0.006) \\
Probability (between) & 0.4457*** \\
 & (0.012) \\
\midrule
Human-AI Idea Similarity (within) & 1.2835*** \\
 & (0.120) \\
Human-AI Idea Similarity (between) & 0.4927*** \\
 & (0.148) \\
\midrule
Seniority & -0.5073*** \\
 & (0.112) \\
Seniority $\times$ Biology & 0.4451*** \\
 & (0.129) \\
Seniority $\times$ Chemistry & 0.4364* \\
 & (0.208) \\
Seniority $\times$ Medicine & 0.3919* \\
 & (0.174) \\
\midrule
\revise{Prior AI engagement} & -0.1605 \\
 & (0.382) \\
\midrule
Observations & 23,615 \\
$R^2$ & 0.396 \\
\bottomrule
\end{tabular}
\fi
\end{table}

\clearpage

\begin{table}[htbp]\centering
\caption{\textbf{Regression results with interactions between field and \revise{rated characteristics}}. Social science is more tolerant of risky (less probable) ideas compared to biomedicine. \revise{The outcome is willingness to pursue. Cluster-robust standard errors in parentheses. Field-level comparisons use social science as the baseline.} *** $p<0.001$, ** $p<0.01$, * $p<0.05$.}
\label{tab:ols_results}
\ifshowdisplayitems
\scriptsize
\renewcommand{\arraystretch}{0.95}
\begin{tabular}{lc}
\toprule
 & \revise{Willingness to pursue} \\
\midrule
Intercept & -0.9717*** \\
 & (0.117) \\
\midrule
Biology & 0.1245*** \\
 & (0.037) \\
Chemistry & 0.1635** \\
 & (0.061) \\
Medicine & 0.2295*** \\
 & (0.050) \\
\midrule
Novelty (within) & 0.1338*** \\
 & (0.014) \\
Novelty (within) $\times$ Biology & -0.0143 \\
 & (0.017) \\
Novelty (within) $\times$ Chemistry & -0.0142 \\
 & (0.027) \\
Novelty (within) $\times$ Medicine & -0.0084 \\
 & (0.021) \\
Novelty (between) & 0.2191*** \\
 & (0.010) \\
\midrule
Feasibility (within) & 0.1316*** \\
 & (0.012) \\
Feasibility (within) $\times$ Biology & -0.0036 \\
 & (0.015) \\
Feasibility (within) $\times$ Chemistry & -0.0183 \\
 & (0.025) \\
Feasibility (within) $\times$ Medicine & 0.0094 \\
 & (0.021) \\
Feasibility (between) & 0.0730*** \\
 & (0.012) \\
\midrule
Probability (within) & 0.2692*** \\
 & (0.013) \\
Probability (within) $\times$ Biology & 0.0584*** \\
 & (0.016) \\
Probability (within) $\times$ Chemistry & 0.0208 \\
 & (0.025) \\
Probability (within) $\times$ Medicine & 0.0632** \\
 & (0.022) \\
Probability (between) & 0.4468*** \\
 & (0.012) \\
\midrule
Human-AI Idea Similarity (within) & 1.1298*** \\
 & (0.232) \\
Human-AI Idea Similarity (within) $\times$ Biology & 0.1499 \\
 & (0.286) \\
Human-AI Idea Similarity (within) $\times$ Chemistry & -0.6233 \\
 & (0.565) \\
Human-AI Idea Similarity (within) $\times$ Medicine & 0.5567 \\
 & (0.377) \\
Human-AI Idea Similarity (between) & 0.4726*** \\
 & (0.148) \\
\midrule
Seniority & -0.1605*** \\
 & (0.049) \\
\midrule
\revise{Prior AI engagement} & -0.1697 \\
 & (0.382) \\
\midrule
Observations & 23,615 \\
$R^2$ & 0.396 \\
\bottomrule
\end{tabular}
\fi
\end{table}

\clearpage

\begin{table}[htbp]\centering
\caption{\textbf{General model vs. specific model for one \revise{rated dimension}}. Test accuracy is reported separately for each evaluation domain
(Biology, Chemistry, Medicine, Social Science), evaluated on held-out
scientist ratings excluded from all training sets. The general model assigns $\alpha_d = 1$ across all
dimensions, while each dimension-specific model sets $\alpha_d = 1$ for a single dimension and $\alpha_d = 0$ otherwise. In order to conduct a fair comparison, we fixed the learning rate to be $3\times10^{-5}$ and the margin weight to be 1. We can simultaneously optimize all three dimensions without the need to train a separate model for each specific dimension. \revise{Models trained on a single dimension transfer poorly to the other dimensions, indicating that the three dimensions are non-redundant, although not necessarily independent. Values are pairwise accuracies.}}
\label{tab:weight_config_all}
\ifshowdisplayitems
\small
\begin{tabular}{llccc}
\toprule
\textbf{Domain} & \textbf{Config} & \textbf{Novelty} & \textbf{Feasibility} & \textbf{Probability}\\
\midrule
\multirow{4}{*}{Training}
 & General          & 0.6845 & 0.6800 & 0.6312 \\
 & Novelty-only     & 0.7109 & 0.4430 & 0.4839 \\
 & Feasibility-only & 0.4760 & 0.6142 & 0.5198 \\
 & Probability-only & 0.6105 & 0.4655 & 0.6313 \\
\midrule
\multirow{4}{*}{Biology}
 & General          & 0.6311 & 0.5667 & 0.6424 \\
 & Novelty-only     & 0.6443 & 0.4489 & 0.4913 \\
 & Feasibility-only & 0.4588 & 0.5707 & 0.5578 \\
 & Probability-only & 0.6212 & 0.4448 & 0.6957 \\
\midrule
\multirow{4}{*}{Chemistry}
 & General          & 0.6283 & 0.5912 & 0.6766 \\
 & Novelty-only     & 0.6626 & 0.4863 & 0.5268 \\
 & Feasibility-only & 0.6077 & 0.5757 & 0.4984 \\
 & Probability-only & 0.5208 & 0.5090 & 0.6350 \\
\midrule
\multirow{4}{*}{Medicine}
 & General          & 0.6218 & 0.6597 & 0.6845 \\
 & Novelty-only     & 0.6176 & 0.4077 & 0.4399 \\
 & Feasibility-only & 0.4637 & 0.6486 & 0.5700 \\
 & Probability-only & 0.6086 & 0.4094 & 0.6513 \\
\midrule
\multirow{4}{*}{Social}
 & General          & 0.6281 & 0.6216 & 0.6361 \\
 & Novelty-only     & 0.6267 & 0.4640 & 0.5025 \\
 & Feasibility-only & 0.4255 & 0.6378 & 0.5192 \\
 & Probability-only & 0.5650 & 0.4164 & 0.6262 \\
\bottomrule
\end{tabular}
\fi
\end{table}

\clearpage

\begin{table}[ht]
  \centering
  \caption{\textbf{Train accuracy \& test accuracy per evaluation domain across dimensions}. Bolded values indicate the better one between the domain-specific model and the general model. We observe that across all dimensions, the domain-specific models for biology and social science are consistently better, in some cases outperforming the general model by nearly 10\% (\revise{5.4 percentage points above} the 56.7\% feasibility accuracy for the general model tested in biology). However, in domains with limited data, such as chemistry and medicine, the general model sometimes performs better. When training data is sufficient (i.e., Biology and Social Science), field-specific models can capture domain-distinctive taste; however, when data is limited, the cross-field general model is still able to capture the underlying logic of scientific judgment. In the study, for a fair comparison, we fixed the learning rate to be $3\times10^{-5}$ and the margin weight to be 1. \revise{Panels: \textbf{a}, novelty; \textbf{b}, feasibility; \textbf{c}, probability. ---, not applicable (domain-specific models are tested only in their own domain).}}
  \label{tab:accuracy_all_dims}
\ifshowdisplayitems
  \begin{subtable}{\linewidth}
    \centering
    \caption{Novelty}
    \label{tab:novelty}
    \begin{tabular}{lccccc}
      \toprule
      \textbf{Domain} & \textbf{Train Acc.}
        & \textbf{Biology} & \textbf{Chemistry} & \textbf{Medicine} & \textbf{Social} \\
      \midrule
      General          & 0.6845 & 0.6311 & \textbf{0.6283} & 0.6218 & 0.6281 \\
      Biology          & 0.7221 & \textbf{0.6864} & ---    & ---    & ---    \\
      Chemistry        & 0.5612 & ---    & 0.6001 & ---    & ---    \\
      Medicine         & 0.6565 & ---    & ---    & \textbf{0.6651} & ---    \\
      Social Science   & 0.6939 & ---    & ---    & ---    & \textbf{0.6434} \\
      \bottomrule
    \end{tabular}
  \end{subtable}

  \vspace{1em}

  \begin{subtable}{\linewidth}
    \centering
    \caption{Feasibility}
    \label{tab:feasibility}
    \begin{tabular}{lccccc}
      \toprule
      \textbf{Domain} & \textbf{Train Acc.}
        & \textbf{Biology} & \textbf{Chemistry} & \textbf{Medicine} & \textbf{Social} \\
      \midrule
      General          & 0.6800 & 0.5667 & \textbf{0.5912} & \textbf{0.6597} & 0.6216 \\
      Biology          & 0.6090 & \textbf{0.6205} & ---    & ---    & ---    \\
      Chemistry        & 0.5612 & ---    & 0.4756 & ---    & ---    \\
      Medicine         & 0.6057 & ---    & ---    & 0.5959 & ---    \\
      Social Science   & 0.5999 & ---    & ---    & ---    & \textbf{0.6218} \\
      \bottomrule
    \end{tabular}
  \end{subtable}

  \vspace{1em}

  \begin{subtable}{\linewidth}
    \centering
    \caption{Probability}
    \label{tab:probability}
    \begin{tabular}{lccccc}
      \toprule
      \textbf{Domain} & \textbf{Train Acc.}
        & \textbf{Biology} & \textbf{Chemistry} & \textbf{Medicine} & \textbf{Social} \\
      \midrule
      General          & 0.6312 & 0.6424 & \textbf{0.6766} & \textbf{0.6845} & 0.6361 \\
      Biology          & 0.6979 & \textbf{0.6716} & ---    & ---    & ---    \\
      Chemistry        & 0.5612 & ---    & 0.6203 & ---    & ---    \\
      Medicine         & 0.6598 & ---    & ---    & 0.6551 & ---    \\
      Social Science   & 0.6828 & ---    & ---    & ---    & \textbf{0.6682} \\
      \bottomrule
    \end{tabular}
  \end{subtable}
\fi
\end{table}

\clearpage

\newpage
\appendix

\setcounter{figure}{0}
\setcounter{table}{0}
\renewcommand{\figurename}{SI Figure}
\renewcommand{\tablename}{SI Table}
\renewcommand{\theHfigure}{SI.\arabic{figure}}
\renewcommand{\theHtable}{SI.\arabic{table}}
\definecolor{boxframe}{RGB}{150,170,200}
\definecolor{boxback}{RGB}{246,248,251}
\newtcolorbox{examplebox}[1]{enhanced, breakable, colback=boxback, colframe=boxframe, coltitle=red, colupper=red, fonttitle=\bfseries\small, title={#1}, boxrule=0.6pt, arc=2pt, left=6pt, right=6pt, top=4pt, bottom=4pt, fontupper=\small, before skip=8pt, after skip=8pt}
\newcommand{\nullhyp}{\textbf{(null)}}
\newcommand{\nonnull}{\textbf{(non-null)}}
\newcommand{\exfield}[1]{\hfill\textit{#1}}

\begin{center}
{\Huge \textbf{Supplementary Information \\[0.1em] for \\[0.5em] ``\revise{Thousands of scientists find AI rarely says no in science}''}}
\end{center}

\vspace{1em}

\tableofcontents

\newpage

\section{Methods}
\label{sec:methods}

\subsection{Data collection}
\label{subsec:data_collection}
We collected 121,640 papers posted after 2023 from six preprint platforms: BioRxiv (68\%; biology), ChemRxiv (3\%; chemistry), MedRxiv (20\%; medical science), PsyArXiv (psychology), EdArXiv (education), and SocArXiv (general social science), with the social science platforms collectively accounting for 9\%. Our dataset consists of full-text empirical papers, each containing at least one extracted human hypothesis (described in detail in Subsection \ref{2.2}). For the hypothesis generation process, we used 26 well-representative LLMs spanning both open-source and commercial models from eight mainstream companies (the full list can be found in Subsection \ref{full_list}). Unlike other hypothesis generation studies \citep{o2025sparks}, which concentrate exclusively on computer science papers, we deliberately excluded papers from arXiv, the most accessible data source. Our rationale is that we find 73\% of papers on arXiv fall into computing and mathematical domains---specifically, computer science (43\%), mathematics (18\%), statistics (5\%), and electronic engineering (7\%). In contrast, hypothesis testing is a practice most deeply embedded in the natural and social sciences. In computing and mathematical fields, much of the work is algorithmic or engineering-oriented, and often does not follow the classical hypothesis-testing framework \citep{cockburn2020threats, denning2013science}. More importantly, full-text papers from arXiv are already largely included in common pretraining corpora, whereas papers from other preprint platforms are typically excluded. This is the case for widely used datasets such as LLaMA \citep{touvron2023llama}, Dolma \citep{soldaini2024dolma}, and Common Crawl--derived corpora (e.g., The Pile \citep{gao2020pile}, RedPajama \citep{weber2024redpajama}, and Common Pile \citep{kandpal2025common}), which underpin many modern LLMs \citep{wu2025mapping, wolfram2025layers}. This discrepancy arises for two main reasons: (1) arXiv provides structured LaTeX source files via accessible storage (e.g., Amazon S3 bucket), whereas platforms such as ChemRxiv, PsyArXiv, and SocArXiv primarily distribute content as PDFs without standardized HTML/XML interfaces; and (2) other platforms present more complex copyright and licensing constraints\footnote{\url{https://blog.dhimmel.com/biorxiv-licenses/}} for large-scale data inclusion in the LLM pretraining. BioRxiv and its sister platform MedRxiv even actively restrict automated crawling\footnote{\url{https://www.biorxiv.org/robots.txt}}. This constraint similarly affects datasets derived from Common Crawl. Our additional context-puzzle-rewriting and leakage detection procedures (Subsection \ref{2.2}) further mitigate the risk of the leakage of human-generated hypotheses, and as shown in the main paper, AI models are not replicating ideas from humans.

We further conducted a data contamination experiment by comparing preprints sampled from all non-arXiv platforms used in the paper against the actual pretraining corpus, for which we use a 10B-token sample from Dolma v1.6 \citep{soldaini2024dolma}. Each paper's full text is segmented into paragraph-level chunks, and chunks shorter than 30 words are discarded. For each valid chunk (paragraph), we extract all 10-gram (10-word) sequences with a two-word sliding window of overlap between each sequence. We then scan Dolma and identify a chunk (paragraph) as matched if more than 50\% of its 10-grams are found. Applying this procedure to all paragraphs from 8,000 random non-arXiv papers in 2024, we find that only 4 paragraphs are matched. Upon closer inspection, however, even these cases are effectively false positives: the matching n-grams consist of common phrasings widely used across Internet text; other adjacent paragraphs from the same paper are not matched; and the ``matching'' occurrences are scattered across the full corpus rather than concentrated in any single source.

\subsection{Name disambiguation}
\label{subsec:name_disambiguation}

Name disambiguation is always a concern in large-scale data for scientific papers. To validate the author matching, we randomly sampled 100 author names in the analysis and their IDs, conducted a manual robustness check, and found a 96\% match rate. Here a match means that the papers attributed to them in the OpenAlex dataset are, with 100\% certainty, their own.

\subsection{The full list of LLMs}
\label{full_list}

The full list of LLMs used in this study to propose hypotheses is provided below. We use instruction-following LLMs throughout.

\begin{itemize}
\item Reasoning models: o3-mini (OpenAI), o4-mini (OpenAI), DeepSeek R1 0528 (DeepSeek), o1 (OpenAI), o3 (OpenAI), o4-mini-deep-research (OpenAI), Tongyi-deep-research (Alibaba)
\item Non-reasoning models (some of which can be configured for reasoning, but were not in this study): GPT-4o (OpenAI), GPT-4o-mini (OpenAI), GPT-4.1 (OpenAI), Grok-3 (xAI), Grok-3-mini (xAI), LLaMA 3.1 8B (Meta), LLaMA 3.1 70B (Meta), LLaMA 3.1 405B (Meta), Gemma 3 4B (Google), Gemma 3 12B (Google), Gemma 3 27B (Google), Phi-4 (Microsoft), Phi-3-mini (Microsoft), Mixtral 8$\times$7B MoE (Mistral AI), Ministral 3B (Mistral AI), DeepSeek-V3 (DeepSeek), Qwen-Turbo (Alibaba), Gemini 2.0 Flash (Google), Gemini 2.0 Flash-Lite (Google)
\end{itemize}

\subsection{Data processing pipeline}
\label{2.2}
\paragraph{Extracting Human Hypotheses} We input the full text of each paper into the advanced reasoning model o3-mini to summarize and extract human hypotheses. Our prompt enforces specific constraints, including the exclusion of inferred content, peripheral assumptions, and vague directional statements (see Subsection \ref{prompts} for details). We exclude all papers from which no human hypotheses could be extracted, as they are likely theoretical or survey-type works. We deliberately avoid a more intuitive alternative---simply searching for the keyword ``hypothesis'' and its variants and summarizing the targeted passages across the paper---because such matches often yield irrelevant content. To accurately identify the core human hypotheses related to the key puzzle, the model must instead consider the paper holistically. In our survey, 98.62\% of scientists reported being generally satisfied with the AI-summarized human hypotheses derived from their own papers.

\paragraph{Extracting Context and Scientific Puzzles} The background of a scientific paper---particularly the introduction and, when available, the related work sections---serves to establish the narrative, build the stage, motivate the study, and lay the groundwork for the readers and following sections. We use this contextual material to approximate the information that builds the core research puzzle for LLMs generating hypotheses. We extract the introduction and related work sections using GROBID\footnote{\url{https://python.langchain.com/docs/integrations/document_loaders/grobid/}} (some papers do not include a distinct ``related work'' section). The two sections minimize the risk of fully disclosing the researchers' own hypotheses, experiments, interpretations, and conclusions, which are typically developed in the following method/result section of the paper. We then prompt the o3-mini model to extract two elements: (1) the core scientific puzzle and (2) the contextual information that sets up the puzzle, specifically factual, reasoning-free statements about the key terms that appear in the puzzle, with a particular focus on avoiding the disclosure of any human hypotheses. During the extraction, we apply few-shot prompting---providing the LLM with several examples of well-formed puzzles and poor ones (e.g., cases that are not true puzzles but rather proposed solutions or results); see Subsection \ref{prompts} for details.

\paragraph{Preventing Human Hypothesis Leakage to LLMs} Sometimes the scientific puzzle itself is essentially the rephrasing of a hypothesis and they cannot be meaningfully separated (e.g., ``does A increase B?''). To prevent data (human hypotheses) leakage during hypothesis generation by LLMs, we include a further leakage detection step. We randomly selected 1000 human hypotheses. Using the GPT-4.1 model, we generated 20 rephrasings for each hypothesis. Note that our rewriting may alter sentence structures, such as transforming declarative sentences into interrogative ones, simulating the variety of ways humans might express the same idea. This resulted in 20000 paraphrased hypotheses. We computed textual similarity using the MPNet model \citep{song2020mpnet} and identified an embedding similarity threshold of 0.82, which captured 95\% of the paraphrased pairs. If \textit{any sentence} in an extracted context and puzzle exceeds the similarity score of 0.82 with the corresponding human hypothesis, we consider it a leakage and will not use the context or puzzle accordingly. In the survey, we find that 99.70\% of scientists are generally satisfied with the AI-summarized context and puzzle from their own paper. We dropped all cases in which authors were unsatisfied with the extracted context, puzzle and hypotheses in the subsequent analysis. \revise{Examples of contexts, puzzles, and human- and AI-generated hypotheses (null and non-null) are given in Boxes 1--3 (Subsection~\ref{sec:nulldef}).}

\subsection{Survey design}
\label{subsec:survey_design}

We prompted LLMs to generate hypotheses for each paper based on its summarized context and stated research puzzle. \revise{The authors' hypotheses were first extracted as a numbered list, and each LLM was asked to generate the same number of hypotheses for that context and puzzle (one response per model and paper), so that the human and model sides have the same ``ceiling'' and null rates are not inflated or deflated by the number of hypotheses generated.} For every paper, this produced a set of hypotheses generated by LLMs, together with the original author's hypotheses, context and puzzle. From the pool of LLM-generated hypotheses, we selected five per paper and sent them to the original author for evaluation, as long as we could access their email\footnote{The case of multiple corresponding authors is not very common in our dataset.}. Each author received a custom set of five most semantically distinguishable hypotheses for their paper, drawn under a stratified rule that partitioned the 26 LLMs into non-reasoning and reasoning\footnote{Reasoning models include agentic deep-research models, since these models reason simultaneously while performing agentic web search.} models (Subsection \ref{full_list}) and required at least two hypotheses from each stratum. Within each stratum, hypotheses were selected to maximize semantic distinguishability from each other, ensuring that the five evaluations represented distinct ideas rather than near-duplicates, making the scientists' judgments easier. This design yielded a nearly even distribution of to-be-judged hypotheses from different models and addressed a concern particularly salient for non-reasoning models, whose generations exhibited substantial within-group similarity, as shown in the main paper. Given the nontrivial nature of the evaluation task, we implemented a comprehension check prior to the rating phase. Participants were required to complete a brief understanding test to ensure they could reliably apply the evaluation criteria: novelty (the extent to which the generated idea introduces new elements beyond the input; rated on a 1--9 scale from not novel to highly novel, with the same scale applied below), feasibility (the extent to which a concrete experiment could be designed to test the proposed idea), and probability (the likelihood that the generated idea is true even without experimental validation). Subsection \ref{survey} provides additional details on the survey. This understanding test filtered out a large amount of low-quality input, with many scientists dropping out of the study at the test. Scientists then assessed the quality of extractions of context, puzzle, and their own ideas, and independently rated each hypothesis along four dimensions: novelty, feasibility, probability, and their favorability (measured by overall adoption intention\revise{, i.e., their reported willingness to pursue the hypothesis in a future paper; ``adoption'' thus refers to this stated willingness, not to actual adoption}). We also excluded all data from authors who indicated that they did not fully remember the content of their paper, and, for authors who consented only partially (e.g., permitting use of their papers in open-source but not closed-source models), we removed the ratings corresponding to the uses they did not consent to. In total, 112,101 scientists received the invitation; 6,749 scientists participated in the crowd-sourced evaluation task, with some participants contributing only partially completed responses, e.g., before the understanding test. This finally yielded 25,139 human-labeled evaluations (each evaluation includes \revise{three rated characteristics and willingness to pursue}) contributed by 5,259 out of all participants. The study was conducted under the University of Chicago Institutional Review Board protocol IRB25-1372, titled ``The Capabilities and Potential of AI for Automating Scientific Idealization: A Large-Scale Human-in-the-Loop Study.''

\subsection{The null-statement classifier}
\label{classify}

We built a pipeline to automatically detect null hypothesis statements. \revise{The same model is applied to AI-generated and extracted human hypotheses (main text Figure \ref{fig:ideation}) and to abstract sentences (main text Figure \ref{macro}), and we refer to it as the null-statement classifier; see Subsection~\ref{sec:nulldef} for validation.} In the preprocessing stage, we used NLTK's TreebankWordTokenizer, a rule-based tokenizer tailored for English text. This tokenizer handles punctuation and contractions (e.g., ``don't'' $\rightarrow$ ``do not'') and respects standard word boundaries more effectively than simple whitespace splitting. After tokenization, we reconstructed each text as a space-separated sequence of tokens to ensure consistency in representation, which is important for downstream vectorization. Unlike more aggressive preprocessing pipelines, we intentionally avoided removing stopwords. This choice preserves subtle but critical linguistic signals---particularly negations such as ``no effect'' or ``not significant''---that are essential for identifying null hypothesis statements.

We then passed the cleaned and tokenized text into a TF-IDF vectorizer, which transforms the corpus into a sparse numerical representation based on term frequency--inverse document frequency weighting. This representation preserves the discriminative power of informative tokens (e.g., ``no'', ``not significant''), allowing them to receive higher weights in the feature space. To construct the training dataset, we generated 1,000 labeled instances of null hypotheses (label 1) and an additional 1,000 non-null examples (label 0) using o3-mini (see Subsection \ref{prompts} for prompt details). Two human annotators independently evaluated 100 generated null hypotheses and 100 generated not-null hypotheses, achieving an agreement rate of 100\%. Following prior work \citep{bao2025there, bao2025division}, we trained an ensemble classifier composed of four widely used text classification models. Specifically, we included two models suited for linearly separable features---Support Vector Machine (with a linear kernel) and Logistic Regression---and two models capable of capturing non-linear relationships---Random Forest (with 100 estimators) and Gradient Boosting Decision Trees (with 100 estimators). This combination enables the model to accommodate potentially diverse feature geometries present in null hypothesis statements. Each classifier produces a probability estimate for the positive class (label = 1). The final prediction is obtained by averaging these probabilities across all classifiers and applying a threshold of 0.5. The task is relatively well-structured, as null hypotheses often contain distinctive lexical patterns (e.g., ``no relationship'', ``no effect''), which are effectively captured by our proposed algorithm. 5-fold cross-validation yields an accuracy of 99.5\%, corresponding to approximately 5 misclassifications per 1,000 samples. Training the classifier on another training set generated by Google Gemini 2.5 Flash achieves 99.9\% agreement with the results based on the o3-mini-generated training set across 1000 random samples (500 null and 500 not-null). When we randomly replaced 50\% of the o3-mini-generated training data with human-written null statements from real-world published papers (identified by the classifier trained on the o3-mini training set), the re-trained classifier demonstrated zero difference in identification on the same test set.

\subsection{The detection of prior AI engagement}
\label{aiuse}

We collect a total of 848,750 papers authored by participating scientists and assign each scientist a prior AI engagement rate, defined as the proportion of their papers involving AI. We primarily adopt the pretrained language model developed by Hao and colleagues to detect AI-engaged research\citep{hao2026artificial}. The authors developed a supervised natural language processing pipeline based on a fine-tuned BERT model that classifies papers using their titles and abstracts. The model was trained in a two-stage procedure, first leveraging coarse labels from explicitly AI-engaged venues and then refining predictions using higher-precision venue-level signals. Separate models trained on titles and abstracts were ensembled to improve robustness (as our null-statement classifier), eliminating the need for manually curated keyword rules. The resulting classifier outputs the probability that a paper incorporates AI, enabling large-scale identification of AI-engaged research across millions of publications. The method was validated against expert-annotated data, where multiple domain experts independently labeled sampled papers with high inter-rater agreement (Fleiss' $\kappa$ = 0.96). Compared to this human ground truth, the model achieved strong performance, reaching an F1-score of approximately 0.875.

To ensure the robustness of our approach, we employ LLMs (specifically, Google Gemini 2.5 Flash) to detect AI engagement, leveraging the growing adoption of LLM-based annotation in the literature \citep{tan2024large}. The model is tasked with determining, based on the title and abstract, whether artificial intelligence or machine learning is involved in the paper. We then examine the agreement between this LLM-based annotation method and Hao's approach \citep{hao2026artificial} across both negative (0: non-AI) and positive (1: AI) classifications. If we treat Hao's method as the ground truth, the two methods exhibit near-perfect agreement in identifying non-AI research, with precision, recall, and F1-scores all exceeding 0.99. While some discrepancies arise in identifying AI-engaged papers, these differences remain within an acceptable range. Google Gemini 2.5 Flash correctly identifies 80.0\% of AI-engaged papers, while misclassifying the remaining 20.0\% as non-AI. Upon closer inspection of these misclassifications, we find that the primary source of disagreement lies in the definition of AI engagement. The LLM-based annotation method classifies a paper as AI-engaged (label 1) only when AI is used as a methodological tool. In contrast, Hao's approach labels papers as AI-engaged not only when AI is used as a method, but also when AI is used as the research topic itself (e.g., ``what should be the new paradigm of regulation for human-AI society in the future?'').

We argue that the latter definition is more appropriate for our research setting. Regardless of whether AI is used as a method or a research subject, it is expected to have implications for AI attitude. Therefore, we retain Hao's method as the primary classification approach in the main analysis. Despite these differences, the two methods demonstrate a high level of overall consistency, achieving an alignment rate of 98.80\%. \revise{Because the classifier captures AI as either a method or a topic, we call it the AI-engaged research classifier and the resulting measure prior AI engagement.}

\subsection{The realistic reference of consistency between independent reviewers}
\label{humanalign}

We do not assume that a perfect reward model would achieve 100\% accuracy in our test set. Such a ceiling is unattainable due to inherent randomness, subjective judgment, and variability in human evaluation. Instead, we benchmark against the level of human agreement observed in real-world peer review. We collected peer reviews from 26,731 submissions across 46 conferences hosted on OpenReview between 2017 and 2025 (see below). The dataset spans multiple disciplines, including computer science, physics, medicine, and the social sciences. Within each conference, we construct pairwise comparisons between papers. For each pair, we consider cases where the reviewer in position $p$ assigns a lower score to one paper and a higher score to the other (excluding ties). The ``human accuracy'' is then operationalized as the probability that a different, non-overlapping reviewer (i.e., the reviewer in another position $q$) agrees with the direction of this preference. To control for positional bias, we randomize the position of the reference reviewer and the non-overlapping reviewer 1,000 times, yielding the human consistency of 61.0\% $\pm$ 0.1\%. Note that this result should be lower than the human agreement upper bound, since reviewers in the same position across different papers are different. \revise{Here a ``position'' is the slot in which OpenReview lists a paper's reviewers (Reviewer 1, 2, 3, \ldots). We ask: if the Reviewer~$p$s of two papers rank them one way, how often do the Reviewer~$q$s agree? Because the reviewers in a given slot differ across papers, this measures agreement between independent reviewers.}

The 46 conferences in our dataset: 1st ContinualAI Unconference, AAAI Conference on Artificial Intelligence 2024 and 2025, Symposium on Advances in Approximate Bayesian Inference 2024, Conference on Language Modeling 2024, Cooking Robotics Workshop 2024, Conference on Robot Learning 2023 and 2024, Conference on Parsimony and Learning 2024, IEEE/CVF Conference on Computer Vision and Pattern Recognition 2023 and 2024, Workshop on Distributed Infrastructure for Common Good 2023, Workshop on Embodiment-Aware Robot Learning 2024, European Conference on Computer Vision 2024, Conference on Empirical Methods in Natural Language Processing 2023, European Space Power Conference 2023, Fast, Low-resource, and Accurate Organ and Pan-cancer Segmentation in Abdomen CT 2023, SIGIR Workshop on Generative Information Retrieval 2024, ACM/IEEE International Conference on Human-Robot Interaction 2023 and 2024, International Conference on Integration of Science and Technology for Sustainable Development 2024, International Conference on Learning Representations 2017 to 2025, International Conference on Machine Learning 2023 and 2024, ACM International Conference on the Theory of Information Retrieval 2024, International Joint Conference on Artificial Intelligence 2024, International Semantic Web Conference 2024, ACM SIGKDD Conference on Knowledge Discovery and Data Mining 2023 and 2024, ACM International Conference on Multimedia 2024, Conference on Neural Information Processing Systems 2021--2024, Neuro-Symbolic Learning and Reasoning in the Era of Large Language Models 2024, Next-generation Data Governance Workshop 2024, Tsinghua University Advanced Machine Learning 2024.

\subsection{Statistical specifications}
\label{stats1}

\paragraph{The Mundlak model with different representations of seniority}

In the main paper, we estimate the relation between the reported willingness to pursue and potential sources of biases and idea quality. We follow the correlated random-effects (Mundlak) tradition to estimate a linear model \citep{schunck2013within}. The logic is that a pooled OLS specification without the Mundlak adjustment may suffer from omitted-variable bias because it does not account for persistent scientist-specific tendencies. For example, some scientists may systematically assign higher ratings overall or be more inclined to pursue AI ideas. To address this concern, we include each idea-level covariate both in deviation-from-scientist-mean form ($\text{rating}_{i,j} - \overline{\text{rating}_{i}}$) and the scientist-level mean $\overline{\text{rating}_{i}}$ of scientist $i$ for idea $j$. This decomposition separates within-scientist variation from between-scientist differences. As a result, the coefficients on the within-scientist components capture how a scientist's willingness to pursue changes when an idea is evaluated as more novel, feasible, likely to be true, or more similar to their own perspectives than is typical for that same scientist (thus ``\textbf{within}'' terms are our primary interest for what kind of quality drives willingness to pursue), while the scientist-level means absorb persistent cross-scientist differences in average evaluation levels.

We extract the main effect of human bias from scientist $i$ on hypothesis $j$ by the following regression:

\begin{equation}
\begin{aligned}
\label{equ1}
\text{Willingness}_{ij} =\;&
\beta_0
+ \beta_1 \bigl(\text{Novelty}_{ij} - \overline{\text{Novelty}}_{i}\bigr)
+ \beta_2 \overline{\text{Novelty}}_{i} \\
&+ \beta_3 \bigl(\text{Feasibility}_{ij} - \overline{\text{Feasibility}}_{i}\bigr)
+ \beta_4 \overline{\text{Feasibility}}_{i} \\
&+ \beta_5 \bigl(\text{Probability}_{ij} - \overline{\text{Probability}}_{i}\bigr)
+ \beta_6 \overline{\text{Probability}}_{i} \\
&+ \beta_7 \bigl(\text{Human-AI Similarity}_{ij} - \overline{\text{Human-AI Similarity}}_{i}\bigr)
+ \beta_8 \overline{\text{Human-AI Similarity}}_{i} \\
&+ \delta_0 \text{Seniority}_{i}
+ \delta_1 \text{Prior AI Engagement}_{i}
+ \delta_2 \text{Field}_{i}
+ \varepsilon_{ij}.
\end{aligned}
\end{equation}

Novelty, feasibility, and probability are included as controls for baseline idea quality; thus $\beta_1$, $\beta_3$, and $\beta_5$ answer what kind of quality drives willingness to pursue. $\beta_7$, $\delta_0$, $\delta_1$, and $\delta_2$ answer what impacts willingness to pursue other than quality. Following a standard parametric uncertainty approach, in the main paper, we reported model-based adjusted predictions (average marginal predictions) with cluster-robust standard errors, holding the empirical distribution of other covariates fixed.

Across the paper, we used log transformations (base $e$) for prior AI engagement, since many participants in our survey had near-zero prior AI engagement---they were not computer scientists. We used the embedding model MPNet \citep{song2020mpnet} to calculate the cosine similarity between human and AI ideas.

We used the within-field yearly citation percentile to represent seniority by default. We also employ two alternative seniority measures---average publication count per year and academic age---and find consistent results across all specifications. Academic age is operationalized by identifying each author's first publication based on their ID in the OpenAlex dataset (name disambiguation in SI section \ref{subsec:name_disambiguation}), with age defined as 2025 minus that first publication year. Academic age is represented on a log\textsubscript{e} scale in regressions, as its distribution is long-tailed---some participants are very junior (fewer than five years of experience). Yearly publication/citation counts are percentilized within each field to (1) ensure comparability across disciplines; (2) account for their long-tailed distribution; and (3) avoid confounding with the cumulative nature of age. Full results of Equation \ref{equ1} with different representations of seniority can be found in Extended Data Table \ref{tab:ols_combined}.

\paragraph{Author fixed effects and ordinal regressions}

We also test an alternative statistical specification, author fixed effects, thereby absorbing variables such as seniority, prior AI engagement, and field, which yields similar signs, magnitudes, and levels of statistical significance compared to the Mundlak approach in the main paper. We estimate the following regression (with $\alpha_i$ representing the author fixed effects):

\begin{equation}
\begin{aligned}
\text{Willingness}_{ij} =\;&
\alpha_i \\
&+ \beta_1 \text{Novelty}_{ij}
+ \beta_2 \text{Feasibility}_{ij}
+ \beta_3 \text{Probability}_{ij}
+ \beta_4 \text{Human\text{-}AI Similarity}_{ij} \\
&+ \varepsilon_{ij}.
\end{aligned}
\end{equation}

Among the three quality dimensions, probability is still the strongest driver of willingness to pursue. Individuals still show a strong preference for ideas that are more similar to their own perspectives (bigger than probability). All coefficients are statistically significant.

We use a linear specification as our main model for ease of interpretation. With nine ordered categories of quality dimensions, the outcome is sufficiently fine-grained that a linear model provides a useful approximation. Here for robustness, we use a cluster-robust ordered logistic regression model (proportional odds model) estimated using the BFGS maximum likelihood optimization method, with Mundlak decomposition to separate within- and between-individual effects to replicate the conclusion in the main paper. Conclusions are consistent.

The regression results of the author-fixed effect model and ordered logistic regression model can be found in SI Table \ref{tab:adoption_combined}.

\paragraph{Seniority by field interaction} We then estimate human bias for scientist \( i \) evaluating hypothesis \( j \) by augmenting Equation \ref{equ1} with interaction terms between seniority and field as shown in Equation \ref{equ2}.

\begin{equation}
\begin{aligned}
\label{equ2}
\text{Willingness}_{ij} =\;&
\beta_0
+ \beta_1 \bigl(\text{Novelty}_{ij} - \overline{\text{Novelty}}_{i}\bigr)
+ \beta_2 \overline{\text{Novelty}}_{i} \\
&+ \beta_3 \bigl(\text{Feasibility}_{ij} - \overline{\text{Feasibility}}_{i}\bigr)
+ \beta_4 \overline{\text{Feasibility}}_{i} \\
&+ \beta_5 \bigl(\text{Probability}_{ij} - \overline{\text{Probability}}_{i}\bigr)
+ \beta_6 \overline{\text{Probability}}_{i} \\
&+ \beta_7 \bigl(\text{Human-AI Similarity}_{ij} - \overline{\text{Human-AI Similarity}}_{i}\bigr)
+ \beta_8 \overline{\text{Human-AI Similarity}}_{i} \\
&+ \delta_0 \text{Seniority}_{i}
+ \delta_1 \text{Prior AI Engagement}_{i}
+ \delta_2 \text{Field}_{i}\\
&+ \gamma_0 \text{Seniority}_{i} \times \text{Field}_{i}\\
&+ \varepsilon_{ij}.
\end{aligned}
\end{equation}

We observe that the field-level differences disappear once this interaction is included. While being senior continues to have a negative effect overall, the interaction terms are positive for all fields except the baseline category (social science), though not large enough to fully offset the baseline negative effect. \revise{These results suggest that lower willingness to pursue AI-generated hypotheses among senior scientists is broadly universal, but is particularly pronounced among social scientists.} Results can be found in Extended Data Table \ref{tab:ols_results34}.

\paragraph{Quality by field interaction} We then estimate human bias for scientist \( i \) evaluating hypothesis \( j \) by augmenting Equation \ref{equ1} with interaction terms between field and within-scientist quality rating differences (parameters \( \gamma_0 \)--\( \gamma_2 \)), as well as a within-scientist preference term capturing alignment between human and AI-generated ideas \( \gamma_3 \). All other statistical specifications remain the same. This specification allows epistemic standards to vary systematically across fields, say, whether a field especially appreciates novel research (Equation \ref{qee}).

\begin{equation}
\begin{aligned}
\label{qee}
\text{Willingness}_{ij} =\;&
\beta_0
+ \beta_1 \bigl(\text{Novelty}_{ij} - \overline{\text{Novelty}}_{i}\bigr)
+ \beta_2 \overline{\text{Novelty}}_{i} \\
&+ \beta_3 \bigl(\text{Feasibility}_{ij} - \overline{\text{Feasibility}}_{i}\bigr)
+ \beta_4 \overline{\text{Feasibility}}_{i} \\
&+ \beta_5 \bigl(\text{Probability}_{ij} - \overline{\text{Probability}}_{i}\bigr)
+ \beta_6 \overline{\text{Probability}}_{i} \\
&+ \beta_7 \bigl(\text{Human\text{-}AI Similarity}_{ij} - \overline{\text{Human\text{-}AI Similarity}}_{i}\bigr)
+ \beta_8 \overline{\text{Human\text{-}AI Similarity}}_{i} \\
&+ \delta_0 \text{Seniority}_{i}
+ \delta_1 \text{Prior AI Engagement}_{i}
+ \delta_2 \text{Field}_{i} \\
&+ \gamma_0 \bigl(\text{Novelty}_{ij} - \overline{\text{Novelty}}_{i}\bigr)\times \text{Field}_{i} \\
&+ \gamma_1 \bigl(\text{Feasibility}_{ij} - \overline{\text{Feasibility}}_{i}\bigr)\times \text{Field}_{i} \\
&+ \gamma_2 \bigl(\text{Probability}_{ij} - \overline{\text{Probability}}_{i}\bigr)\times \text{Field}_{i} \\
&+ \gamma_3 \bigl(\text{Human\text{-}AI Similarity}_{ij} - \overline{\text{Human\text{-}AI Similarity}}_{i}\bigr) \times \text{Field}_{i} \\
&+ \varepsilon_{ij}
\end{aligned}
\end{equation}

As shown in Extended Data Table \ref{tab:ols_results}, the regression results are consistent with those reported in the main paper. The newly introduced interaction terms yield an interesting pattern: we do not find meaningful field-level differences in how novelty or feasibility or human-AI idea alignment affect willingness to pursue, but we do observe significant heterogeneity across fields in the effect of probability. The interaction results show that the positive effect of within-author probability is significantly stronger in biology and medicine than in social science, but not significantly different in chemistry. Since the baseline effect is already positive and significant ($0.269$, $p < 0.001$), the positive interaction terms for biology ($0.058$, $p < 0.001$) and medicine ($0.063$, $p = 0.003$) imply that these fields are even more willing to pursue ideas that are judged as more probable. This suggests that, compared with social science, biology and medicine exhibit a stronger preference for relatively ``safe'' research directions. We further include interaction terms between field, within-author rating differences, and seniority in SI Table \ref{tab:ols_results_m2}. Across all specifications, our main conclusions remain robust.

\subsection{Subtopic-level distribution analysis}
\label{app:label_nonres}

Preprint platforms require authors to assign a subtopic label to each submission (see below). We collected these labels and compared their distribution across respondents (based on their papers) with that of the full corpus (121,640 papers). The two distributions are closely aligned (Pearson's $r$ = 0.98; Spearman's $\rho$ = 0.96; Mean absolute error = 0.01). Each subtopic label is followed by two numbers: the first is its proportion among respondents, the second its proportion in the full corpus.

\begin{itemize}
    \item \textbf{Biology} (20 subtopics): Animal Behavior and Cognition (1.93\%, 1.93\%); Biochemistry and Molecular Biology (12.48\%, 11.04\%); Bioengineering and Synthetic Biology (2.06\%, 2.44\%); Bioinformatics and Computational Systems Biology (6.79\%, 7.52\%); Biophysics (1.10\%, 0.74\%); Cancer Biology (5.90\%, 5.85\%); Cell Biology (7.91\%, 6.32\%); Developmental Biology (4.46\%, 3.75\%); Ecology (3.55\%, 4.32\%); Evolutionary Biology (2.85\%, 3.04\%); Genetics and Genomics (6.42\%, 6.93\%); Immunology (5.26\%, 4.29\%); Microbiology (9.20\%, 9.72\%); Neuroscience (13.61\%, 13.26\%); Paleontology (0.11\%, 0.09\%); Pathology and Disease Biology (7.96\%, 9.16\%); Pharmacology and Toxicology (1.06\%, 0.85\%); Physiology and Organismal Biology (4.41\%, 4.86\%); Plant Biology (2.47\%, 3.64\%); Scientific Communication and Education (0.47\%, 0.23\%).
    \item \textbf{Medicine} (40 subtopics): Addiction Medicine (1.74\%, 0.87\%); Allergy, Immunology, and Rheumatology (1.74\%, 1.76\%); Anesthesia and Pain Medicine (0.68\%, 0.89\%); Cardiovascular Medicine (7.82\%, 11.35\%); Dentistry and Oral Medicine (0.57\%, 0.52\%); Dermatology (0.27\%, 0.51\%); Emergency and Critical Care Medicine (0.92\%, 1.24\%); Endocrinology and Metabolic Disease (2.50\%, 3.10\%); Epidemiology and Public Global Health (10.65\%, 8.52\%); Forensic Medicine (0.27\%, 0.10\%); Gastroenterology and Hepatology (1.06\%, 1.74\%); Genetic and Genomic Medicine (9.89\%, 7.43\%); Geriatric and Palliative Medicine (1.82\%, 1.87\%); Health Informatics and Digital Health (4.51\%, 3.43\%); Health Policy, Economics, and Systems (1.28\%, 1.45\%); Hematology (0.79\%, 0.59\%); Infectious Diseases (11.65\%, 14.60\%); Medical Education (1.22\%, 0.89\%); Medical Ethics (0.14\%, 0.08\%); Nephrology (0.84\%, 1.26\%); Neurology (8.53\%, 8.37\%); Nursing (0.65\%, 0.28\%); Nutrition (1.06\%, 1.34\%); Obstetrics, Gynecology, and Reproductive Health (3.01\%, 3.35\%); Occupational and Environmental Health (1.09\%, 0.78\%); Oncology (5.62\%, 6.10\%); Ophthalmology (0.81\%, 1.01\%); Orthopedics (1.17\%, 0.82\%); Otolaryngology (0.27\%, 0.29\%); Pathology and Laboratory Medicine (1.20\%, 0.72\%); Pediatrics (1.39\%, 2.08\%); Pharmacology, Therapeutics, and Toxicology (0.62\%, 0.48\%); Primary Care Research (0.52\%, 0.70\%); Psychiatry and Clinical Psychology (7.52\%, 5.66\%); Radiology and Imaging (1.36\%, 1.27\%); Respiratory Medicine (1.63\%, 2.02\%); Sports Medicine and Rehabilitation (2.04\%, 1.45\%); Surgery (0.43\%, 0.52\%); Transplantation (0.46\%, 0.32\%); Urology (0.27\%, 0.22\%).
    \item \textbf{Chemistry} (14 subtopics): Agriculture and Food Chemistry (0.96\%, 1.21\%); Analytical Chemistry (4.60\%, 4.77\%); Biological and Medicinal Chemistry (11.47\%, 15.17\%); Catalysis (7.23\%, 7.01\%); Chemical Education (0.90\%, 0.80\%); Chemical Engineering and Industrial Chemistry (0.66\%, 1.61\%); Earth, Space, and Environmental Chemistry (3.41\%, 4.60\%); Energy and Electrochemistry (8.00\%, 9.31\%); Inorganic and Organometallic Chemistry (5.97\%, 5.69\%); Materials Science and Nanoscience (18.94\%, 18.16\%); Organic Chemistry (14.04\%, 11.95\%); Physical Chemistry (6.27\%, 5.06\%); Polymer Science (4.24\%, 3.39\%); Theoretical and Computational Chemistry (13.32\%, 11.26\%).
    \item \textbf{Social sciences} (43 subtopics): Behavior Analysis (1.82\%, 1.46\%); Biological Psychology and Cognitive Neuroscience (6.35\%, 6.48\%); Clinical and Counseling Psychology (2.77\%, 4.50\%); Cognitive Psychology, Cognition, and Perception (18.32\%, 14.61\%); Developmental Psychology (2.54\%, 3.31\%); Health and Community Psychology (1.56\%, 2.97\%); Linguistics and Psycholinguistics (5.39\%, 4.43\%); Social, Personality, and Cultural Psychology (23.01\%, 22.99\%); Anthropology (0.25\%, 0.13\%); Communication and Media Studies (2.13\%, 1.70\%); Criminology and Criminal Justice (0.61\%, 0.45\%); Demography and Population Studies (0.84\%, 0.54\%); Economics (1.62\%, 1.81\%); Family, Life Course, and Society (0.94\%, 0.88\%); Gender and Sexuality Studies (0.68\%, 1.21\%); Geography, Urban Studies, and Planning (0.25\%, 0.80\%); Health, Medicine, and Society (1.13\%, 2.63\%); History (0.80\%, 1.44\%); Inequality, Stratification, and Mobility (1.31\%, 0.86\%); Law and Legal Studies (1.02\%, 0.95\%); Library and Information Science (0.10\%, 0.06\%); Metascience and Science and Technology Studies (8.59\%, 8.87\%); Political Science and International Relations (2.93\%, 2.52\%); Public Policy and Public Administration (0.88\%, 0.50\%); Race, Ethnicity, and Migration (1.05\%, 0.37\%); Religion (1.04\%, 1.16\%); Research Methods, Statistics, and Measurement (5.35\%, 3.90\%); Social Work and Social Policy (0.02\%, 0.17\%); Sociology (general) (1.33\%, 1.33\%); Work, Organizations, and Occupations (0.90\%, 1.29\%); Adult, Vocational, and Continuing Education (0.00\%, 0.06\%); Curriculum and Instruction (0.00\%, 0.00\%); Early Childhood Education (0.08\%, 0.06\%); Educational Leadership, Administration, and Policy (0.10\%, 0.19\%); Educational Psychology (1.88\%, 2.26\%); Educational Technology and Online Learning (0.27\%, 0.22\%); Higher Education (0.18\%, 0.45\%); International and Comparative Education (0.00\%, 0.00\%); K--12 Education (0.06\%, 0.15\%); Language, Literacy, and Bilingual Education (0.59\%, 0.77\%); Science and Mathematics Education (1.19\%, 1.16\%); Special and Inclusive Education (0.08\%, 0.15\%); Teacher Education and Professional Development (0.06\%, 0.19\%).
\end{itemize}

\subsection{Robustness to sample selection and response bias}
\label{app:selection}

The estimation sample consists of researchers who responded to the survey, and
respondents need not be representative of the population from which they were
drawn. If the propensity to respond is correlated with both the covariates and
the outcome, the coefficients in the main result may be biased.

We assess this using an auxiliary frame covering the full set of contacted researchers, respondents and non-respondents alike. For every individual in the frame, we observe field, average annual publications, average annual citations, academic age, and prior AI engagement. The frame therefore permits a direct comparison of respondents and non-respondents on pre-treatment observables, and allows us to reweight the estimation sample so that its covariate distribution matches that of the full contacted population.

Let $S_i = 1$ denote that researcher $i$ appears in the sample and
$X_i$ is the vector of pre-treatment observables. The reweighting procedure below
maintains selection on observables,
\begin{equation}
  \label{eq:soo}
  Y_i \perp S_i \mid X_i .
\end{equation}

We estimate the response propensity $\hat{p}(X_i) = \Pr(S_i = 1 \mid X_i)$ on
the full frame by logistic regression, fit \emph{separately within each field}
so that the selection mechanism is allowed to differ across disciplines, and
include second-order terms (squares and pairwise interactions of the
covariates) so that the propensity is not restricted to be linear in $X$.
Predicted propensities are truncated to $[0.02, 0.98]$. Each sampled researcher
then receives the stabilised weight
\begin{equation}
  w_i \;=\; \frac{\Pr(S=1)}{\hat{p}(X_i)},
\end{equation}
winsorised at the $99$th percentile and normalised to mean one, which reweights
the respondents back to the full contacted population: researchers of a type
that respond rarely receive proportionally larger weight. Weights are
constructed at the researcher level and applied to all ideas evaluated by that
researcher. The main outcome equation (Equation \ref{equ1}) is then re-estimated by weighted least squares, with all other features of the main specification left unchanged (e.g., standard errors are clustered at the same level).

SI Table~\ref{tab:selection} reports the outcome equation with and without
weighting. All coefficients of substantive interest are essentially unchanged in sign, magnitude, and significance classification under reweighting.

\subsection{Implementation of novelty checkers}
\label{novelty}

We evaluate several mainstream novelty evaluation methods in the main paper. All approaches essentially operationalize novelty as the degree of deviation between a generated idea $g$ and its corresponding context and puzzle $c$, capturing how unexpected the generation is with respect to the input.

\paragraph{Semantic Similarity}
This is the most intuitive way: we encode both the context/puzzle $c$ and generated idea $g$ into dense vector representations using a pretrained sentence embedding model \citep{song2020mpnet}. Novelty is defined as the inverse cosine similarity:
\begin{equation}
\text{Novelty}_{\text{sem}}(g, c) = 1 - \cos(\mathbf{e}_g, \mathbf{e}_c)
\end{equation}
where $\mathbf{e}_g$ and $\mathbf{e}_c$ denote the embeddings of $g$ and $c$, respectively. Lower similarity corresponds to higher novelty.

\paragraph{n-gram Novelty (2/3-gram)}
We quantify the lexical novelty of generated ideas using bi-grams and tri-grams relative to the original context and puzzle text. Let $\mathcal{N}_n(x)$ denote the set of $n$-grams extracted from sequence $x$.

We define $n$-gram novelty as the complement of the Jaccard similarity between the generated hypothesis and its corresponding context and puzzle:
\begin{equation}
\text{Novelty}_n(g, c)
=
1-
\frac{|\mathcal{N}_n(g) \cap \mathcal{N}_n(c)|}
{|\mathcal{N}_n(g) \cup \mathcal{N}_n(c)|},
\qquad n \in \{2,3\}.
\end{equation}
This formulation captures the extent to which the generated hypothesis introduces lexical combinations that do not already appear in the original input. Higher values indicate greater lexical divergence from the sources.

\paragraph{Entropy-based Novelty (Length-normalized Cross-Entropy)}
To capture distributional deviation relative to the original context and puzzle, we measure novelty using length-normalized cross-entropy. For each generated hypothesis $g$, we compare its token sequence against the empirical token distribution induced by its context and puzzle text $c$. Let $P_{c}(w)$ denote the empirical probability of token $w$ as estimated from $c$.

We define distributional novelty as:
\begin{equation}
\text{Novelty}_{\mathrm{CE}}(g, c)
=
-\frac{1}{|g|}
\sum_{w \in g}
\log \max(P_{c}(w), \epsilon),
\end{equation}
where $|g|$ is the number of tokens in $g$ (length normalization). To handle unseen tokens, we apply smoothing by using $\max(P_{c}(w), \epsilon)$, $\epsilon = 10^{-8}$.

This measure quantifies the average token-level surprisal of the generated hypothesis under the token distribution of its paired context and puzzle. Intuitively, it captures how difficult it is to explain the generated text using the lexical distribution of the original input. Higher values indicate that the generated hypothesis uses fewer expected tokens relative to the input and therefore exhibits greater distributional novelty.

\paragraph{Natural Language Inference (NLI)}
We use an NLI model, DeBERTa \citep{he2020deberta}, to evaluate the relationship between the context/puzzle and the idea generated from it. Let $p_e$ and $p_c$ denote the probabilities of entailment and contradiction, respectively. A high $p_e$ indicates that the generated idea is well supported by, and can be inferred from, the context/puzzle, whereas a high $p_c$ indicates that the idea conflicts with or contradicts it. We define:
\begin{equation}
\text{Derivation}_{\text{NLI}}(g, c) =p_e - p_c
\end{equation}
A higher derivation score implies lower novelty.

\paragraph{Contextual Perplexity}

We compute the conditional perplexity of each generated idea given its corresponding context/puzzle. Concretely, we concatenate the context/puzzle and the generated idea into a single sequence, and mask out the context/puzzle tokens by setting the ignore index to -100 so that the loss is computed only over the generated portion. This setup ensures that the model evaluates the likelihood of the ideas conditioned on the preceding context/puzzle, rather than in isolation. The resulting metric---obtained by exponentiating the cross-entropy loss---captures how ``surprised'' the model is by the generated ideas: lower perplexity indicates that the answer is more predictable given the context/puzzle, while higher perplexity suggests greater novelty or unexpectedness. We first want to see whether different LLMs yield similar results and we consider four similar-sized models: Qwen-7B, Mistral-7B-v0.1, LLaMA-2-7B, and DeepSeek-LLM-7B-base. Empirically, we find that the conditional perplexity scores produced by these four models are highly correlated with one another, indicating strong agreement across evaluators (SI Table \ref{cond_ppl}). We therefore aggregate them by taking the mean perplexity as a unified measure of novelty. However, despite this internal consistency, the resulting mean perplexity-based metric exhibits little to no correlation with human judgments of novelty.

\paragraph{GraphEval-GNN \citep{feng2025grapheval}}

GraphEval is a lightweight graph-based framework for LLM-driven idea evaluation. Its key insight is that directly prompting an LLM to judge a complex idea yields biased and prompt-sensitive scores; the idea is instead represented as a node (or a set of nodes) and evaluated through graph algorithms over a similarity graph. In the original design, a small prompted LLM first decomposes each (long-text) idea into fine-grained viewpoints (atomic claims or facts). Because each idea in our setting is a single AI-generated hypothesis---typically one to three sentences, far shorter than the paper abstracts decomposed in the original paper of GraphEval---we thus treat each hypothesis as an atomic unit. Concretely, (i) each hypothesis is embedded with BERT into a single node; (ii) nodes are connected via top-5 cosine-similarity edges across hypotheses, forming a hypothesis-similarity graph; (iii) a two-layer weighted GNN learns node representations through neighborhood aggregation and predicts the hypothesis-level label.

Following the original design, we classify the raw scientist ratings into the four review-decision classes used by GraphEval\footnote{In our implementation, we found that the original 1--9 scale classes were too fine-grained for the model to learn effectively.} (reject, poster, oral, spotlight) by binning within-field quantiles rather than at fixed thresholds. Concretely, within each field we take the empirical $25$th, $50$th, and $75$th percentiles of the raw novelty ratings as cut points $c_1 \le c_2 \le c_3$ (rounded to the nearest integer on the $1$--$9$ scale), and assign a rating $r$ to reject if $r \le c_1$, poster if $c_1 < r \le c_2$, oral if $c_2 < r \le c_3$, and spotlight if $r > c_3$ (a spotlight hypothesis is one rated among the most novel \emph{for its field}, as shown in the original paper). Where the discreteness of the scale and the central concentration of ratings collapse two cut points onto the same integer, we shift one cut point by a single rating value so that all four classes remain non-empty. Quantile binning keeps the four classes comparably sized within each field---avoiding the severe imbalance that fixed thresholds would induce given that ratings cluster near the middle of the scale---so that macro-averaged metrics are not dominated by a sparse class, and it aligns the labels with the within-field, relative notion of quality used throughout our analysis.

We fit GraphEval-GNN transductively within each field: all hypotheses in the field, both train and test, are embedded into a single graph, with training hypotheses carrying these binned labels and test hypotheses being unlabeled, so that the model is supervised only by the rated training split. GraphEval-GNN has modest data requirements ($300$ training and $50$ test ideas in the original paper), so our sample is more than sufficient for training. For each hypothesis, GraphEval-GNN outputs a probability distribution over the four classes, which we collapse into a scalar novelty score by summing the probabilities of the three more-novel classes:

\begin{equation}
\mathrm{Score} = P(\text{Spotlight}) + P(\text{Oral}) + P(\text{Poster}).
\end{equation}

\subsection{Training reward models on human ratings}
\label{reward}

\paragraph{Basic setup} We first convert the original ratings into pairwise comparisons among five hypotheses evaluated by the same scientist (denoted $h_1$, $h_2$, $h_3$, $h_4$, and $h_5$), where each pair indicates which hypothesis is preferred along a given dimension. Scientists' own ideas are added to the prompt to represent their judgment perspectives. To prevent overfitting, we deliberately avoid using all possible pairs. For example, if ($h_1$, $h_2$), ($h_2$, $h_3$), and ($h_1$, $h_3$) are all included, the relation between $h_1$ and $h_3$ can be inferred by transitivity (if $h_1>h_2$ and $h_2>h_3$, then $h_1>h_3$); in our experiments, including such inferable pairs induced strong overfitting (training accuracy near 90\% but test accuracy below 60\%). We therefore retain only pairs whose relations cannot be deduced from others, namely the consecutive pairs ($h_1$, $h_2$), ($h_2$, $h_3$), ($h_3$, $h_4$), and ($h_4$, $h_5$).

The reward model maps a hypothesis \(y\) to a vector of dimension-specific rewards,
\[
r(y)=(r_1(y),r_2(y),\dots,r_D(y)),
\]
where \(D=3\) in our setting, corresponding to novelty, feasibility, and probability. Given two hypotheses \(y_a\) and \(y_b\), we adopt the Bradley--Terry (BT) framework to model preference based on reward differences. For each dimension \(d\), we define a preference sign \(s_d \in \{+1,-1,0\}\), where \(s_d=+1\) if \(y_a\) is preferred to \(y_b\), \(s_d=-1\) if \(y_b\) is preferred to \(y_a\), and \(s_d=0\) if the comparison is tied\footnote{In experiments we found that excluding tie pairs in training raises the accuracy, thus we excluded them.}. Using this sign, we define the signed reward difference as \(\Delta_d = s_d \bigl(r_d(y_a) - r_d(y_b)\bigr)\), which is positive when the model assigns a higher reward to the human-preferred hypothesis and negative otherwise. The per-dimension BT loss is then defined as
\[
\mathcal{L}^{\mathrm{BT}}_d = - \log \sigma\!\left(\Delta_d - \lambda m_d\right).
\]

$\sigma$ is the sigmoid function to map any real-valued number ($\Delta_d - \lambda m_d$) into a range between 0 and 1. To account for not only the direction but also the strength of preference, we further incorporate the absolute score gap \(m_d \ge 0\) between the two hypotheses on dimension \(d\) as a margin term, where the strength is controlled by $\lambda$.

We combine losses across dimensions using a weighted sum,
\[
\mathcal{L}_{\mathrm{multi\text{-}BT}} = \sum_{d=1}^{D} \alpha_d \mathcal{L}^{\mathrm{BT}}_d.
\]

\paragraph{Evaluation setup}

As shown in Extended Data Table \ref{tab:llm_judge_accuracy}, we report the full results for LLM-as-a-judge and SOTA reward models\footnote{Note that \texttt{nicolinho/QRM-Gemma-2-27B} was ranked No.~4 on RewardBench as of May 2026. We included it because the No.~2 model, \texttt{Databricks-Mosaic-Research/PGRM}, was unavailable by May 2026.} on our held-out human-labeled dataset. We construct pairwise comparisons from human judgments by selecting instances where scientists assign a higher score to one AI-generated hypothesis over another (no tie). A model's prediction is considered correct if it agrees with the human preference, and we report simple accuracy.

For reward models, the evaluation follows their standard usage: each model assigns scalar scores to both hypotheses, which we then convert into a directional preference (i.e., A $>$ B or A $<$ B) and compare against the human judgment. This procedure is applied consistently to both our trained models and all SOTA reward models. We evaluate LLM-as-a-judge under the same framework for consistency, using three representative models as reported in the main paper: Gemini 2.5 Flash, DeepSeek R1, and OpenAI's o4-mini Deep Research model. Under this setup, all three LLM judges perform sometimes worse than random chance. This appears to be driven by a central tendency bias, as they frequently assign similar or neutral scores to both hypotheses, despite the absence of ties in our dataset.

When prompted instead with a direct comparative question (e.g., ``Which hypothesis is better?''), LLM-as-a-judge performance improves (as there is no tie option for output). However, this improvement is unstable and difficult to interpret. We observe pronounced positional and framing biases: swapping the order of hypotheses, or rephrasing the prompt (e.g., ``Is A better/worse than B?'' vs. ``Is B worse/better than A?''), leads to significant fluctuations in their evaluation. These sensitivities have also been observed in other ambiguous judgment tasks \citep{wu2025automatically}, raising concerns about whether the observed gains, although still lower than our trained model, reflect genuine evaluative capability or artifacts of prompt design. Meanwhile, across all settings, matching human experts' probability of assigning truth to hypotheses was the easiest dimension on which LLMs performed.

\begin{figure}[p]
\phantomsection
\addcontentsline{toc}{subsection}{SI Figure~1: Test accuracy vs.\ training step for candidate models}
    \centering
    \includegraphics[width=1\textwidth]{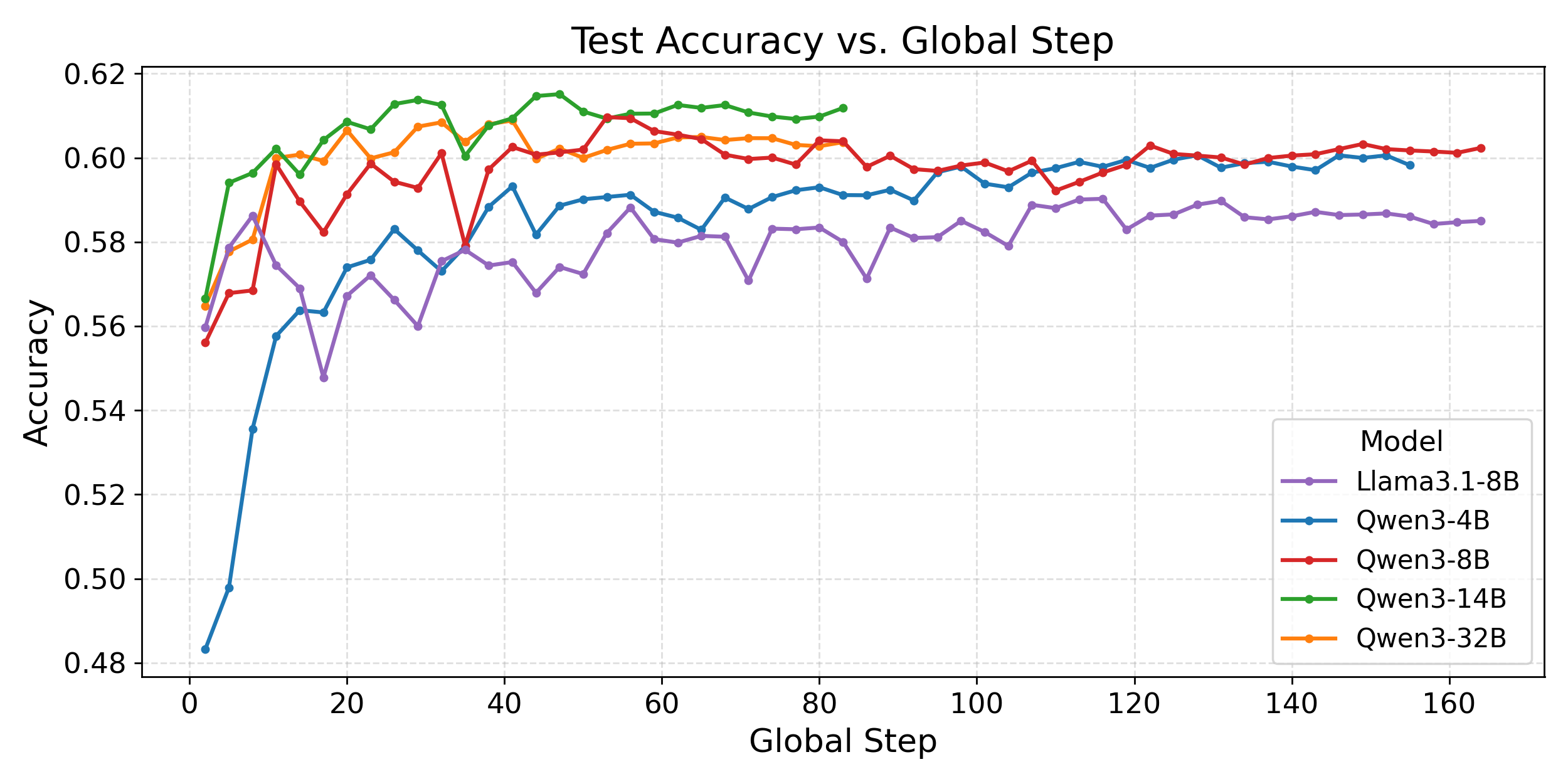}
    \caption{\textbf{Test accuracy versus global training step for five candidate models.} All models are trained using 2 GPUs except Qwen3-14B and Qwen3-32B, which are
trained using 4 GPUs. As a result, each global step for these two models
corresponds to twice the amount of data processed compared to the 2-GPU training,
accounting for their half global steps compared to others.}
    \label{fig:model_comparison_accuracy}
\end{figure}

\paragraph{Qwen3-14B is the optimal model} SI Figure~\ref{fig:model_comparison_accuracy} presents test accuracy as a function of
global training step for five candidate models. Qwen3-14B achieves the highest
test accuracy throughout training, reaching approximately 0.615 at peak and
stabilizing around 0.610. Qwen3-32B converges to a similar range near 0.604,
offering no substantial accuracy gain over Qwen3-14B. Llama-3.1-8B records the lowest final accuracy, stabilizing near 0.585.

Regarding hardware, all models are trained on NVIDIA A100 80GB PCIe GPUs except Qwen3-32B, which
requires NVIDIA H200 NVL GPUs due to its need for larger GPU memory. Given that Qwen3-32B
achieves no substantial improvement over Qwen3-14B in test accuracy while
demanding more advanced hardware, Qwen3-14B has been proven to be the most practical
choice. It delivers the best accuracy among all candidates while remaining
trainable on 4 A100 GPUs. Accordingly, all subsequent experiments are conducted
using Qwen3-14B.

SI Figure \ref{fig:model_comparison_accuracy} is provided to justify the selection of the base model (Qwen3-14B). In order to conduct a fair comparison, we fixed the learning rate to be $3\times10^{-5}$ and the margin weight to be 1\footnote{Note that these parameters, here and throughout Section \ref{reward}, are set uniformly for fair comparison, but they are not necessarily the optimal parameter configuration for model performance. We provide a separate paragraph later that specifically discusses the optimal parameters for each trained model.}.

\paragraph{Models are able to handle three quality dimensions simultaneously} Extended Data Table~\ref{tab:weight_config_all} presents training and test accuracy across four
evaluation domains and three dimensions, comparing a general model trained jointly
on all three quality dimensions ($\alpha_d = 1$ for all dimensions---novelty, feasibility, and probability) against three
dimension-specific models, each trained with $\alpha_d$ only switched-on on a
single dimension (Novelty-only, Feasibility-only, or Probability-only), using
Qwen3-14B with learning rate $3\times10^{-5}$ and margin weight $= 1$.

The training accuracy results reveal that the general model learns all three
dimensions in a balanced way, with training and test scores remaining close, indicating
no signs of overfitting. \revise{This shows that a 14B model can learn the three targets together; it does not by itself establish that expert judgments on these dimensions are independent, because a high-capacity model could also learn highly correlated targets without a trade-off. Models trained on a single dimension transfer poorly to the others in most cases (e.g., novelty-only models score 0.41--0.49 on feasibility), so we describe the dimensions as non-redundant but not necessarily independent.} The test results reveal a consistent pattern across domains and dimensions.
On \textbf{Feasibility}, the general model achieves scores comparable to the Feasibility-only model across all four domains
(Biology: 0.5667 vs.\ 0.5707; Chemistry: 0.5912 vs.\ 0.5757; Medicine: 0.6597
vs.\ 0.6486; Social: 0.6216 vs.\ 0.6378), suggesting that dimension-specific
training yields no clear advantage. On \textbf{Novelty} and \textbf{Probability}, a similar pattern appears as \textbf{Feasibility}.
\revise{Together, these findings indicate that, as a practical matter, a single jointly trained model suffices for all three dimensions without sacrifice on any individual dimension.}

\paragraph{Domain-specific models are necessary if the data is enough}

As shown in Extended Data Table \ref{tab:accuracy_all_dims}, we trained five models: four domain-specific models (Biology, Social Science, Chemistry, and Medicine) and one general model trained on the full dataset. We tested these models in the specific domains. Recall that the data is imbalanced---biology and social science have more preprints than medicine and chemistry. Biology, which accounts for the largest share of the data, outperforms the general model in domain-specific evaluation, suggesting that specialization is particularly effective when abundant in-domain data are available. Social Science, which has a more moderate share of the data, also exceeds the general model, albeit by a smaller margin. This indicates that each field indeed contains nuanced evaluative tastes that can be captured through specialization in model training. By contrast, Medicine and Chemistry have considerably fewer examples, and their domain-specific models do not consistently surpass the general model, which benefits from training on pooled and significantly larger datasets across all fields. Taken together, these results suggest that domain-specific training is preferable when sufficient in-domain data are available, while general models are more robust when in-domain data are limited.

\paragraph{Hyperparameter search}

Since the training objective involves two interacting hyperparameters, the
learning rate and the margin weight, their joint effect on model performance is
difficult to predict a priori. We therefore conduct a grid search over 80 trials,
combining five learning rate values ($\{1, 3, 5, 7, 9\}$) with four learning rate scales
($\{10^{-3}, 10^{-4}, 10^{-5}, 10^{-6}\}$) and four margin weight values
($\{0, 0.1, 0.5, 1\}$), evaluating each configuration on the general model
trained across all domains and dimensions on a separate validation set. This search also serves to examine the robustness of performance across the explored ranges.

The optimal hyperparameter configuration varies across scientific domains in the testing process. Biology model performs best with a learning rate of $2\times10^{-5}$ and a margin weight of $1.0$, the Chemistry model with a learning rate of $5\times10^{-6}$ and a margin weight of $0.1$, the Social Science model with a learning rate of $3\times10^{-5}$ and a margin weight of $1.0$, and the Medicine model with a learning rate of $2\times10^{-5}$ and a margin weight of $0.0$. The margin weight does not always help but it is necessary in some cases of training.

\subsection{Simulation models}
\label{simlation}

\revise{\paragraph{Overview and scope} We develop a stylized simulation to illustrate the potential consequences of uncritical AI use. The model builds on two empirical findings, drawn from survey experiments and observational data: a contraction in the field's ``radius'' and a suppression of null statements.}

\revise{A field faces many questions, each with an unknown true effect $\theta_j$, and each study $i$ reports an estimate $\hat\theta_{i,j} = \theta_j + B_j + \varepsilon_{i,j}$. Part of the error is random noise unique to each study, which averages out as studies accumulate; the rest, a shared bias $B_j$ plus noise correlated across studies that use the same tools and data, never averages out. AI acts as \textit{a fast but error-correlated instrument}: it produces more studies, shrinking random noise, but makes them more alike, enlarging shared error. Null suppression adds a further bias. If the results researchers could obtain form a bell curve around the true effect, leaving part of the results below the significance threshold (nulls) invisible (e.g., unpublished) pushes the visible (e.g., published) average upward. This distortion barely affects large effects but badly skews small ones, the most common type, and the one that requires careful investigation. A single dial, the AI share $a \in [0,1]$, increases the number of studies but also increases error correlation, field-wide blind spots, and null suppression. Measuring \textit{net knowledge} as one minus the squared gap between the field's estimates and the truth, the simulation illustrates three patterns. Net knowledge peaks at intermediate AI use, and earlier in mature fields, where extra studies add little. Beyond that peak, shared error grows faster than additional studies can reduce random error. Null suppression steepens the decline: at full automation, net knowledge is 0.56 with suppression versus 0.76 without.}

\paragraph{A field as a collective estimating a space of quantities}

A research field faces a space of questions it could ask, such as the
magnitude of each of many drug effects, or the prevalence of each of many social regularities, then studies them one by
one. We write the true magnitudes as $\theta_1,\dots,\theta_m$, all non-negative\footnote{Negative impacts are included here as well: we study only the magnitude of the effect, not its direction.}, drawn from a single
distribution in which most effects are small and a few are large:
\[
\theta_j\sim|\mathcal N(0,\omega^2)|,
\]
where $\omega$ sets how large real effects in the field tend to be.

Each study contributes a noisy measurement of whichever question it takes up. With significant AI
use, a study of question $j$ reports
\[
\hat\theta_{i,j} \;=\; \theta_j + \underbrace{B_j}_{\text{shared bias}} + \underbrace{\varepsilon_{i,j}}_{\text{error}} .
\]
Here $\varepsilon_{i,j}\sim\mathcal N(0,\sigma^2)$ with
$\mathrm{Corr}(\varepsilon_{i,j},\varepsilon_{i'})=\rho$ for any two studies ($i$ and $i'$) of the same
question, so that $\rho$ is the share of the error variance that all studies hold
in common and $\sigma$ sets the scale of a single study's measurement error. The two components above behave in opposite ways:

\begin{itemize}\setlength{\itemsep}{1pt}
\item \textbf{Idiosyncratic error}, a fraction $1-\rho$ of the error variance---each study's own
noise. Only this uncorrelated part averages out over enough studies.
\item \textbf{Shared bias} $B_j$, together with the \emph{correlated} share $\rho$ of the error variance---the component everyone gets wrong in the same direction, or in similar ways. This \emph{never}
averages away.
\item The shared bias decomposes into two parts, $B_j = B_h + \Bsel(z_j)$:
a question-independent \emph{blindspot} $B_h\ge 0$ (shared methods,
data and models make the whole field wrong in the same direction, regardless of
which question is asked), and a question-dependent \emph{selection} term
$\Bsel(z_j)$ arising from suppressed nulls, derived in \eqref{eq:bsel} below.
Both scale with AI adoption.
\end{itemize}

What the field learns about question $j$ from its $n_j$ studies decomposes cleanly:
\begin{equation}\label{eq:mse}
\MSE_j \;=\; \underbrace{B_j^2}_{\text{shared bias (will not average out)}}
\;+\; \sigma^2\,\underbrace{\rho}_{\text{correlation floor (will not either)}}
\;+\; \sigma^2\,\underbrace{\frac{1-\rho}{n_j}}_{\text{the \emph{only} piece more studies shrink}} .
\end{equation}
where MSE$_j$ represents the mean squared error of the pooled collective estimates of question $j$, i.e., how badly the field gets one question wrong.

\textbf{Greater throughput, generated by AI, can reduce the last term and no other.} A corpus of
correlated studies on one question is worth only $\neff = n_j/[1+(n_j-1)\rho] \to 1/\rho$ independent
ones: at $\rho=0.20$, even a million studies of one question are worth five. The field's total error
is the average of \eqref{eq:mse} over its questions, $\MSE=\tfrac1m\sum_j\MSE_j$ (i.e., how badly the field gets all questions wrong), and net knowledge is
$V=1-\MSE$.

\paragraph{\revise{Selective publication around a decision threshold}}

\revise{The modeled distribution is the sampling distribution of study estimates, $\hat\theta_{i,j}\sim\mathcal N(\theta_j,\sigma^2)$. A statement is classified as null when it falls below a decision threshold, and nulls and significant statements are the two regions of this single sampling distribution on either side of the threshold; they are equal in mass only when the true effect lies at the threshold. We keep every statement above the threshold and keep a null only with probability $1-s$, where $s$ is the suppression rate. At $s=0$ the distribution is intact; at $s=1$ nothing below the threshold survives. Selective publication conditional on significance inflates the average of what gets visible to other scientists (e.g., published), the well-known exaggeration (Type~M) error.}
Writing $z_j=\theta_j/\sigma$ for how far the question's true effect sits above
zero in noise units, the displacement \revise{for a threshold at zero} is
\begin{equation}\label{eq:bsel}
\boxed{\ \Bsel(z_j)\;=\;\sigma\,\frac{s\,\varphi(z_j)}{1-s\,\Phi(-z_j)}\,,\qquad \varphi(0)=\tfrac{1}{\sqrt{2\pi}}\approx0.40\ }
\end{equation}
\revise{More generally, let a statement be null when $\hat\theta_{i,j} < c\,\sigma$ for a significance threshold $c$ (e.g., $c=1.645$ for a one-sided test at $\alpha=0.05$). The selection bias is then
\begin{equation*}
\Bsel(z_j; c) = \sigma \, \frac{s\,\varphi(z_j - c)}{1 - s\,\Phi(c - z_j)},
\end{equation*}
which reduces to \eqref{eq:bsel} at $c=0$ and satisfies $\Bsel(z_j;c)=\Bsel(z_j-c;0)$. A nonzero threshold is therefore equivalent to shifting true effects by $c$ relative to the decision boundary, and the threshold at zero entails no loss of generality.} Two further consequences follow from the same cut: the
visible pool also \emph{narrows} (it drops its lower tail), so the literature looks more precise
exactly as it becomes more displaced; and $\Bsel$ does not depend on $n_j$. No amount of
throughput recovers statements that were never visible or published. Human researchers already \revise{suppress part of these results}, known as the file-drawer problem\citep{rosenthal1979file, chen2025geographical}; \revise{in the model, AI use raises the suppression rate}. The total shared bias on a question is $B_j = B_h + \Bsel(z_j)$.

\paragraph{Where the cut bites hardest}

The displacement depends on where the true effect of the question sits. At \revise{$s=0.85$ (the illustrative AI null-suppression rate used in the model)}:

\begin{itemize}\setlength{\itemsep}{1pt}
\item \textbf{Large true effects survive almost untouched.} At $z_j=2$ the displacement is
$0.05\,\sigma$, at $z_j=3$ only $0.004\,\sigma$: there was almost nothing \revise{below the threshold} to delete,
so removing it changes almost nothing.
\item \textbf{Small effects are distorted most.} At $z_j=0$ the displacement is
\revise{$0.59\,\sigma$---the two regions were balanced, and suppressing one destroys the cancellation}. The
questions a field gets most wrong are exactly those that possess a true effect that is smallest and most questionable, and by
construction those are most of its questions.
\end{itemize}

\paragraph{AI as a fast but error-correlated instrument}

AI is an instrument that is far faster, but whose errors are also more highly correlated. Adoption is
captured by a single parameter $a\in[0,1]$. Increasing $a$ raises throughput $n$---the sole
benefit, insofar as more studies shrink the averageable term $\sigma^2(1-\rho)/n$---but it also raises
three harms: \revise{(1) the cross-study error correlation $\rho$ (shared methods, data and models make errors
more alike), (2) the shared blindspot $B_h$, and (3) the null-suppression rate $s$}. All three scale with
adoption and fall on the parts of Equation \eqref{eq:mse} that throughput cannot reach: speed addresses only
the first error type and leaves the three harms untouched.

\paragraph{Three patterns}

The parameters used in the model can be found in SI Table \ref{tab:params}. As we discuss below,
changing the numbers slides the hump, but it remains a hump, leaning away from AI in mature
fields, and falling faster when nulls are buried.

\textbf{(1) AI's benefit follows an inverted-U shape.} Net knowledge $V$ (1 minus MSE, so that higher
is better) traces a hump as the AI share $a$ rises from $0$ to $1$:

\begin{itemize}\setlength{\itemsep}{1pt}
\item \textbf{Up-slope} (low $a$): few studies, low correlation---added AI throughput cancels
random noise, and knowledge rises.
\item \textbf{Peak} ($a=a^\ast$): the marginal noise-cancelling benefit exactly balances the marginal
growth in bias and correlation.
\item \textbf{Down-slope} (high $a$): the three harms accumulate in the irreducible part of the error;
additional speed buys almost nothing, and knowledge falls.
\end{itemize}

The optimum $a^\ast$ is therefore \emph{interior}---a human/AI mix. The only benefit enters through
$\sigma^2(1-\rho)/n$, whose marginal value decays like $1/n^2$; the three harms grow with no such
damping. The benefit necessarily dominates while $n$ is small (the up-slope) and the harms once $n$
is large (the down-slope). Equivalently: switch throughput off and only harms remain, so the optimum
sits at $a=0$; switch all harms off and only the benefit remains, so it sits at $a=1$; with both
active, the peak lies in between.

\textbf{(2) \revise{In mature fields the optimum lies at lower AI use}.} Young fields have few studies, so AI's
throughput gains are valuable. Mature fields already have many, so additional studies add little
and the ``everyone looks alike'' harm dominates. The more studies a field already has (a maturity
factor $\tau$), the lower its optimal $a^\ast$: raising $\tau$ lifts $n$ at every $a$, so the marginal value of additional throughput ($\sim\!1/n^2$) is smaller everywhere, while the blindspot and null-suppression harms are untouched by $\tau$; the balance therefore tips at a smaller $a$.

\textbf{(3) \revise{Suppressing more nulls lowers net knowledge further}.} The more aggressively AI use suppresses
nulls (higher $s$ at full automation), the further \emph{left} the optimum shifts and the
\emph{deeper} full-AI science falls. The selection bias $\Bsel$ rises with $s$ and is
\emph{independent of $n$}, so extra suppression adds harm that no amount of throughput can offset,
lowering $V$ at every high-$a$ point\revise{; raising $s$ at full automation lowers $V(1)$ and shifts $a^\ast$ leftward}. Comparing suppression off against on
is the sharpest statement of the point: with the file drawer switched off, full automation still
leaves $V(1)=0.76$. Switching it on brings that to \revise{$0.56$}.

\subsection{Supporting evidence for null results in statistical analysis}
\label{sec:equivalence}

Across the paper we report three principal null results:

\begin{itemize}
  \item Once idea quality is controlled for, prior AI engagement is not associated with
        \revise{willingness to pursue AI-generated hypotheses} (Figure~\ref{fig:judgment}d in the main text).
  \item Field-level attitude differences are driven by senior authors' attitudes;
        among junior authors we detect no statistically significant field-level
        variation in \revise{willingness to pursue AI-generated hypotheses} (Figure~\ref{fig:judgment}f in the main text).
  \item Seniority predicts \revise{a residual reluctance} to pursue AI-generated ideas,
        but we find no corresponding effect on the quality evaluation of those
        ideas (Extended Data Figure \ref{fig:percep}).
\end{itemize}

Because a non-significant coefficient is not itself evidence of absence, we provide additional evidence that the null results are not artifacts.

\paragraph{Prior AI engagement does not predict willingness to pursue}

A one-SD increase in prior AI engagement corresponded to a
$-0.007$-point change in willingness to pursue (90\% CI $[-0.031,\,0.018]$; two-sided
$p=.66$). This is the coefficient in the main text
expressed in a different unit rather than a different estimate: the raw
coefficient reported there ($-0.1697$) is rescaled by the standard deviation of
log prior AI engagement ($0.038$) so that the estimate and the equivalence bounds share
the units of the willingness scale. We further conducted two one-sided tests against an
equivalence bound of $\pm 0.192$, i.e., $0.1$~SD of the willingness scale
($\mathrm{SD}=1.92$). Equivalence was supported ($p_{\text{TOST}}<.001$), and the
interval in fact excludes effects larger than $0.031$ points ($0.016$~SD)---a
bound roughly six times tighter than the one we pre-specified. The same conclusion holds against an alternative arbitrary bound of $\pm 0.150$, so it does not depend on how the bound is chosen. The unadjusted association was similarly bounded
($\beta = 0.033$, 90\% CI $[0.001,\,0.066]$). Once idea quality is controlled
for, prior AI engagement therefore does not predict whether a scientist is willing to pursue an AI-generated idea, at any magnitude we would regard as substantively meaningful within the observed range of prior AI engagement. Full details can be found in SI Table~\ref{tab:tost_ai_use}.

\paragraph{Field differences in willingness to pursue are concentrated among senior scientists}
Field differences in willingness to pursue are concentrated among established scientists. Because the substantive claim is that these differences emerge with seniority, we tested it directly as a field~$\times$~seniority interaction, estimating gaps within terciles of respondents' seniority (citation percentile by default) rather than extrapolating a linear interaction to a seniority of zero, a point outside the observed support (minimum $=0.006$). The interaction was significant, $F = 4.08$, $p = .007$, and survived a wild cluster bootstrap-$t$ ($p = .010$); a fully flexible tercile specification pointed the same way, $F = 2.08$, $p = .053$. The gradient is monotone. The largest absolute field gap grows from $0.077$ scale points ($0.04$~SD) in the junior tercile to $0.374$ ($0.19$~SD) in the senior tercile, and the number of significant pairwise contrasts rises from zero of six to three of six. Estimated directly, the medicine-social gap widens by $0.336$ points across the two strata ($p = .006$). At the junior end the $90\%$ intervals place every field difference below $0.224$ points ($0.117$~SD)---beneath the lower limit of the largest senior gap ($0.233$)---so the disciplines that diverge sharply among senior scientists are, at the start of a career, bounded within a range that excludes differences of that magnitude. We read the junior-end result as an upper bound: field differences in willingness to pursue are not a fixed property of disciplines but something that accumulates with standing in them (i.e., seniority). Full details can be found in SI Table \ref{tab:field_gaps_by_status}.

\paragraph{Seniority predicts willingness to pursue, not evaluation}

Seniority was associated with a markedly lower willingness to pursue AI-generated
ideas ($-0.119$~SD per unit of seniority, $p = .001$), but its
association with how those same ideas were evaluated was non-significant in
every dimension. Re-estimating all four outcomes on a
common right-hand side and in standard-deviation units, so that the slopes are
directly comparable, the quality associations are novelty $-0.038$~SD,
$p = .27$; feasibility $+0.034$~SD, $p = .32$; probability $-0.050$~SD,
$p = .12$.
All three dimensions are non-significant under either specification.
Because a contrast of significance
levels is not itself a test, we estimated a stacked model in which all four
outcomes are standardized and fully interacted with outcome identity, yielding a
joint covariance matrix over the four seniority slopes. The slopes differ
significantly across outcomes, $F= 6.16$, $p < .001$, and each
quality dimension differs from willingness to pursue individually (novelty $\Delta = 0.081$,
$p = .033$; feasibility $\Delta = 0.154$, $p < .001$; probability $\Delta = 0.069$, $p = .017$). This joint test, not the pattern of
significance levels, is the basis of the claim. As a sensitivity check, taking
the seniority effect on willingness to pursue as an equivalence bound ($\pm 0.119$~SD), all
three quality dimensions were statistically equivalent to zero (largest
$p_{\text{TOST}} = .018$), holding down to a bound of $0.104$~SD; this bound is
estimated from the same data and does not hold against the smaller
quality-adjusted seniority coefficient of Extended Data
Table~\ref{tab:ols_combined}. We characterize the result as a difference in magnitude---the seniority gap in
willingness to pursue is roughly two to three and a half times any corresponding
difference in assessment. Senior researchers decline AI-generated ideas at rates their own
quality ratings do not account for. Full details can be found in SI
Table~\ref{tab:seniority_dissociation}.

\begin{revblock}
\subsection{Defining and validating substantive null hypotheses}
\label{sec:nulldef}

\paragraph{What we measure} Every alternative hypothesis is paired with a statistical null, whether or not that null is written down, and the absence of an explicit null does not mean the researcher failed to formulate one. This default null is present for human and AI hypotheses alike, and it is not what we measure or intend. By ``negation'' in ideation we meant to measure a substantive null: a hypothesis whose predicted outcome is itself the absence of an effect, relationship, or difference, so that the researcher's theoretical bet (or one of their bets) is on the null. A paper that tests ``A increases B'' has formulated a statistical null; a paper that proposes ``A has no effect on B'' as its expectation has formulated a substantive one. These differ in how a study is designed, powered, and interpreted. Conversely, an explicit null written merely as the formal counterpart of a stated alternative (``H$_0$: no difference'') is not a substantive null in our sense. This is mainly safeguarded by our extraction prompts on human papers, which emphasize the core hypotheses of each study's research design rather than peripheral, result-level descriptions (Supplementary Information \ref{prompts}). These extractions were also validated by the original authors. We don't see signals such as ``H0'' appearing in extracted texts.

\paragraph{How extraction handles different cases} (1) \textit{A paper presents multiple hypotheses}: all core human hypotheses are extracted as a numbered list, and each model generates the same number of hypotheses. (2) \textit{Papers without explicit hypotheses} are excluded from the survey. (3) \textit{Explicit non-null hypotheses paired with explicit null hypotheses} do not reflect how authors state their substantive hypotheses, and they appear rarely in abstracts of published papers. As discussed above, authors' substantive hypotheses typically state only one side: the side on which their argument rests. These pairs do occur in peripheral reports of experiments, but our prompt explicitly instructs the model not to extract them and we do not use them. (4) \textit{Papers whose main substantive hypothesis is a null}: in the survey, a paper may contain one or more hypotheses, and we count it in this category whenever at least one of its extracted hypotheses is a null hypothesis. An LLM is counted as proposing the null if at least one of its proposed hypotheses is the null hypothesis. All extracted hypotheses are verified by the authors in the survey.

\paragraph{Validation and the limits of text} Two graduate students independently annotated 100 null statements from abstracts in our corpus (for example, ``exposure to cadmium has no effect on the balance between vertical and horizontal microbial transmission''). They agreed in all cases on whether a sentence is a null statement, but disagreed substantially on whether a statement originated at the ideation stage (a hypothesis) or the reporting stage (a finding). Because abstracts are written after the results are known, an a priori null hypothesis and a post hoc non-significant result are often phrased identically; bidirectional questions (``whether A affects B'' under certain conditions) commit to neither, and undisclosed post hoc hypothesizing is common, which is widely known as ``HARKing'' \citep{rubin2017does}. For the corpus measure we therefore do not assign a stage to each detected statement; a paper counts as significant only if none of its abstract sentences is a null statement of any kind.

\paragraph{Direction leakage from contexts and puzzles} We randomly sampled 300 pairs, each consisting of extracted human hypotheses and their corresponding context and puzzle (i.e., the input from which the LLM generates hypotheses). Two graduate students independently assessed whether the context and puzzle revealed the direction of the human hypotheses or any clue. One annotator judged that 92\% of the pairs showed no leakage, while the other judged 82\%. Three factors explain both the overall low leakage rate and the gap between the two annotators: \textbf{1. Pre-filtering.} We had already removed many puzzles that could reveal the human hypothesis using a textual similarity filter (Supplementary Information \ref{2.2}). Our rewriting step converts directional claims into neutral questions (e.g., ``A increases B'' $\rightarrow$ ``Does A increase B?''). We compute the semantic similarity between each original hypothesis and its rewritten form. If the similarity exceeds a predefined threshold (0.82, calibrated on 20 rewrites for each of 1{,}000 hypotheses), we flag the rewrite as leaking information and discard the corresponding context/puzzle. \textbf{2. Prompt constraints.} The context-extraction prompt prohibited revealing the human scientists' answer to the puzzle (Supplementary Information \ref{prompts}). We first extract the hypotheses, and then provide them to the LLM \textit{together with} the introduction and related work of the original paper so that the LLM can extract the context without leaking the hypotheses. \textbf{3. Concentrated disagreement.} Nearly all cases flagged as leakage were of a single type, to which the two annotators applied different levels of strictness: puzzles phrased as goal-oriented questions. For example, the puzzle ``How can senescent tumor cells remaining after chemotherapy be effectively cleared to improve treatment outcomes?'' corresponds to the hypothesis that ``IAP antagonists can kill senescent tumor cells when xxx (an interaction effect or a condition).'' One annotator treated such puzzles as simple statements of a research goal. The other flagged them on the grounds that a question of the form ``how can we enhance/promote/attenuate\ldots'' already signals the hypothesized effect. Such framing, however, reveals mainly the direction of the effect rather than the hypothesis's key variables. Moreover, the extraction prompt explicitly prohibits leaking the human answer, and our similarity-based filter is designed to catch remaining overlap.

\paragraph{Conditional rates} Among contexts and puzzles for which any model proposed a null hypothesis, humans also proposed one in roughly 4\% of cases, compared with 0.8\% overall. Among contexts and puzzles for which humans proposed a null, at least one LLM did so in 9.4\% of cases and no model did in the remaining 90.6\%. When a context and puzzle lends itself to null-directed research (in human sense), LLMs are thus more likely than usual to propose a null (nearly all LLMs do so in fewer than 0.4\% of cases overall), but agreement with humans remains very low and is unrelated to model size (SI Table~\ref{tab:null_overlap}).

\paragraph{Estimates from the literature} We also turn to prior literature. The rates produced by our classifier agree closely with independent estimates obtained under different methods. We find that 74\% of papers in medicine and 77\% in social science (2010--2025) are significant-only, matching an earlier estimate of 77\% in political science (2010--2024) \citep{briggs2026null}. Other studies report that about 16\% of hypothesis-testing papers across disciplines report negative statements \citep{fanelli2010positive}, and about 32\% of empirical psychology abstracts contain at least one negative statement \citep{aczel2018quantifying}. Most of these studies, however, do not specifically separate a priori null hypotheses from post hoc non-significant reports; all of them count the surface statement. Further evidence shows that this blurring is pervasive in human writing. In 72\% of non-significant findings mentioned in abstracts of three leading psychology journals, authors concluded that the effect was absent \citep{aczel2018quantifying}. Of 137 negative statements in the same abstracts, 23\% inferred from a non-significant result that the effect was absent in the population, 48\% described the absence of a difference in the observed sample, and only 18\% correctly reported mere non-significance \citep{aczel2018quantifying}. Earlier work found that non-significant results were interpreted as claims of no effect in about 60\% of cases \citep{hoekstra2006probability}. ``No effect'' could be a hypothesis, on equal footing with ``an effect exists.'' Treating ``failed to detect'' as direct justification for ``no effect'' skips a crucial step: asking whether the measurement was capable of detecting the effect in the first place. In fact, many of the 135 ``no effect'' claims in economics rest on estimates that cannot rule out meaningful effects \citep{fitzgerald2024need}. Absence may originate at the ideation stage, whereas non-significance may arise during experimentation. The same surface wording thus carries different meanings, and authors themselves rarely distinguish between them.

\paragraph{Examples} Boxes 1--3 show contexts and puzzles for which a human or a model proposed a null hypothesis, examples of substantive and non-substantive nulls, and null statements in published abstracts.

\begin{examplebox}{Box 1a.\ One model proposed both null and non-null hypotheses \exfield{Chemistry}}
\textbf{Context.} Solution-processed metal halide perovskites now rival silicon in power conversion
efficiency, and film quality is a key determinant of device performance. Films are typically deposited
from a solution of a lead salt and an organic ammonium salt and then thermally annealed. Lead iodide is
the most common lead source, but lead salts with other counterions, such as chloride and acetate, have
also been used. Larger, well-crystallized grains are generally associated with longer charge-carrier
lifetimes and higher efficiencies, yet how the counterion influences film formation remains poorly
understood.

\smallskip
\textbf{Puzzle.} How does the counterion of the lead precursor affect perovskite film formation and the
optoelectronic properties of the resulting films?

\smallskip
\begin{itemize}[leftmargin=1.2em, itemsep=2pt, topsep=2pt]
  \item \textbf{Human, hypothesis 1} \nonnull: Acetate-based precursors yield larger grains than
        iodide-based precursors because the slower release of the acetate byproduct delays nucleation.
  \item \textbf{Human, hypothesis 2} \nonnull: Residual acetate at grain boundaries passivates defects
        and extends charge-carrier lifetime.
  \item \textbf{Ministral 3B, hypothesis 1} \nonnull: Counterions with higher volatility accelerate
        nucleation and reduce final grain size.
  \item \textbf{Ministral 3B, hypothesis 2} \nullhyp: When annealing temperature and duration are held
        constant, counterion identity has no effect on grain size.
  \item \textbf{GPT-4.1, hypothesis 1} \nonnull: Chloride-containing precursors slow crystallization
        and yield more uniform films with fewer pinholes.
  \item \textbf{GPT-4.1, hypothesis 2} \nonnull: The counterion alters the intermediate solvate phase,
        which in turn sets the preferred crystal orientation of the final film.
\end{itemize}
\textit{Each model generates as many hypotheses as the authors stated (here, two). A response counts as
proposing a null if at least one of its hypotheses is null; Ministral 3B's response therefore counts as
null, GPT-4.1's does not.}
\end{examplebox}

\begin{examplebox}{Box 1b.\ Human proposed a substantive null; no model did \exfield{Social science}}
\textbf{Context.} Affective polarization, the dislike and distrust that partisans feel toward the opposing
party as opposed to disagreement over policy, has risen sharply in the United States over recent decades.
It is commonly measured with feeling thermometers and trait ratings of out-party members, and social media
is frequently blamed for its growth. Because platforms rank content algorithmically, what users see
depends partly on platform design and not only on whom they choose to follow. Observational studies find
that users exposed to more cross-cutting content tend to view the out-party more warmly, but experimental
evidence on how feed content shapes partisan animosity remains scarce.

\smallskip
\textbf{Puzzle.} What is the relationship between the ideological composition of users' feeds over a short term (e.g., six months) and their feelings toward the opposing party?

\smallskip
\begin{itemize}[leftmargin=1.2em, itemsep=2pt, topsep=2pt]
  \item \textbf{Human} \nullhyp: Changing the ideological composition of users' feeds over 6 months does not alter their affective polarization, because partisan animosity is rooted in stable identities that a short-term change in feeds cannot shift.
  \item \textbf{GPT-4o} \nonnull: Increasing the share of counter-attitudinal content in users' feeds in six months reduces affective polarization by raising perceived warmth toward out-party members.
  \item \textbf{o3} \nonnull: The effect of feed composition depends on prior partisan strength: more
        cross-cutting exposure reduces polarization among weak partisans but increases it among strong
        partisans through reactance.
\end{itemize}
\end{examplebox}

\begin{examplebox}{Box 1c.\ A model proposed a substantive null; the human did not \exfield{Biology}}
\textbf{Context.} The fruit fly \textit{Drosophila melanogaster} harbors a simple gut microbiota
dominated by \textit{Lactobacillus} and \textit{Acetobacter} species, and germ-free (axenic) flies are
easy to generate, making it a tractable system for studying host--microbe interactions. Gut bacteria
contribute to host nutrition, metabolism, and intestinal homeostasis. Yet studies comparing axenic and
colonized flies have produced conflicting results: gut bacteria have been reported to shorten lifespan,
to extend it, and to have no effect.

\smallskip
\textbf{Puzzle.} What explains the conflicting reports on how gut bacteria affect lifespan in
\textit{Drosophila}?

\smallskip
\begin{itemize}[leftmargin=1.2em, itemsep=2pt, topsep=2pt]
  \item \textbf{Human} \nonnull: Gut bacteria shorten lifespan on protein-rich diets but extend it on
        protein-poor diets, through bacteria-dependent increases in protein intake.
  \item \textbf{Gemma 3 12B} \nullhyp: Once dietary nutrient intake is equalized between axenic and
        colonized flies, the presence of gut bacteria has no effect on lifespan.
  \item \textbf{o4-mini} \nonnull: Gut bacteria shorten lifespan by accelerating age-related intestinal
        barrier dysfunction.
\end{itemize}
\end{examplebox}

\begin{examplebox}{Box 2.\ Substantive versus non-substantive null hypotheses}
\begin{itemize}[leftmargin=1.2em, itemsep=2pt, topsep=2pt]
  \item \textbf{Non-substantive null, not counted:} ``H$_0$: There is no difference in 30-day mortality between
        the two treatment arms.'' The paper's key idea is H$_1$; H$_0$ is only its formal counterpart.
  \item \textbf{Substantive null, counted:} ``We expect that bilingualism has no effect on
        executive-function performance once socioeconomic status is controlled.'' The theoretical bet
        itself is on absence.
\end{itemize}
\end{examplebox}

\begin{examplebox}{Box 3a.\ Null statements in a published abstract (adapted from ref.~\citep{hyde2005gender})}
The differences model, which argues that males and females are vastly different psychologically, dominates the popular media. Here, the author advances a very different view, the gender similarities hypothesis, which holds that males and females are of no difference on most, but not all, psychological variables. Results from a review of 46 meta-analyses support the gender similarities hypothesis. Gender differences can vary substantially in magnitude at different ages and depend on the context in which measurement occurs. Overinflated claims of gender differences carry substantial costs in areas such as the workplace and relationships.
\end{examplebox}

\begin{examplebox}{Box 3b.\ Null statements in a published abstract (reproduced from ref.~\citep{przybylski2019violent})}
\ldots There was no evidence for a critical tipping point relating violent game engagement to aggressive behaviour. Sensitivity and exploratory analyses indicated these null effects extended across multiple operationalizations of violent game engagement and when the focus was on another behavioural outcome, namely, prosocial behaviour. The discussion presents an interpretation of this pattern of effects in terms of both the ongoing scientific and policy debates around violent video games, and emerging standards for robust evidence-based policy concerning young people's technology use.
\end{examplebox}

\subsection{Robustness of the null-hypothesis gap}
\label{sec:nullrobust}

\paragraph{Richer inputs} We randomly sampled 10000 papers from our preprint platforms and, besides the context and puzzle, built two richer inputs: the Introduction alone, and the Introduction plus Related Work (when Related Work is available), both extracted with GROBID. Both variants were sanitized by removing every sentence whose similarity to the extracted human hypothesis exceeded 0.82 and prompting GPT-4.1 to rewrite the remaining text into a coherent passage. Each model then generated hypotheses under all three inputs. To test whether outputs move when the input changes, we embedded hypotheses with MPNet, projected them to 2D and 3D with UMAP, and, for each pair of conditions with coordinate matrices $A$ and $B$ (rows matched by paper), computed: the paired Euclidean distance $\lVert A_i-B_i\rVert_2$ against the median distance between non-matching papers; a permutation test shuffling the rows of $B$ 1{,}000 times ($p=(\#\{\text{null}\le\text{observed}\}+1)/1{,}001$); retrieval of each paper's counterpart (Top-1, Top-10 and mean reciprocal rank (MRR)); Procrustes disparity; per-dimension Pearson and Spearman correlations; and the Jaccard similarity of 10-nearest-neighbour sets. Outputs were highly stable.

\begin{itemize}
    \item \textbf{Paired distance.} The median paired distance for the same paper across conditions is only around 0.008, compared with a baseline distance of roughly 11--13 for non-matching pairs, giving a \textit{distance ratio} no greater than 0.001. Every comparison yields a permutation $p$-value of 0.001, the minimum attainable with 1{,}000 permutations.
    \item \textbf{Retrieval.} When retrieving the same paper from the other condition, Top-1 accuracy is 0.83--0.99 in 2D (0.86--0.99 in 3D), and MRR is 0.86--0.99.
    \item \textbf{Global structure.} Procrustes disparity is 0.008--0.079 in 2D and 0.006--0.071 in 3D. All per-dimension Pearson and Spearman correlations are at least 0.95.
    \item \textbf{Local structure.} The 10-nearest-neighbour Jaccard similarity is 0.70--0.95 in 2D and 0.72--0.96 in 3D.
\end{itemize}

Across models, OpenAI models are systematically the most stable. o3-mini, GPT-4.1, GPT-4o, o3, and o4-mini reach Top-1 accuracy of about 0.97--0.99. Gemma 3 4B and Gemma 3 27B are the least stable, with Top-1 accuracy of about 0.83--0.90 and 10-nearest-neighbour Jaccard of about 0.70--0.80, though still far above chance. Across conditions, the difference between Introduction and Introduction + Related Work is generally slightly smaller than the difference between either of these and the context and puzzle, but the gap is small. Metrics in 3D are generally slightly better aligned than in 2D, likely because projecting to 2D forces variance into fewer dimensions. Note that the mean paired distance (0.05--0.71) is substantially larger than the median, indicating that the differences are driven mainly by a small number of outlier papers.

Null rates (SI Table~\ref{tab:expand_null}) differed across inputs only for Gemma 3 12B ($\chi^2$ test, $p=0.0063$), an effect that does not survive Benjamini--Hochberg correction ($q\approx0.11$). Expanding the input does not systematically change the tendency to propose nulls.

\paragraph{AI polishing of abstracts} To test whether AI-assisted writing alone could produce the post-2022 corpus trend, we sampled 100000 arXiv abstracts from 2019, before commercial LLMs were available, and asked Claude Opus 5.5 (medium) to ``polish the text without changing its content''. Polishing did not change the rate of null statements, suggesting that AI-assisted writing per se is unlikely to be the main driver.

\paragraph{Journal articles without preprint versions}
\label{sec:si_published}
From OpenAlex we randomly sampled articles in social science, medicine, biology and chemistry published in 2026 for which no preprint could be found by title on any preprint server; these peer-reviewed papers are very unlikely to be in any model's pretraining data. Because only titles and abstracts can be legally used, o3-mini first judged whether a context and puzzle and the corresponding human hypothesis could be recovered from this text without the context puzzle revealing the hypothesis (prompt see Supplementary Information \ref{prompts}); if so it performed the extraction, and otherwise the paper was discarded. This yielded 10{,}000 contexts and puzzles from which the models proposed hypotheses. Author email addresses are unavailable for these papers from OpenAlex, and abstracts omit much detail, so we use this sample only as a robustness check. Human hypotheses predicted a null in roughly 1.0\% of cases, whereas every LLM did so in at most 0.35\% (SI Table~\ref{tab:null_published}). Reasoning models' hypotheses remained significantly more dispersed than those of non-reasoning models (0.896 vs.\ 0.695 in 3D and 1.565 vs.\ 1.213 in 2D; MPNet embeddings, displacement vectors, UMAP).

\paragraph{Base versus instruction-tuned models} If null suppression were driven mainly by post-training, as creative homogenization appears to be \citep{li2026narrative}, base models should propose substantially more nulls than their instruction-tuned counterparts. We compared three base models with their instruction-tuned counterparts but found no consistent evidence that base models propose a higher proportion of null hypotheses. On the OpenAlex data mentioned above, the instruction-tuned versions of Qwen3-4B and Ministral-3B indeed proposed substantially fewer null hypotheses than their base models (71.1\% and 91.0\% fewer, respectively). For OLMo-3-7B, however, the relationship reversed: the instruction-tuned model proposed 3\% more null hypotheses than its base counterpart. This proportion may depend on the composition of the context puzzle pool, and different post-training methods may affect it differently, but overall we observe no unified pattern. A pretraining and compression account, in which sparse, long-tail knowledge degrades disproportionately \citep{wu2025missing,sun2024head,shumailov2024ai}, remains a possible explanation, as do sycophancy and instruction-following priors \citep{sharma2024towards,cheng2026elephant}.

\subsection{Robustness of hypothesis-space and similarity measures}
\label{sec:si_space}

We tested the robustness of the hypothesis space (the space spanned by the ends, i.e., the generated hypotheses, relative to the start, i.e., the shared context and puzzle) by replacing the general-purpose MPNet with SPECTER2 \citep{Singh2022SciRepEvalAM}, an embedding model trained with citation-based contrastive learning on scientific documents, and by using different dimensionality-reduction methods (UMAP, t-SNE and PCA). Because pairwise similarity among hypotheses derived from the same paper partly reflects shared terminology, we also reran the hypothesis-similarity analysis (Fig.~\ref{fig:ideation}b) using SPECTER2 and displacement vectors rather than raw vectors or MPNet.

We treated each reasoning model, each non-reasoning model and the human authors as distinct regions and measured each region's extent as the standard deviation along each dimension (Extended Data Fig.~\ref{fig:diverse}a,b). Reasoning models consistently traverse a larger space than non-reasoning models. Reasoning models and humans do not differ significantly. These results were robust to alternative dimensionality-reduction methods (PCA, 2D: 0.15 vs.\ 0.09, $p = 0.001$; 3D: 0.15 vs.\ 0.09, $p < 0.001$; t-SNE, 2D: 38.1 vs.\ 17.8, $p = 0.004$; 3D: 16.1 vs.\ 7.5, $p < 0.001$) and to the alternative embedding model, with SPECTER2 combined with UMAP yielding the same conclusions. Across all settings, non-reasoning models show the strongest within-group homogeneity (Extended Data Fig.~\ref{fig:diverse}c--f). Hypotheses generated by models from the same provider are no more similar to one another than those from different providers, possibly because model architectures, post-training methods and pretraining corpora have largely converged \citep{wu2025automatically,wolfram2025layers}.

Throughout the analysis, we used three dimensionality-reduction methods (PCA, t-SNE and UMAP), because variance spread across all 768 dimensions would otherwise obscure any statistically meaningful differences. We used three methods as each preserves a different aspect of high-dimensional embedding geometry. PCA is a linear, deterministic projection that retains the directions of maximal global variance; t-SNE is nonlinear and emphasizes local neighborhood structure at the cost of distorting global distances; and UMAP is also nonlinear but better balances local and global structure. They are arguably the three most widely used dimensionality reduction methods, and results that hold across them are therefore unlikely to reflect the assumptions of any single projection. For t-SNE, we first reduced the embeddings to 50 dimensions with PCA, following the standard practice \citep{van2008visualizing}. This preprocessing step serves two purposes. First, it suppresses noise carried by low-variance directions, which would otherwise distort the pairwise distances from which t-SNE computes its neighborhood affinities. Second, it mitigates distance concentration in very high-dimensional spaces, which makes nearest-neighbor structure less reliable. The subsequent t-SNE projection therefore reflects the dominant structure of the embedding space rather than residual noise.

We also recomputed human--AI idea similarity in the regression of willingness to pursue as (1) the cosine similarity between SPECTER2 embeddings and (2) the cosine similarity between SPECTER2 displacement vectors. All conclusions are unchanged (SI Table~\ref{tab:ols_pursue}).

\subsection{Pretraining coverage of field inputs}
\label{sec:si_dolma}

One natural explanation for the weaker performance of LLMs in the social sciences is that social-science content is underrepresented in pretraining data. If so, models would have less exposure to the vocabulary, constructs, and argumentative conventions of these fields. To test this, we examined how well the context-and-puzzle inputs used for hypothesis generation are covered by Dolma-v1.6-sample, a publicly available sample of approximately 10 billion tokens from the Dolma pretraining corpus \citep{soldaini2024dolma}. We randomly sampled 100 papers from each of the four fields in our study. For $n$-grams of length 3, 4, and 5, we computed two measures:

\textbf{hit\_n:} the proportion of $n$-grams in the input (context and puzzle) that occur at least once in the Dolma sample.

\textbf{logcount\_n:} the mean of $\log_{10}(\mathrm{count}+1)$ across all $n$-grams in the input. Unmatched $n$-grams contribute zero.

Higher values indicate greater lexical coverage and higher occurrence frequency, respectively. SI Table \ref{tab:dolma} reports the field-level means of these paper-level measures.

Social-science inputs score higher than biology and chemistry on all six measures. They rank second only to medicine on five measures and first on trigram log frequency. These results provide no evidence that social-science task inputs are less well covered by this corpus. If anything, the language of social-science inputs is more familiar to the corpus than that of biology or chemistry inputs. This measure is certainly confounded by how frequently these n-grams occur in everyday use, a point we discuss in hypothesis probability below. The underrepresentation explanation therefore does not appear to account for the performance gap \emph{at the level of lexical coverage in this corpus}, and models share pretraining corpora \citep{wu2025mapping, wolfram2025layers}.

Note that these overlaps differ from the 10-gram overlap analysis in Supplementary Information \ref{subsec:data_collection}, which we conducted to study data leakage. There, we extracted 10-gram (10-word) sequences with a two-word overlap between consecutive sequences of a paper. We then scanned Dolma and marked a chunk (paragraph) in the paper as matched if more than 50\% of its 10-grams were found.

We propose that a more plausible explanation lies in the epistemic structure of the social sciences. These fields are theoretically pluralistic: the same phenomenon can often be explained by several competing frameworks, such as rational-choice, institutional, cultural, or psychological accounts. Which explanation applies often depends on context, including the population, the historical period, the institutional setting, and the measurement choices. Theories also change over time, partly because the people being studied respond to the categories and findings used to describe them \citep{hacking1995looping}. As a result, a given research puzzle rarely has a single canonical answer. The hypothesis a particular author proposes reflects a specific theoretical commitment and interpretive stance, and other experts might reasonably frame it differently.

\textbf{Social scientists disagree more often.} Bibliometric evidence is consistent with this picture. Analyzing more than four million scientific articles, Lamers and colleagues \citep{lamers2021disagreement} found that expressed disagreement in citation sentences was highest in the social sciences and humanities among the five broad fields they examined, with substantial variation across subfields. In our setting, this pluralism has a direct consequence. A generated hypothesis may be coherent, well grounded, and defensible within one theoretical tradition, yet still fail to match the original authors' hypothesis or to satisfy an expert evaluator who works within another tradition. Lower ratings in the social sciences may therefore partly reflect weaker convergence between plausible answers and the single designated reference answer, instead of a weaker ability to reason about social phenomena. Our own data likewise show that fields weight quality dimensions differently: biology and medicine place significantly more weight on \textit{probability} than the social sciences do (Figure~\ref{fig:judgment}h; Extended Data Table~\ref{tab:ols_results}).

Mohseni, Herrmann, and Orona \citep{mohseni2026hard} offer an alternative account: social science may appear harder to predict because people already intuit much of human behavior, so only hypotheses that go beyond common sense get published. On this account, LLMs that absorb folk psychology from human text should produce hypotheses that are obvious but plausible, and therefore high in probability. \textbf{Our data contradict this prediction}. Social-science hypotheses were rated lowest of all fields on probability, as well as on novelty and feasibility (main text Figure \ref{fig:judgment}i--k). The problem is therefore not obviousness; the pattern is more consistent with the absence of a stable consensus for models to reproduce in pluralistic fields.

Taken together, theoretical pluralism and disagreement offer plausible structural explanations for the lower ratings in social science. Our lexical-coverage check does not support simple underrepresentation in the sampled corpus as a sufficient explanation, but it cannot identify the cause of the rating gap. Nor does the folk-knowledge account explain the low probability ratings. Compared with willingness to pursue, ratings also seem less susceptible to common sources of bias such as seniority. Competing theoretical perspectives may instead reduce agreement on what counts as the right hypothesis.

\subsection{Automated judges rating human- and AI-generated hypotheses}
\label{sec:si_judgehuman}

In the main LLM-as-a-judge experiment (main text Figure \ref{fig:autoeval}), the judge receives the scientist's own idea as a persona, so it cannot rate that idea itself. We therefore sampled 5{,}000 contexts and puzzles and had the judge, given the context and puzzle alone, rate the human-generated and AI-generated hypotheses separately with the same rubric. On feasibility and probability, human hypotheses scored only 2--4\% higher than AI hypotheses, a difference that is neither substantial nor statistically significant. On novelty, AI hypotheses scored substantially higher: 7.302 vs.\ 6.195 under DeepSeek-R1 (18\% higher) and 6.671 vs.\ 4.361 under Gemini 2.5 Flash (53\% higher), $p<0.001$ in both cases. These scores document a tendency of the automated judges; they do not establish that AI hypotheses are more novel than human ones. Complementary evidence from a human study is reported in ref.~\citep{si2025ideas}.

\subsection{Preprint representativeness and publication matching}
\label{sec:si_preprint}

By title matching, 47\% of the biology, 41\% of the medicine, 37\% of the chemistry and 17\% of the social-science preprints in our corpus posted between 2023 and 2025 have already been published in peer-reviewed journals. These figures are right-censored, because many 2025 preprints have not completed peer review, and are likely underestimates, because titles often change during revision. In the other direction, among 100{,}000 randomly sampled published papers in OpenAlex, a traceable preprint or working-paper version exists for 25\% in economics (including working papers whose link could be retrieved by LLM agents), 13\% in biology, 10\% in psychology, 8\% in chemistry and 7\% in medicine, again likely underestimates. A large-scale linkage of about 105 million publications \citep{rzayeva2025preprinting} reports, for 2021--2022, about 9\% in the biological sciences, 4--5\% in the medical and health sciences, 5--6\% in the chemical sciences, 7\% in psychology and 4\% in economics (excluding working papers). Our estimates are somewhat higher, which we attribute to two factors. First, our sample covers more recent years, during which preprints have become increasingly widespread. Second, our sample is restricted to research articles, whereas their denominator also includes publication types that are rarely or never preprinted; case reports and narrative reviews, for example, cannot be posted on medRxiv. Our proportions are therefore naturally higher.

\subsection{Cross-field comparability of ratings}
\label{sec:si_comparability}

A common rubric and comprehension test do not by themselves establish identical evaluative standards across disciplines. Our main analyses, however, do not rely on cross-field comparability of raw scores: the regression includes field indicators and decomposes each rating into within- and between-scientist components, and author-fixed-effect and ordered-logit specifications yield the same within-author associations (SI Table~\ref{tab:adoption_combined}). Reward-model training and evaluation use only comparisons among hypotheses rated by the same scientist, which are invariant to any strictly increasing transformation of that scientist's scale, and GraphEval labels are defined by within-field percentiles. Ratings on the three quality dimensions are likely more objective than willingness to pursue, as they are insensitive to common sources of bias such as seniority (Extended Data Figure~\ref{fig:percep}).
\end{revblock}

\newpage
\section{Survey and Prompts}
\label{sec:supplement}

\subsection{Prompts}
\label{prompts}

\begin{tcolorbox}[colback=white,colframe=black,title=Hypothesis Extraction Prompt,breakable]
\ttfamily
You are reading a paper. Your task is to extract **relevant, explicit scientific hypotheses** proposed by the authors.

Extraction Instructions:

1. Present the hypotheses as a **numbered list**, with **one hypothesis per line**.

2. **Exclude** background/factual claims, method descriptions, presumptions, mathematical prerequisites/preconditions, and narrow inferences/interpretations drawn from tables and figures.

3. **Apply strict selection criteria - do not over-generate**. Extract only those hypotheses that the authors explicitly motivate and place at the core of the paper's main argument (typically introduced early, e.g., in the Introduction or thereafter). Omit minor and peripheral hypotheses confined to specific method, experiment, or result subsections.

4. **Exclude vague directional statements** such as "xxx is a valuable model for studying xxx"; "The proposed model provides a promising and crucial direction for xxx"; and "This technology can be applied to improve xxx," as these are summaries of the paper's overarching narrative rather than hypotheses.

5. Extract hypotheses based strictly on the **raw content**. Do not generate or infer hypotheses on your own. Preserve the **original meaning** of the authors' hypotheses.

6. If no relevant, explicit hypotheses are found, output an empty string "".

Strictly format your output as (with nothing else):

1. <Hypothesis>

2. ...

----- PAPER STARTS -----

f\{text\}

----- PAPER ENDS -----
\end{tcolorbox}

\begin{tcolorbox}[colback=white,colframe=black,title=Context/Puzzle Extraction Prompt,breakable]
\ttfamily
You are a helpful research assistant. You will be given the introduction of a scientific paper. Your task is to identify and extract two kinds of structured information from the text:

1. The **broad scientific context** of the work:

**Context** consists of only *factual*, *non-speculative, non-reasoning* statements. These include big pictures and related works, etc. This should read like what a researcher might see *before* they propose a theory.

You extract hypothesis-agnostic explanations or definitions of key terms and concepts that appear in the material.

These should help a reader understand the **context**, and should not overlap with the **reasoning or hypotheses**.

Think of this as building a small supporting knowledge base. All contextual snippets must be grounded in the text, though you may rewrite or clarify them for precision and readability.

2. At the **end of each context**, add a sentence that explicitly states *the core question or puzzle* that the paper addresses and can be derived from the context.

Focus on the **high-level picture** of the work.

**Avoid**:

  - Specific hypotheses or findings

  - Technical/experimental details, methods, or datasets

  - Any mention of author-proposed solutions

Your output should take the form of **describing the context then proposing a question.**

Below are some illustrative examples of the puzzle:

Bad: Integration of hypergraph-based models that represent RNA-peptide relationships will significantly improve the prediction of ncPEPs in diverse cancer subtypes compared to existing deep learning approaches.

Good: What can we do to improve deep learning models so they can accurately and robustly predict cancer-associated noncoding peptides (ncPEPs)?

Bad: Declining androgen levels during aging reduce suppression of gastric ILC2s, increasing cytokine production and susceptibility to gastric disease.

Good: How do androgen levels influence the development of gastric diseases?

Bad: Multi-modal sensor fusion improves real-time activity recognition in smart homes by resolving ambiguities in unimodal data and merging them using a deep LSTM model trained on the CASAS dataset.

Good: How to best integrate information from different individual sensory inputs to recognize real-time activity in smart homes?

Now, based on the given introduction, write the human-hypothesis-agnostic context and puzzle:

Introduction: \{text\}

Human hypotheses: \{text\}
\end{tcolorbox}

\begin{tcolorbox}[colback=white,colframe=black,title=Hypothesis Generation Prompt,breakable]
\ttfamily
You are an experienced scientist. You are writing a paper about the following scientific puzzle.
Your task is to propose **\{num\_hypotheses\} relevant, publishable, novel, and feasible** hypotheses that have not appeared in the literature to investigate the puzzle.

Instructions:

1. Present the hypotheses as a **numbered list**, with **one hypothesis per line**.

2. **Be concise**: exclude **background/factual claims or methodological/experimental descriptions**.

3. Keep the hypotheses **relevant** to the core of the puzzle.

4. **Do not generate vague directional statements** such as "xxx is a valuable model for studying xxx"; "The proposed model provides a promising and crucial direction for xxx"; and "This technology can be applied to improve xxx", as these are summaries of the paper's overarching narrative rather than hypotheses.

5. **Be creative** - reach beyond your existing knowledge base to propose **untested** ideas.

6. Ensure that the generated hypotheses are **clear, specified, well-reasoned, valid, and actionable**.

7. Ensure that **exactly \{num\_hypotheses\} hypotheses** are proposed.

----- CONTEXTUAL PUZZLE STARTS -----

f\{text\}

----- CONTEXTUAL PUZZLE ENDS -----

Strictly format your output as (with nothing else):

1. <Hypothesis>

2. <Hypothesis>

3. ...

Now, your output of the hypotheses:
\end{tcolorbox}

\begin{tcolorbox}[colback=white,colframe=black,title=Model Evaluation Prompt,breakable]
\ttfamily
You are an experienced scientist who is judging ideas (hypotheses) proposed from the same context and puzzle as your own paper.

**Given the context and puzzle of your own paper**, you will judge the novelty, empirical feasibility, and likelihood of being true of the given hypotheses.

A hypothesis here can be a testable idea, or more broadly, a guiding assumption in research methods, design, and direction.

Please use the following definitions:

<Novelty>: To what extent do the generated hypotheses - based on the given context - generalize and introduce new, unseen, and interesting ideas?

Given the context and puzzle: "Students in urban high schools consistently underperform on standardized science exams compared to their suburban counterparts. While factors such as <funding>, <class size>, and <teacher experience> have been well studied, they do not fully explain the persistent performance gap. The puzzle: Why do students in urban high schools continue to underperform in science despite comparable teacher credentials and similar classroom resources?"

Hypothesis A: Urban students' lower science achievement may stem from prolonged exposure to high-information-density environments in urban areas, which trains the brain to favor a "fast-response cognitive mode" while weakening the "slow-reasoning circuits" essential for scientific thinking.

Hypothesis B: Larger class sizes and the associated lower individual attention from teachers lead to decreased academic performance in science subjects among students in urban areas, compared to their peers in suburban areas.

Hypothesis A is thus MORE NOVEL than hypothesis B, as class size has been studied in the provided context.

<Empirical feasibility>: How empirically feasible are the hypotheses? Relations between variables in an empirically feasible hypothesis should be operationalizable, measurable, implementable, and testable.

"Even in a society where all individuals share a unified collective consciousness, social conflict would still persist" thus is LESS FEASIBLE than "More English listening practices could improve the English writing scores among high-school students".

<Likelihood of being true>: How believable the hypotheses sound, based on existing knowledge or internal logic? A probable (likely true) hypothesis should be intuitive.

"Regular physical exercise reduces the risk of cardiovascular disease in adults" thus is MORE PROBABLE than "Drinking two liters of soda per day enhances memory retention in older adults".

Please do not evaluate format, writing style or grammar except where it prevents understanding.

Please rate the novelty on a scale from **1 to 9**.

9: Highly novel: The hypothesis generalizes and stands out significantly from the given context following an unconventional reasoning path. If empirically validated, it could unlock a range of new, impactful, and interesting implications.

5: Moderately novel: The hypothesis introduces some new ideas, but they are expected given the context.

1: Not novel: The hypothesis is essentially a rephrasing of the content in the provided context, with no meaningful innovation.

Please rate the likelihood of being true on a scale from **1 to 9**.

9: Highly probable: The hypothesis makes sense. It is highly likely to be true, even without empirical validation.

5: Moderately probable: The hypothesis has an equal chance of being true or false; it requires experimental validation.

1: Not probable: The hypothesis is incoherent, logically flawed, and very likely to be false.

Please rate the empirical feasibility on a scale from **1 to 9**.

9: Highly feasible: The hypothesis clearly states relationships between variables, including null or negative relations (e.g., A has no relationship with B). The variables are operationalizable, and I can readily envision experiments to test the hypothesis.

5: Moderately feasible: The hypothesis refers to some identifiable variables and implies a general relationship, but the variables lack clear definitions or operationalization. I can roughly outline a direction for empirical tests, though the experimental design would need refinement of details.

1: Not feasible: The hypothesis is infeasible. The variables are neither measurable nor operationalizable, and they lack clarity. This makes it difficult to specify their relation empirically or design a testable experiment.

**Return ONLY a JSON object in the following format**:

\{\{

  novelty: <int, 1 to 9>,

  feasibility: <int, 1 to 9>,

  likelihood: <int, 1 to 9>

\}\}

CONTEXT AND PUZZLE:

\{context\_puzzle\}

<SCIENTIST PERSONA>: Your own perspective on this context and puzzle can be summarized as the following hypotheses:

\{idea\} (not always provided to LLMs)

HYPOTHESIS TO BE JUDGED:

\{hypothesis\}
\end{tcolorbox}

\begin{tcolorbox}[colback=white,colframe=black,title=Model Evaluation with Explicit Requirement of Extreme Scores,breakable]
\ttfamily
<Same as the model evaluation prompt>

**Assign an explicit low (1) or high (9) rating when a quality dimension is clearly poor or excellent.**
\end{tcolorbox}

\begin{tcolorbox}[colback=white,colframe=black,title=Model Evaluation with Direct Comparison,breakable]
\ttfamily
\textbf{System Prompt:}\\
You are an experienced scientist who is judging ideas (hypotheses) proposed from the same context and puzzle as your own paper.

\vspace{1em}
\textbf{User Prompt:}

You will be given:
\begin{enumerate}
    \item the \textbf{context}: background information and the research puzzle of the paper
    \item two proposed hypotheses: \textbf{Hypothesis A} and \textbf{Hypothesis B} proposed based on the given context
\end{enumerate}

\vspace{0.5em}
Your task:\\
Compare the two hypotheses on novelty according to the following criteria:\\
\textit{``To what extent do the generated hypotheses - based on the given context - generalize and introduce new, unseen, and interesting ideas?''}

[Replace novelty with other dimensions:]

Feasibility: \textit{``How empirically feasible are the hypotheses? Relations between variables in an empirically feasible hypothesis should be operationalizable, measurable, implementable, and testable.''}

Probability: \textit{``How believable the hypotheses sound, based on existing knowledge or internal logic? A probable (likely true) hypothesis should be intuitive.''}

\vspace{0.5em}
Context:\\
\{context\}

\vspace{0.5em}
Hypothesis A:\\
\{left\_resp\}

\vspace{0.5em}
Hypothesis B:\\
\{right\_resp\}

\vspace{0.5em}
Respond with ONLY one word: \textbf{``left''} if Hypothesis A is better, \textbf{``right''} if Hypothesis B is better.
\end{tcolorbox}

\begin{tcolorbox}[colback=white,colframe=black,title=Reward Model Training Prompt,breakable]
\ttfamily
criteria\_definitions = \{

    "probability": (

        "How believable the hypotheses sound, based on existing knowledge or internal logic? A probable (likely true) hypothesis should be intuitive."

    ),

    "novelty": (

        "To what extent do the generated hypotheses - based on the given context - generalize and introduce new, unseen, and interesting ideas?"

    ),

    "feasibility": (

        "How empirically feasible are the hypotheses? Relations between variables in an empirically feasible hypothesis should be operationalizable, measurable, implementable, and testable."

    )

\}

You are an experienced scientist evaluating a scientific hypothesis.

Criterion:
\{criterion\_definition\}

Context and puzzle:
\{context\_puzzle\}

Your own perspective on this context and puzzle can be summarized as the following hypotheses:

\{idea\}

Please evaluate the assistant's hypothesis with respect to this criterion:

\{hypotheses to be judged\}
\end{tcolorbox}

\begin{tcolorbox}[colback=white,colframe=black,title=Generation of Null Hypotheses,breakable]
\ttfamily
You are an experienced scientist.

null: Based on the following information, write one possible NULL hypothesis: A null hypothesis is a default assumption that there is no effect, no difference, or no relationship between variables. It is the hypothesis a study seeks to test or potentially reject.

not\_null: Based on the following information, write one possible ALTERNATIVE (non\_null) hypothesis: An alternative hypothesis proposes that there is an effect, a difference, or a relationship between variables. It is what researchers hope to support through evidence.
\end{tcolorbox}

\begin{tcolorbox}[colback=white,colframe=black,title=Robustness Check for the Detection of AI Engagement,breakable]
\ttfamily
You are an experienced scientist.

Does the following paper involve AI and machine learning?
Return **ONLY 1 (yes) or 0 (no)**.

Text:
\{text\}
\end{tcolorbox}

\begin{tcolorbox}[colback=white,colframe=black,title=Extracting Key Hypotheses from Published Abstracts,breakable]
\ttfamily
You are an experienced scientist. You will be given the abstract of a published scientific paper. Your task is to determine whether the **key hypotheses** proposed by the authors can be recovered from the text.

**Hypotheses**: Extract only the **explicit scientific hypotheses** that the authors place at the core of the paper's main argument. Each hypothesis should be a self-contained, testable statement. Do not include background facts, methodological details, or secondary findings.

If you can recover at least one explicit hypothesis, format your output strictly as follows (with nothing else):\\
Hypotheses:\\
1. <Hypothesis>\\
2. <Hypothesis>\\
...

If you cannot (e.g., the text contains no explicit hypotheses), return:\\
Hypotheses: ""

Text:\\
\{text\}
\end{tcolorbox}

\begin{tcolorbox}[colback=white,colframe=black,title=Recovering Context and Research Puzzle Given Extracted Hypotheses and Published Abstracts,breakable]
\ttfamily
You are an experienced scientist. You will be given the abstract of a published scientific paper, together with the **key hypotheses** that have already been extracted from it. Your task is to determine whether the **broad scientific context** and the **core research puzzle** can be recovered from the text **without leaking any of the given hypotheses**.

1. **Context**: The context should contain only *factual*, *non-speculative* statements without reasoning, such as the big picture of the field. It should read like what a researcher would know *before* proposing a theory.

2. **Puzzle**: At the **end of the context**, add one sentence that explicitly states *the core question or puzzle* the paper addresses, which should follow from the context. Focus on the **high-level picture** of the work.

**Avoid hypothesis leakage**: The given hypotheses are provided **only so that you know what to exclude**. The context and puzzle must be **hypothesis-agnostic**. They must not state, paraphrase, imply, or hint at any of the given hypotheses, the paper's findings, or the authors' proposed solutions. This includes the specific mechanisms, variables, relationships, or directions of effect named in the hypotheses. A reader who sees only the context and puzzle should not be able to infer any of the hypotheses.

Before finalizing, check every sentence of the context and the puzzle against each given hypothesis. Remove or rewrite any sentence that overlaps with a hypothesis.

If you can recover both components without leakage, format your output strictly as follows (with nothing else):
\begin{verbatim}
Context and puzzle: <Context>. <Puzzle>
\end{verbatim}

If you cannot (e.g., the context and puzzle cannot be separated from the hypotheses), return:\\
Context and puzzle: ""

Text:\\
\{text\}

Hypotheses:\\
\{hypotheses\}
\end{tcolorbox}

\subsection{The pilot study, the recruitment email, and the survey}
\label{survey}

\begin{figure}[p]
\phantomsection
\addcontentsline{toc}{subsection}{SI Figure~2: The recruitment email}
    \centering
    \includegraphics[width=1\textwidth]{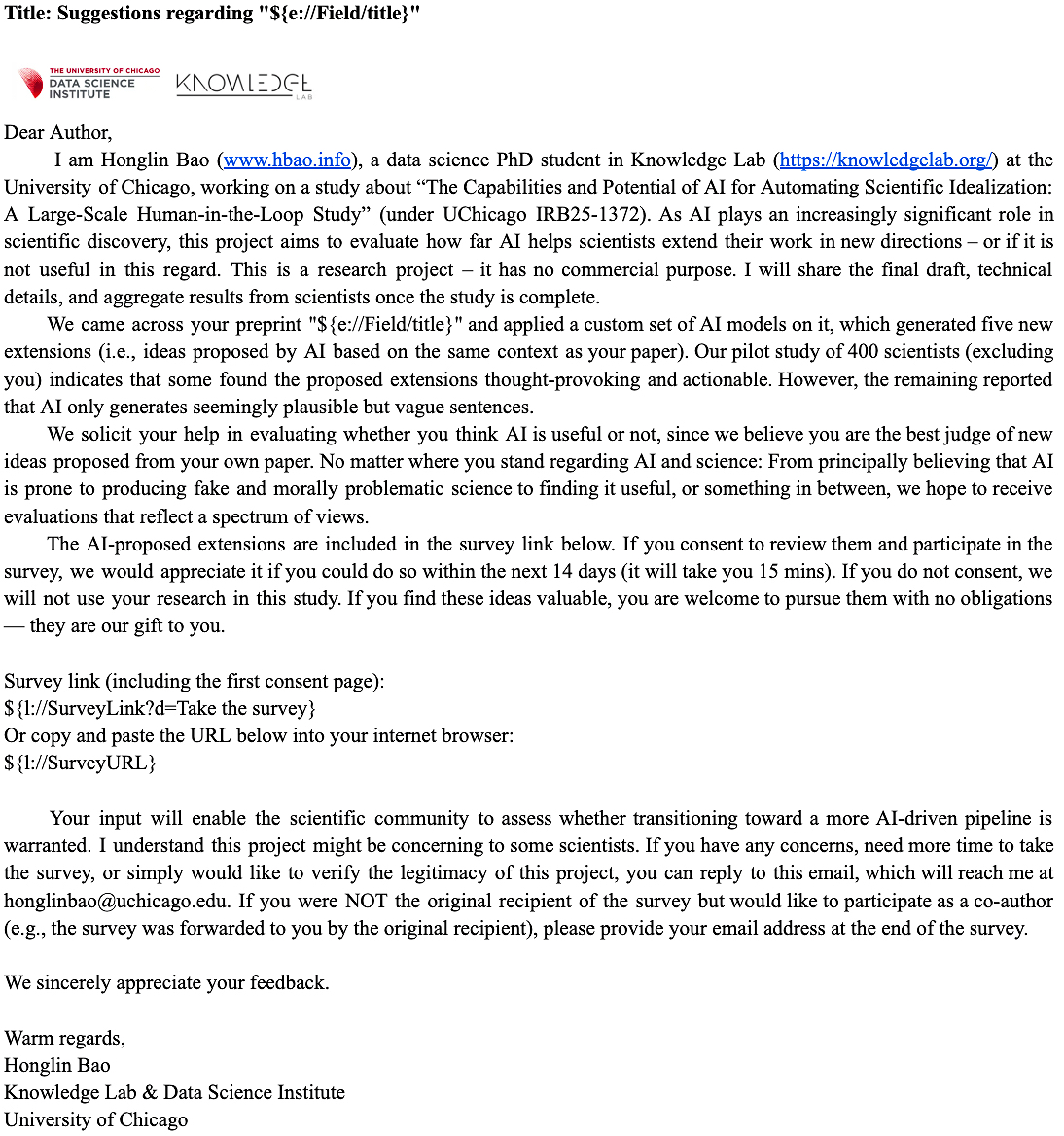}
    \caption{\textbf{The recruitment email.}}
    \label{fig:email}
\end{figure}

We administered the survey on Qualtrics, distributing the first pilot study to a random sample of 400 recipients. Among them, 241 reported being satisfied or at least content with the AI-generated ideas, indicating a generally positive attitude. The remaining respondents expressed skepticism, describing the ideas as superficially plausible yet lacking substantive coherence. The recruitment email (SI Figure \ref{fig:email})\footnote{This is the final version. We revised it from slightly different earlier versions of the email after observing that some scientists strongly disliked AI and responded with inappropriate messages, while some were highly enthusiastic about AI and sent messages offering praise and seeking further collaboration with us---something we had not anticipated during the pilot study. This is actually exactly our intention: to cover a wide spectrum of positions.} and the survey are provided below.

\vspace{1em}
\textbf{The survey}:

Thank you for participating! We are a group of researchers at the Knowledge Lab and Data Science Institute, University of Chicago. We are studying whether AI can help scientists extend their work in new directions, and especially, propose novel, empirically feasible, and likely true scientific hypotheses.

In this survey, you will see a few AI-generated hypotheses that were developed based on your own preprint ``\verb|${e://Field/title}|''. Each will come with a short context (e.g., definitions of concepts and the scientific puzzle discussed in your paper). We greatly value your expert feedback on them---particularly regarding their novelty, empirical feasibility, and likelihood of being true. Your input will enable the entire scientific community to assess whether transitioning toward a more AI-driven pipeline is warranted.

Expected time: around 15 minutes. You will evaluate 5 hypotheses, each judged across three dimensions, and respond to several related questions.

\begin{itemize}[leftmargin=1.5em]
    \item You may withdraw at any time by closing the survey.
    \item Participation is voluntary and has no impact on your publication record.
    \item Your participation in this study does not involve any risk to you beyond that of everyday life. However, you may feel mild frustration or discomfort if AI-generated hypotheses misrepresent your work or seem poor in quality. Please note that some AI-generated hypotheses may be intentionally weak, as these serve as valuable data points for evaluation purposes.
    \item Your ratings and the fine-tuned machine learning models derived from them will be shared openly with the research community after proper de-identification. We wish to assure you that this project is intended solely for scientific research, with no commercial objectives.
    \item If you would like the final draft, technical details, aggregate results, and AI-generated hypotheses later, you can opt in at the end.
\end{itemize}

This project is conducted under the University of Chicago IRB25-1372, titled ``The Capabilities and Potential of AI for Automating Scientific Idealization: A Large-Scale Human-in-the-Loop Study''. If you have any questions about your rights as a participant in this research, feel you have been harmed, or wish to discuss other study-related concerns with someone who is not part of the research team, you can contact the University of Chicago Social \& Behavioral Sciences Institutional Review Board (IRB) Office by phone at (773) 702-2915, or by email at \href{mailto:sbs-irb@uchicago.edu}{sbs-irb@uchicago.edu}.

\textbf{Project Lead}\\
Honglin Bao: PhD student at UChicago Data Science Institute.\\
\href{mailto:honglinbao@uchicago.edu}{honglinbao@uchicago.edu}\\
James A. Evans: Max Palevsky Professor at UChicago Data Science Institute and Sociology.\\
\href{mailto:jevans@uchicago.edu}{jevans@uchicago.edu}

\vspace{1em}
\begin{itemize}[label={}, leftmargin=1em]
    \item \radio{I have read the information above and agree to participate.}
    \item \radio{I do not agree to participate.}
\end{itemize}

\vspace{1em}
\noindent\rule{\linewidth}{0.4pt}

\textbf{Are you an author of the following paper?}

\verb|${e://Field/title}|

\begin{itemize}[label={}, leftmargin=1em]
   \item \radio{Yes}
   \item \radio{No}
\end{itemize}

(If the answer is no, the survey will automatically end)
\vspace{1em}

\textbf{How familiar are you with this paper's content?}

\begin{itemize}[label={}, leftmargin=1em]
   \item \radio{$1$ = I almost forgot the paper's content}
   \item \radio{$2$ = I can only remember part of the content}
   \item \radio{$3$ = I remember the content, but I am not quite confident about my judgment}
   \item \radio{$4$ = I was deeply involved in this research. I remember this piece. I am confident about my judgment}
   \item \radio{$5$ = I supervised/led this research. I remember this piece very well. I am very confident about my judgment}
\end{itemize}

(We only used ratings from authors with a confidence score above (and including) 3)

\noindent\rule{\linewidth}{0.4pt}

\vspace{1em}
\textbf{Context review}

Before you evaluate the AI-generated scientific hypotheses, we would like to know to what extent the context, the puzzle, and the extracted hypotheses reflect your own paper.

Please review the information below and answer the questions.

\textbf{Paper:}\\
\verb|${e://Field/title}|

\textbf{Context and puzzle:}\\
\verb|${e://Field/context_puzzle}|

\textbf{Extracted human hypotheses:}\\
\verb|${e://Field/human}|

\vspace{1em}
\textbf{How well do you feel the extracted context and puzzle represent what you actually did in your paper, e.g., the basic concepts and topics you explored?}

\begin{itemize}[label={}, leftmargin=1em]
   \item \radio{Very good. They are indeed what I did.}
   \item \radio{Good. They are generally fit.}
   \item \radio{They are OK with some deviations from what I did.}
   \item \radio{They are hallucinations.}
\end{itemize}

(Only 0.3\% of scientists chose the fourth option ``hallucinations'')

\textbf{How well do you think the extracted hypotheses capture the hypotheses, explicitly or implicitly, included in your own paper?}

\begin{itemize}[label={}, leftmargin=1em]
   \item \radio{Very good. They are indeed what I did.}
   \item \radio{Good. They are generally fit.}
   \item \radio{They are OK with some deviations from what I did.}
   \item \radio{They are hallucinations.}
\end{itemize}

(Only 1.4\% of scientists chose the fourth option ``hallucinations'')

\vspace{1em}
\noindent\rule{\linewidth}{0.4pt}

\textbf{Instructions}

In this study, you will judge the novelty, empirical feasibility, and probability (likelihood of being true) of the given hypotheses in relation to your paper.

A hypothesis here can be a testable idea, or more broadly, a guiding assumption in research methods, design, and direction. You will see a set of hypotheses that may be rephrased directly from your own paper or proposed by AI based on the content of your paper.

\textbf{Please use the following definitions:}

\begin{itemize}[leftmargin=1.5em]
   \item \textbf{Novelty}: To what extent do the generated hypotheses---based on the given context---generalize and introduce new, unseen, and interesting ideas?
   \item \textbf{Empirical feasibility}: How empirically feasible are the hypotheses? Relations between variables in an empirically feasible hypothesis should be operationalizable, measurable, implementable, and testable.
   \item \textbf{Probability (likelihood of being true)}: How believable the hypotheses sound, based on existing knowledge or internal logic? A probable (likely true) hypothesis should be intuitive.
\end{itemize}

Please do not evaluate format, writing style or grammar except where it prevents understanding.

To familiarize you with the evaluation criteria, please answer the following practice questions.

\vspace{1em}
\noindent\rule{\linewidth}{0.4pt}

\textbf{Practice questions}

\paragraph{Which hypothesis is more \textbf{probable}, i.e., likely to be true?}

\begin{itemize}[label={}, leftmargin=1em]
   \item \radio{Regular physical exercise reduces the risk of cardiovascular disease in adults.}
   \item \radio{Drinking two liters of soda per day enhances memory retention in older adults.}
\end{itemize}

(The correct answer: the first one)

\paragraph{Which hypothesis is more \textbf{novel}?}

Context: Students in urban high schools consistently underperform on standardized science exams compared to their suburban counterparts. While factors such as \underline{funding}, \underline{class size}, and \underline{teacher experience} have been well studied, they do not fully explain the persistent performance gap.

The puzzle: Why do students in urban high schools continue to underperform in science despite comparable teacher credentials and similar classroom resources?

\begin{itemize}[label={}, leftmargin=1em]
   \item \radio{Urban students' lower science achievement may stem from prolonged exposure to high-information-density environments in urban areas, which trains the brain to favor a ``fast-response cognitive mode'' while weakening the ``slow-reasoning circuits'' essential for scientific thinking.}
   \item \radio{Larger class sizes and the associated lower individual attention from teachers lead to decreased academic performance in science subjects among students in urban areas, compared to their peers in suburban areas.}
\end{itemize}

(The correct answer: the first one, as class size has been well studied in the provided context)

\paragraph{Which hypothesis is more \textbf{empirically feasible and testable}?}

\begin{itemize}[label={}, leftmargin=1em]
   \item \radio{Even in a society where all individuals share a unified collective consciousness, social conflict would still persist.}
   \item \radio{More English listening practice could improve the English writing scores among high-school students.}
\end{itemize}

(The correct answer: the second one. The survey will not continue unless participants fill in all the right answers)

\vspace{1em}
\noindent\rule{\linewidth}{0.4pt}

\textbf{Hypothesis evaluation}

Given the context and puzzle, please evaluate the following hypothesis generated by AI. You will assess three dimensions independently, followed by an overall favorability rating of adoption (i.e., the extent to which the hypothesis is worth pursuing). AI models did not have access to your paper's conclusions, though they may still have proposed similar ideas.

\textbf{Hypothesis (5 in Total):}
\verb|${lm://Field/1 to 5}|

\vspace{1em}
\textbf{Please rate the \emph{novelty} on a scale from 1 to 9.}

\textit{Rating Guide:}
\begin{itemize}[leftmargin=1.5em]
   \item \textbf{9 -- Highly novel}: The hypothesis generalizes and stands out significantly from the given context following an unconventional reasoning path. If empirically validated, it could unlock a range of new, impactful, and interesting implications.
   \item \textbf{5 -- Moderately novel}: The hypothesis introduces some new ideas, but they are expected given the context.
   \item \textbf{1 -- Not novel}: The hypothesis is essentially a rephrasing of the content in the provided context, with no meaningful innovation.
\end{itemize}

\ratingscalenine{1 (Not novel)}{9 (Highly novel)}

\vspace{1em}
\textbf{Please rate the \emph{probability}, i.e., likelihood of being true, on a scale from 1 to 9.}

\textit{Rating Guide:}
\begin{itemize}[leftmargin=1.5em]
   \item \textbf{9 -- Highly probable}: The hypothesis makes sense. It is highly likely to be true, even without empirical validation.
   \item \textbf{5 -- Moderately probable}: The hypothesis has an equal chance of being true or false; it requires experimental validation.
   \item \textbf{1 -- Not probable}: The hypothesis is incoherent, logically flawed, and very likely to be false.
\end{itemize}

\ratingscalenine{1 (Not probable)}{9 (Highly probable)}

\vspace{1em}
\textbf{Please rate the \emph{empirical feasibility} on a scale from 1 to 9.}

\textit{Rating Guide:}
\begin{itemize}[leftmargin=1.5em]
   \item \textbf{9 -- Highly feasible}: The hypothesis clearly states relationships between variables, including null or negative relations (e.g., ``A has no relationship with B''). The variables are operationalizable, and I can readily envision experiments to test the hypothesis.
   \item \textbf{5 -- Moderately feasible}: The hypothesis refers to some identifiable variables and implies a general relationship, but the variables lack clear definitions or operationalization. I can roughly outline a direction for empirical tests, though the experimental design would need refinement of details.
   \item \textbf{1 -- Not feasible}: The hypothesis is infeasible. The variables are neither measurable nor operationalizable, and they lack clarity. This makes it difficult to specify their relation empirically or design a testable experiment.
\end{itemize}

\ratingscalenine{1 (Not feasible)}{9 (Highly feasible)}

\vspace{1em}
(The order of the quality-dimension questions and the hypotheses is randomized)
\vspace{1em}

\vspace{1em}
\textbf{After reviewing all five AI-generated hypotheses, how likely would you be to study the following one for a new research paper?} This could be your overall favorability of adoption and the judgment of the overall ``quality'' of hypotheses.

\verb|${lm://Field/1 to 5}| for five hypotheses

\ratingscaleseven{1 (Not at all)}{7 (Very much)}

(The order of the hypotheses is randomized)

\vspace{1em}
\noindent\rule{\linewidth}{0.4pt}
\vspace{1em}

\textbf{Follow-up}

If you are not the original recipient of this survey but still complete it (for example, if it was forwarded by your coauthor), please provide your email address below.

\begin{itemize}[label={}, leftmargin=1em]
   \item \checkbx{I am NOT the original recipient. This is my email: \underline{\hspace{6cm}}}
   \item \checkbx{I am the original recipient.}
\end{itemize}

We will follow up with you to share the aggregate results once the study is complete.

\begin{itemize}[label={}, leftmargin=1em]
   \item \checkbx{I am interested in the aggregate result.}
   \item \checkbx{I am not interested in the aggregate result.}
\end{itemize}

Any other comments or suggestions?

\underline{\hspace{6cm}}

\vspace{1em}
Thank you for your time and expertise. Your judgments will help us evaluate and improve automated scientific discovery systems. We appreciate your contribution!

\clearpage

\section{SI Tables and Figures}

\begin{table}[!ht]
\phantomsection
\addcontentsline{toc}{subsection}{SI Table~1: Pearson correlation of conditional perplexity across models}
\centering
\caption{\textbf{Pearson correlation of conditional perplexity across judge models}. This table shows the model perplexity of AI-generated ideas conditioned on the context and puzzle across 7B models. The perplexities these models produce are highly correlated with one another, but not with human judges' assessments of novelty.}
\label{cond_ppl}
\begin{tabular}{lcccc}
\toprule
 & Qwen2-7B & Mistral-7B & Llama2-7B & Deepseek-llm-7b-base \\
\midrule
Qwen2-7B             & 1.000000 & 0.869664 & 0.842063 & 0.919653 \\
Mistral-7B           & 0.869664 & 1.000000 & 0.936273 & 0.900076 \\
Llama2-7B            & 0.842063 & 0.936273 & 1.000000 & 0.882319 \\
Deepseek-llm-7b-base & 0.919653 & 0.900076 & 0.882319 & 1.000000 \\
\bottomrule
\end{tabular}
\end{table}

\clearpage
\begin{table}[ht]
\phantomsection
\addcontentsline{toc}{subsection}{SI Table~2: Simulation model parameters}
\centering
\caption{\textbf{Simulation model parameters.} \revise{Parameter values are illustrative. They encode the direction suggested by our findings (more null suppression with AI), not calibrated magnitudes.}}
\label{tab:params}
\begin{tabular}{ccl}
\toprule
Parameter & Value & Description\\
\midrule
$\omega$     & 1.0    & scale of the true-effect law $\theta_j\sim|\mathcal N(0,\omega^2)|$  \\
$m$          & 20{,}000 & size of the sampled truth space $\{\theta_j\}$ \\
seed         & 0      & random seed for drawing the truth space \\
$n_0$     & $3$    & baseline number of studies at AI use $a{=}0$\\
$n_1$     & $30$   & number of studies at $a{=}1$\\
$\rho_0$  & $0.01$ & error correlation for human research\\
$\rho_1$  & $0.20$ & error correlation under full AI automation\\
$s_0$     & \revise{$0.75$} & null-suppression rate for human research\\
$s_1$     & \revise{$0.85$} & null-suppression rate under full AI automation\\
$B_{h}$   & $0.10$ & blind-spot bias at $a$=1\\
$\sigma$  & 1.0  & noise scale \\
$\tau$    & 1.0  & field maturity factor (baseline; varied as [0.6, 1.2, 1.8]) \\
\bottomrule
\end{tabular}
\end{table}

\clearpage
\begin{table}[htbp]
\phantomsection
\addcontentsline{toc}{subsection}{SI Table~3: Willingness to pursue with author fixed-effects vs.\ ordered logit models}
\centering
\caption{\textbf{Willingness to pursue with author fixed-effects vs.\ ordered logit models}. The major conclusions are robust to Extended Data Table \ref{tab:ols_combined}.}
\label{tab:adoption_combined}
\begin{tabular}{lcc}
\toprule
 & (1) Author FE (OLS) & (2) Ordered Logit \\
\midrule
Novelty (within)                    & 0.121***  & 0.164***  \\
                                    & (0.006)   & (0.008)   \\
Novelty (between)                   & ---       & 0.291***  \\
                                    &           & (0.014)   \\
Feasibility (within)                & 0.129***  & 0.186***  \\
                                    & (0.006)   & (0.006)   \\
Feasibility (between)               & ---       & 0.111***  \\
                                    &           & (0.015)   \\
Probability (within)                & 0.313***  & 0.407***  \\
                                    & (0.006)   & (0.009)   \\
Probability (between)               & ---       & 0.579***  \\
                                    &           & (0.016)   \\
Human-AI idea similarity (within)   & 1.281***  & 1.649***  \\
                                    & (0.119)   & (0.154)   \\
Human-AI idea similarity (between)  & ---       & 0.508**   \\
                                    &           & (0.189)   \\
Seniority                           & ---       & -0.189**  \\
                                    &           & (0.062)   \\
Prior AI engagement (log)           & ---       & -0.212    \\
                                    &           & (0.463)   \\
Chemistry$^{a}$                     & ---       & 0.051     \\
                                    &           & (0.070)   \\
Medicine$^{a}$                      & ---       & 0.135**   \\
                                    &           & (0.052)   \\
Social science$^{a}$                & ---       & -0.152*** \\
                                    &           & (0.047)   \\
Constant                            & 0.000     & ---       \\
                                    & (0.000)   &           \\
\midrule
Observations    & 23,855   & 23,615  \\
$R^2$           & 0.304    & ---     \\
Adj. $R^2$      & 0.304    & ---     \\
Log-Likelihood  & -37,039  & -37,077 \\
AIC             & 74,090   & 74,190  \\
BIC             & 74,130   & 74,350  \\
\bottomrule
\multicolumn{3}{l}{\footnotesize Standard errors in parentheses. *** p<0.001, ** p<0.01, * p<0.05.} \\
\multicolumn{3}{l}{\footnotesize $^{a}$ Biology is the baseline field.} \\
\multicolumn{3}{l}{\footnotesize Column (1) uses the author-fixed effect model, so only within-author} \\
\multicolumn{3}{l}{\footnotesize variation is identified; between-author and time-invariant controls drop out.} \\
\end{tabular}
\end{table}

\clearpage
\begin{table}[htbp]
\phantomsection
\addcontentsline{toc}{subsection}{SI Table~4: Regression results with all tested interactions between seniority, field, and quality dimensions}
\centering
\caption{\textbf{Regression results with all tested interactions between seniority, field, and quality dimensions}. Conclusions of Extended Data Tables \ref{tab:ols_results34} and \ref{tab:ols_results} are robust. *** $p<0.001$, ** $p<0.01$, * $p<0.05$. Field-level comparisons use social science as the baseline.}
\label{tab:ols_results_m2}
\scriptsize\renewcommand{\arraystretch}{0.85}
\begin{tabular}{lc}
\toprule
 & Willingness to pursue \\
\midrule
Intercept & -0.8070*** \\
 & (0.126) \\
\midrule
Biology & -0.0983 \\
 & (0.076) \\
Chemistry & -0.0545 \\
 & (0.125) \\
Medicine & 0.0340 \\
 & (0.101) \\
\midrule
Novelty (within) & 0.1338*** \\
 & (0.014) \\
Novelty (within) $\times$ Biology & -0.0143 \\
 & (0.017) \\
Novelty (within) $\times$ Chemistry & -0.0142 \\
 & (0.027) \\
Novelty (within) $\times$ Medicine & -0.0084 \\
 & (0.021) \\
Novelty (between) & 0.2178*** \\
 & (0.010) \\
\midrule
Feasibility (within) & 0.1316*** \\
 & (0.012) \\
Feasibility (within) $\times$ Biology & -0.0036 \\
 & (0.015) \\
Feasibility (within) $\times$ Chemistry & -0.0183 \\
 & (0.025) \\
Feasibility (within) $\times$ Medicine & 0.0094 \\
 & (0.021) \\
Feasibility (between) & 0.0742*** \\
 & (0.012) \\
\midrule
Probability (within) & 0.2692*** \\
 & (0.013) \\
Probability (within) $\times$ Biology & 0.0584*** \\
 & (0.016) \\
Probability (within) $\times$ Chemistry & 0.0208 \\
 & (0.025) \\
Probability (within) $\times$ Medicine & 0.0632** \\
 & (0.022) \\
Probability (between) & 0.4457*** \\
 & (0.012) \\
\midrule
Human-AI Idea Similarity (within) & 1.1298*** \\
 & (0.232) \\
Human-AI Idea Similarity (within) $\times$ Biology & 0.1499 \\
 & (0.286) \\
Human-AI Idea Similarity (within) $\times$ Chemistry & -0.6233 \\
 & (0.565) \\
Human-AI Idea Similarity (within) $\times$ Medicine & 0.5567 \\
 & (0.377) \\
Human-AI Idea Similarity (between) & 0.4927*** \\
 & (0.148) \\
\midrule
Seniority & -0.5073*** \\
 & (0.112) \\
Seniority $\times$ Biology & 0.4451*** \\
 & (0.129) \\
Seniority $\times$ Chemistry & 0.4364* \\
 & (0.208) \\
Seniority $\times$ Medicine & 0.3919* \\
 & (0.174) \\
\midrule
Prior AI Engagement (log) & -0.1605 \\
 & (0.382) \\
\midrule
Observations & 23,615 \\
$R^2$ & 0.397 \\
\bottomrule
\end{tabular}
\end{table}

\clearpage
\begin{table}[!h]
\phantomsection
\addcontentsline{toc}{subsection}{SI Table~5: Nonresponse-adjusted estimates of willingness to pursue}
\centering
\caption{\textbf{Nonresponse-adjusted estimates of willingness to pursue}. Major conclusions in the main text are robust.}
\label{tab:selection}
\begin{tabular}{lcc}
\toprule
 & (1) & (2) \\
 & Unweighted & Inverse probability weighted \\
\midrule
Novelty (within)              & 0.123$^{***}$ & 0.122$^{***}$ \\
                              & (0.006) & (0.006) \\
Novelty (between)             & 0.219$^{***}$ & 0.218$^{***}$ \\
                              & (0.010) & (0.010) \\
Feasibility (within)          & 0.130$^{***}$ & 0.129$^{***}$ \\
                              & (0.006) & (0.006) \\
Feasibility (between)         & 0.073$^{***}$ & 0.073$^{***}$ \\
                              & (0.012) & (0.012) \\
Probability (within)          & 0.314$^{***}$ & 0.315$^{***}$ \\
                              & (0.006) & (0.006) \\
Probability (between)         & 0.447$^{***}$ & 0.448$^{***}$ \\
                              & (0.012) & (0.012) \\
Idea similarity (within)      & 1.284$^{***}$ & 1.272$^{***}$ \\
                              & (0.120) & (0.120) \\
Idea similarity (between)     & 0.473$^{***}$ & 0.493$^{***}$ \\
                              & (0.148) & (0.147) \\
Seniority                     & $-0.161$$^{***}$ & $-0.157$$^{***}$ \\
                              & (0.049) & (0.049) \\
Prior AI engagement           & $-0.170$ & $-0.232$ \\
                              & (0.382) & (0.384) \\
\midrule
Biology                       & 0.125$^{***}$ & 0.128$^{***}$ \\
                              & (0.037) & (0.037) \\
Chemistry                     & 0.164$^{**}$ & 0.164$^{**}$ \\
                              & (0.061) & (0.061) \\
Medicine                      & 0.230$^{***}$ & 0.233$^{***}$ \\
                              & (0.050) & (0.050) \\
Constant                      & $-0.972$$^{***}$ & $-0.987$$^{***}$ \\
                              & (0.117) & (0.117) \\
\midrule
Observations (ideas)          & 23{,}615 & 23{,}615 \\
\bottomrule
\end{tabular}
\begin{minipage}{0.78\textwidth}
\vspace{0.5em}
\footnotesize\emph{Notes:} Social science is the omitted field category (the reference). Continuous evaluation measures are decomposed into researcher-specific means (between) and deviations from those means (within). Standard errors are clustered in parentheses. Model 1 (the unweighted model) replicates the main result in the paper (Extended Data Table \ref{tab:ols_combined} Model 1). $^{***}p<0.001$, $^{**}p<0.01$, $^{*}p<0.05$.
\end{minipage}
\end{table}

\clearpage
\phantomsection
\addcontentsline{toc}{subsection}{SI Table~6: Equivalence tests for the effect of prior AI engagement on willingness to pursue}
\begin{longtable}{lccc}
\caption{\textbf{Equivalence tests for the effect of prior AI engagement on willingness to pursue}. The absence of an effect is statistically supported, not merely unrejected. It complements the non-significant prior-AI-engagement coefficient reported in Extended Data Table~\ref{tab:ols_combined} and Figure~\ref{fig:judgment}d.}
\label{tab:tost_ai_use} \\
\toprule
 & (1) Unadjusted & (2) Adjusted & (3) Adjusted \\
\midrule
\endfirsthead
\multicolumn{4}{l}{\tablename\ \thetable{} -- continued from previous page} \\
\toprule
 & (1) Unadjusted & (2) Adjusted & (3) Adjusted \\
\midrule
\endhead
\midrule
\multicolumn{4}{r}{\textit{Continued on next page}} \\
\endfoot
\bottomrule
\endlastfoot
Prior AI engagement (per 1 SD) & $0.033$           & $-0.007$           & $-0.007$ \\
                             & $(0.020)$         & $(0.015)$          & $(0.015)$ \\
90\% confidence interval     & $[0.001,\,0.066]$ & $[-0.031,\,0.018]$ & $[-0.031,\,0.018]$ \\
$p$ (two-sided)              & $0.089$           & $0.657$            & $0.657$ \\
Equivalence bound $\Delta$   & $\pm 0.192$       & $\pm 0.192$        & $\pm 0.150$ \\
$p_{\text{TOST}}$            & $<0.001$          & $<0.001$           & $<0.001$ \\
Equivalence supported        & Yes$^{\dagger}$   & Yes$^{\dagger}$    & Yes$^{\dagger}$ \\
Smallest supportive $\Delta$ & $0.066$           & $0.031$            & $0.031$ \\
\midrule
Covariates                   & No     & Yes    & Yes \\
Observations (ideas)         & 23,615 & 23,615 & 23,615 \\
\midrule
\multicolumn{4}{p{0.95\textwidth}}{\footnotesize The dependent variable is willingness
to pursue. Coefficients are rescaled to a one-standard-deviation change in log
prior AI engagement, so that estimates and equivalence bounds share the same units (this rescaling, and not a change of specification, is why the adjusted
estimate reads $-0.007$ here and $-0.1697$ in Extended Data
Table~\ref{tab:ols_combined}: the two are the same coefficient, the former
multiplied by $\mathrm{SD}(\log\text{ prior AI engagement}) = 0.0383887$). Clustered standard errors are in parentheses. Adjusted models include field indicators, seniority, and the within- and between-researcher components of
novelty, feasibility, probability, and human-AI idea similarity. Model~1 drops all covariates and is reported only to
show that the bound is not an artifact of conditioning. Models~2 and 3 are the
same fitted model evaluated
against two different equivalence bounds, reported to show that the conclusion
does not depend on the choice of bound. Equivalence is assessed by two one-sided
tests (TOST) at $\alpha = 0.05$, so the reported intervals are 90\%
($=1-2\alpha$) rather than 95\%. The bound $\Delta = 0.192$ is $0.1$ standard
deviations of the willingness scale ($\mathrm{SD} = 1.92$). ``Smallest supportive $\Delta$'' is the tightest bound at which equivalence would
still be supported.} \\
\end{longtable}

\clearpage
\phantomsection
\addcontentsline{toc}{subsection}{SI Table~7: Field differences in willingness to pursue across the seniority distribution}
\begin{longtable}{lccc}
\caption{\textbf{Field differences in willingness to pursue across the seniority distribution}. Field gaps emerge with seniority rather than being present throughout, consistent with the interaction reported in Extended Data Table~\ref{tab:ols_results34} and Figure~\ref{fig:judgment}f.}
\label{tab:field_gaps_by_status} \\
\toprule
Contrast & Junior (T1) & Senior (T3) & Change (T3$-$T1) \\
\midrule
\endfirsthead
\multicolumn{4}{@{}l}{\textit{SI Table~7 (continued)}} \\
\toprule
Contrast & Junior (T1) & Senior (T3) & Change (T3$-$T1) \\
\midrule
\endhead
\midrule
\multicolumn{4}{r@{}}{\textit{Continued on next page}} \\
\endfoot
\bottomrule
\addlinespace[2pt]
\multicolumn{4}{@{}p{\textwidth}@{}}{\footnotesize \textit{Notes.} OLS contrasts in willingness to pursue (SD $= 1.92$); standard errors clustered in parentheses. $N = 23,615$ ratings. All models control for prior AI engagement and the within- and between-respondent ratings of novelty, feasibility, success probability and AI-human idea similarity. Seniority is the respondent's citation percentile entered as terciles; T1 and T3 denote the junior and senior terciles. Gaps are estimated within strata rather than by extrapolating a linear interaction to a citation percentile of zero, which lies outside the observed support (minimum $= 0.006$). The field~$\times$~seniority interaction is significant, $F = 4.08$, $p = 0.007$, and survives a wild cluster bootstrap-$t$ with the null imposed and Rademacher weights ($B = 999$, $p = 0.010$); a fully flexible tercile specification points the same way, $F = 2.08$, $p = 0.053$. In the junior tercile, the $90\%$ intervals place every field difference below $0.224$ points ($0.117$~SD) and beneath the lower limit of the largest senior gap ($0.233$). We report this as an upper bound: at $80\%$ power the design detects differences of $0.161$~SD, so it cannot adjudicate margins tighter than that. $^{*}p<.05$, $^{**}p<.01$, $^{***}p<.001$ (two-sided).}
\endlastfoot
Bio $-$ Chem & -0.069 (0.094) & -0.039 (0.090) & 0.030 (0.130) \\
Bio $-$ Med & -0.077 (0.072) & -0.126 (0.074) & -0.048 (0.103) \\
Bio $-$ Soc & -0.040 (0.065) & 0.248$^{***}$ (0.062) & 0.288$^{**}$ (0.089) \\
Chem $-$ Med & -0.008 (0.110) & -0.086 (0.107) & -0.078 (0.154) \\
Chem $-$ Soc & 0.030 (0.105) & 0.288$^{**}$ (0.099) & 0.258 (0.145) \\
Med $-$ Soc & 0.038 (0.087) & 0.374$^{***}$ (0.086) & 0.336$^{**}$ (0.121) \\
\addlinespace
\midrule
Largest $|$gap$|$ & 0.077 & 0.374 & 0.336 \\
Significant contrasts & 0 of 6 & 3 of 6 & 2 of 6 \\
$90\%$ interval excludes above & 0.224 & --- & --- \\
\end{longtable}

\clearpage
\phantomsection
\addcontentsline{toc}{subsection}{SI Table~8: Seniority predicts willingness to pursue but not quality evaluation}
\begin{longtable}{lcccc}
\caption{\textbf{Seniority predicts willingness to pursue but not quality evaluation}. The seniority effect is specific to willingness to pursue and does not extend to how ideas are assessed, supporting the dissociation shown in Extended Data Figure~\ref{fig:percep}.}
\label{tab:seniority_dissociation} \\
\toprule
 & (1) Willingness & (2) Novelty & (3) Feasibility & (4) Probability \\
\midrule
\endfirsthead
\multicolumn{5}{l}{\tablename\ \thetable{} -- continued from previous page} \\
\toprule
 & (1) Willingness & (2) Novelty & (3) Feasibility & (4) Probability \\
\midrule
\endhead
\midrule
\multicolumn{5}{r}{\textit{Continued on next page}} \\
\endfoot
\bottomrule
\endlastfoot
\multicolumn{5}{l}{\textit{Panel A. Effect of seniority (standard-deviation units)}} \\[2pt]
Seniority (citation pct.) & $-0.119^{***}$ & $-0.038$  & $0.034$   & $-0.050$ \\
                          & $(0.036)$     & $(0.035)$ & $(0.034)$ & $(0.033)$ \\
90\% confidence interval  & $[-0.179,\,-0.060]$ & $[-0.096,\,0.019]$ & $[-0.022,\,0.091]$ & $[-0.104,\,0.003]$ \\
$p$ (two-sided)           & $0.001$ & $0.275$ & $0.317$ & $0.123$ \\
\midrule
\multicolumn{5}{l}{\textit{Panel B. Difference from the willingness-to-pursue slope}} \\[2pt]
$\Delta$ vs.\ willingness & --- & $0.081^{*}$ & $0.154^{***}$ & $0.069^{*}$ \\
                          &     & $(0.038)$   & $(0.036)$     & $(0.029)$ \\
$p$ (two-sided)           & --- & $0.033$     & $<0.001$      & $0.017$ \\[2pt]
\multicolumn{5}{l}{Joint test that all three differences are zero: $F = 6.16$, $p < 0.001$} \\
\midrule
\multicolumn{5}{l}{\textit{Panel C. Equivalence tests against $\Delta = \pm 0.119$ SD}} \\[2pt]
$p_{\text{TOST}}$         & --- & $0.010$ & $0.007$ & $0.018$ \\
Equivalence supported     & --- & Yes$^{\dagger}$ & Yes$^{\dagger}$ & Yes$^{\dagger}$ \\
Smallest supportive bound & --- & $0.096$ & $0.091$ & $0.104$ \\
\midrule
Observations (ideas) & 23,615 & 23,615 & 23,615 & 23,615 \\
\midrule
\multicolumn{5}{p{0.95\textwidth}}{\footnotesize Each outcome is standardized so
that coefficients are directly comparable across columns; all four models share
an identical right-hand side comprising seniority (citation percentile by default), prior AI engagement, field indicators, and the within- and
between-researcher components of human-AI idea similarity, with social science as the omitted category (the reference). Standard
errors are clustered and reported in parentheses. Panel~B is
estimated from a single stacked model in which the four standardized outcomes
are fully interacted with outcome identity; this reproduces the column-wise
slopes in Panel~A exactly while supplying the joint covariance needed to test
cross-outcome contrasts. Panel~C reports two one-sided tests against an
equivalence bound set at the magnitude of the seniority effect on willingness to pursue
($0.119$~SD), so that the null hypothesis being rejected is that seniority moves
quality ratings as much as it moves willingness to pursue; ``smallest supportive bound'' is
the tightest bound at which equivalence would still hold. The bound is anchored to the willingness-to-pursue slope in Panel~A, since it is estimated with the same right-hand side as the quality columns it is compared against. Panel~C is a sensitivity check rather than the primary test, which is Panel~B; because the bound is estimated from the same sample, equivalence does not hold against a bound set by the quality-adjusted seniority coefficient. Equivalence is assessed at $\alpha = 0.05$, so 90\% ($=1-2\alpha$) intervals are reported throughout. $^{***}p<0.001$, $^{**}p<0.01$, $^{*}p<0.05$.} \\
\end{longtable}

\clearpage
\begin{table}[htbp]
\phantomsection
\addcontentsline{toc}{subsection}{SI Table~9: Null-hypothesis rates under three inputs}
\centering
\caption{\revise{\textbf{Null-hypothesis rates under three inputs.} Rates are sorted by the context-and-puzzle (C.P.) setting. C.P., context and puzzle as input; Intro., Introduction as input; I.R., Introduction and Related Work as input; the next three columns are differences between settings, and the last column is the $p$-value of a $\chi^2$ test of independence across the three settings. $n$ = 10000 papers. All model rates are low and well below the human rate.}}
\label{tab:expand_null}
\begin{revblock}
\small
\setlength{\tabcolsep}{5pt}
\begin{tabular}{lrrrrrrr}
\toprule
Model & C.P. & Intro. & I.R. & Intro. vs C.P. & Intro. vs I.R. & I.R. vs C.P. & $p$-value \\
\midrule
Human null hypothesis   & 0.0103 & 0.0103 & 0.0103 & 0.0000    & 0.0000    & 0.0000    & 1.0000 \\
\midrule
Llama 3.1 8B Instruct   & 0.0033 & 0.0040 & 0.0026 & 0.0007    & $-$0.0007 & $-$0.0014 & 0.2254 \\
Gemma 3 12B Instruct    & 0.0032 & 0.0026 & 0.0052 & $-$0.0006 & 0.0020    & 0.0026    & 0.0063 \\
Gemini 2.5 Flash-Lite   & 0.0026 & 0.0024 & 0.0034 & $-$0.0002 & 0.0008    & 0.0010    & 0.3668 \\
Gemini 2.5 Flash        & 0.0024 & 0.0022 & 0.0027 & $-$0.0002 & 0.0003    & 0.0005    & 0.7704 \\
DeepSeek-V3 (0324)      & 0.0020 & 0.0014 & 0.0011 & $-$0.0006 & $-$0.0009 & $-$0.0003 & 0.2461 \\
Gemma 3 27B Instruct    & 0.0016 & 0.0020 & 0.0022 & 0.0004    & 0.0006    & 0.0002    & 0.6165 \\
Llama 3.1 70B Instruct  & 0.0015 & 0.0013 & 0.0023 & $-$0.0002 & 0.0008    & 0.0010    & 0.1921 \\
GPT-4o mini             & 0.0010 & 0.0007 & 0.0005 & $-$0.0003 & $-$0.0005 & $-$0.0002 & 0.4214 \\
Phi-4                   & 0.0009 & 0.0008 & 0.0008 & $-$0.0001 & $-$0.0001 & 0.0000    & 0.9608 \\
Gemma 3 4B Instruct     & 0.0007 & 0.0012 & 0.0018 & 0.0005    & 0.0011    & 0.0006    & 0.0852 \\
GPT-4.1                 & 0.0006 & 0.0009 & 0.0010 & 0.0003    & 0.0004    & 0.0001    & 0.5943 \\
DeepSeek-R1 (0528)      & 0.0005 & 0.0007 & 0.0003 & 0.0002    & $-$0.0002 & $-$0.0004 & 0.4491 \\
o3                      & 0.0002 & 0.0000 & 0.0001 & $-$0.0002 & $-$0.0001 & 0.0001    & 0.3678 \\
GPT-4o                  & 0.0002 & 0.0003 & 0.0005 & 0.0001    & 0.0003    & 0.0002    & 0.4965 \\
o4-mini                 & 0.0002 & 0.0002 & 0.0003 & 0.0000    & 0.0001    & 0.0001    & 0.8668 \\
o3-mini                 & 0.0001 & 0.0000 & 0.0000 & $-$0.0001 & $-$0.0001 & 0.0000    & 0.3679 \\
o1                      & 0.0001 & 0.0001 & 0.0001 & 0.0000    & 0.0000    & 0.0000    & 1.0000 \\
Deep Research           & 0.0001 & 0.0001 & 0.0001 & 0.0000    & 0.0000    & 0.0000    & 1.0000 \\
\bottomrule
\end{tabular}
\end{revblock}
\end{table}

\clearpage
\begin{table}[htbp]
\phantomsection
\addcontentsline{toc}{subsection}{SI Table~10: Null-hypothesis rates on journal articles without preprints}
\centering
\caption{\revise{\textbf{Share of hypotheses predicting a null result on 2026 journal articles without a preprint version}, by source ($n$ = 10{,}000 contexts and puzzles).}}
\label{tab:null_published}
\begin{revblock}
\begin{tabular}{lr}
\toprule
Source & Null rate \\
\midrule
Human hypothesis        & 0.0101 \\
\midrule
o1                      & 0.0000 \\
Deep Research           & 0.0000 \\
o3-mini                 & 0.0001 \\
o4-mini                 & 0.0002 \\
o3                      & 0.0002 \\
GPT-4o                  & 0.0002 \\
DeepSeek-R1 (0528)      & 0.0005 \\
GPT-4.1                 & 0.0006 \\
Gemma 3 4B Instruct     & 0.0006 \\
Phi-4                   & 0.0007 \\
GPT-4o mini             & 0.0010 \\
Gemma 3 27B Instruct    & 0.0014 \\
Llama 3.1 70B Instruct  & 0.0015 \\
DeepSeek-V3 (0324)      & 0.0016 \\
Gemini 2.5 Flash-Lite   & 0.0022 \\
Gemini 2.5 Flash        & 0.0025 \\
Llama 3.1 8B Instruct   & 0.0029 \\
Gemma 3 12B Instruct    & 0.0035 \\
\bottomrule
\end{tabular}
\end{revblock}
\end{table}

\clearpage
\begin{table}[htbp]
\phantomsection
\addcontentsline{toc}{subsection}{SI Table~11: Model null hypotheses when the human hypothesis is null}
\centering
\caption{\revise{\textbf{Share of contexts and puzzles with a human null hypothesis for which each model also proposed a null hypothesis.}}}
\label{tab:null_overlap}
\begin{revblock}
\small
\begin{tabular}{rlr}
\toprule Rank & Model & Overlap with humans' null hypotheses \\ \midrule
1 & Ministral 3B & 1.85\% \\
2 & Mixtral 8$\times$7B MoE & 1.05\% \\
3 & Gemma 3 12B & 0.99\% \\
4 & Llama 3.1 405B & 0.92\% \\
5 & Phi-3-mini & 0.90\% \\
6 & Gemma 3 27B & 0.77\% \\
7 & Grok-3 mini & 0.71\% \\
8 & o3 & 0.57\% \\
9 & Gemini 2.0 Flash & 0.51\% \\
10 & o3-mini & 0.50\% \\
11 & Gemini 2.0 Flash-Lite & 0.50\% \\
12 & GPT-4o & 0.48\% \\
13 & o1 & 0.40\% \\
14 & GPT-4.1 & 0.40\% \\
15 & Qwen-Turbo & 0.40\% \\
16 & Llama 3.1 8B & 0.40\% \\
17 & o4-mini & 0.31\% \\
18 & Deep Research & 0.31\% \\
19 & Grok-3 & 0.30\% \\
20 & DeepSeek-V3 & 0.29\% \\
21 & DeepSeek-R1 (0528) & 0.20\% \\
22 & Gemma 3 4B & 0.20\% \\
23 & GPT-4o mini & 0.19\% \\
24 & Phi-4 & 0.00\% \\
25 & Llama 3.1 70B & 0.00\% \\
\bottomrule
\end{tabular}
\end{revblock}
\end{table}

\clearpage
\begin{table}[htbp]
\phantomsection
\addcontentsline{toc}{subsection}{SI Table~12: Lexical coverage of field inputs in a pretraining corpus}
\centering
\caption{\revise{\textbf{Lexical coverage of context-and-puzzle inputs in Dolma-v1.6-sample}, by field (100 random papers per field; field-level means of paper-level measures). hit$_n$, share of the input's $n$-grams found at least once; logcount$_n$, mean $\log_{10}(\mathrm{count}+1)$ over the input's $n$-grams.}}
\label{tab:dolma}
\begin{revblock}
\small
\setlength{\tabcolsep}{5pt}
\begin{tabular}{lrrrrrr}\toprule
Field & hit$_3$ & hit$_4$ & hit$_5$ & logcount$_3$ & logcount$_4$ & logcount$_5$ \\ \midrule
Biology & 0.660 & 0.352 & 0.155 & 1.344 & 0.508 & 0.171 \\
Chemistry & 0.649 & 0.343 & 0.144 & 1.332 & 0.501 & 0.165 \\
Medicine & 0.722 & 0.407 & 0.182 & 1.493 & 0.589 & 0.204 \\
Social science & 0.714 & 0.386 & 0.163 & 1.508 & 0.553 & 0.174 \\
\bottomrule\end{tabular}
\end{revblock}
\end{table}

\clearpage
\begin{table}[htbp]
\phantomsection
\addcontentsline{toc}{subsection}{SI Table~13: Alternative measures of human--AI idea similarity}
\centering
\caption{\revise{\textbf{Willingness to pursue under alternative measures of human--AI idea similarity.} Model 1 uses SPECTER2 embeddings; Model 2 uses SPECTER2 displacement vectors (the shared context-and-puzzle embedding subtracted). Cluster-robust standard errors (CRV1) in parentheses. ``Within'' and ``between'' denote the within- and between-scientist components of each variable. Social science is the reference field. $^{*}p<0.05$; $^{**}p<0.01$; $^{***}p<0.001$.}}
\label{tab:ols_pursue}
\begin{revblock}
\small
\begin{tabular}{lcc}
\toprule
 & Model 1 (SPECTER2) & Model 2 (Displacement + SPECTER2) \\
\midrule
\multicolumn{3}{l}{\textit{Field (ref.\ = Social science)}} \\
\quad Biology            & 0.121$^{***}$ & 0.143$^{***}$ \\
                         & (0.038)       & (0.037) \\
\quad Chemistry          & 0.177$^{**}$  & 0.171$^{**}$ \\
                         & (0.061)       & (0.061) \\
\quad Medicine           & 0.210$^{***}$ & 0.232$^{***}$ \\
                         & (0.049)       & (0.049) \\[3pt]
\multicolumn{3}{l}{\textit{Novelty}} \\
\quad Within             & 0.121$^{***}$ & 0.117$^{***}$ \\
                         & (0.006)       & (0.006) \\
\quad Between            & 0.221$^{***}$ & 0.221$^{***}$ \\
                         & (0.010)       & (0.010) \\[3pt]
\multicolumn{3}{l}{\textit{Feasibility}} \\
\quad Within             & 0.131$^{***}$ & 0.133$^{***}$ \\
                         & (0.006)       & (0.006) \\
\quad Between            & 0.070$^{***}$ & 0.074$^{***}$ \\
                         & (0.011)       & (0.011) \\[3pt]
\multicolumn{3}{l}{\textit{Probability}} \\
\quad Within             & 0.312$^{***}$ & 0.317$^{***}$ \\
                         & (0.006)       & (0.006) \\
\quad Between            & 0.445$^{***}$ & 0.445$^{***}$ \\
                         & (0.012)       & (0.012) \\[3pt]
\multicolumn{3}{l}{\textit{AI--human idea similarity}} \\
\quad Within             & 7.909$^{***}$ & 0.679$^{***}$ \\
                         & (0.702)       & (0.115) \\
\quad Between            & 2.889$^{***}$ & 0.112 \\
                         & (0.827)       & (0.129) \\[3pt]
Academic age             & $-0.038^{*}$  & $-0.038^{*}$ \\
                         & (0.017)       & (0.017) \\
Prior AI engagement      & $-0.304$      & $-0.371$ \\
                         & (0.385)       & (0.387) \\
Constant                 & $-3.299^{***}$ & $-0.687^{***}$ \\
                         & (0.761)        & (0.089) \\
\midrule
Observations             & 23,910 & 23,865 \\
$R^2$                    & 0.394  & 0.392 \\
Adjusted $R^2$           & 0.394  & 0.392 \\
F-statistic              & 895.0$^{***}$ & 879.4$^{***}$ \\
\bottomrule
\end{tabular}
\end{revblock}
\end{table}

\clearpage

\clearpage
\clearpage
\printrefsegment{si}{\revise{Supplementary references}}

\end{document}